\documentclass[prx,twocolumn,superscriptaddress]{revtex4-1}

 \usepackage{lineno,hyperref} 
 \usepackage{verbdef}
\usepackage{graphicx}
\usepackage{upgreek}
\usepackage{subfigure}

\usepackage{color,colortbl} 
\definecolor{Gray}{gray}{0.9}
\usepackage{siunitx}
\usepackage{soul}
\sethlcolor{yellow}

\usepackage{pgfplots}
\usetikzlibrary{pgfplots.groupplots}

\usepackage{cancel}

 \usepackage{amsmath}
 \usepackage{physics}
\usepackage{float}
\usepackage{amssymb}

\newcommand{\msr}{$\mu$SR }

\begin{document}
\sisetup{range-phrase=--}



\title{Observation of a charge-neutral muon-polaron  complex in antiferromagnetic Cr$_2$O$_3$}



\author{M. H. Dehn}
\email[]{mdehn@triumf.ca}
\affiliation{Department of Physics and Astronomy, University of British Columbia,
Vancouver, BC V6T 1Z1, Canada }
\affiliation{Stewart Blusson Quantum Matter Institute, University of British Columbia, Vancouver, BC V6T 1Z4, Canada}
\affiliation{\textsc{Triumf},  Vancouver, BC V6T 2A3, Canada}

\author{J. K. Shenton}
\affiliation{Department of Materials, ETH Z\"urich, CH-8093 Z\"urich, Switzerland}

\author{S. Holenstein}
\affiliation{Physik-Institut der Universit\"at Z\"urich,  CH-8057, Z\"urich, Switzerland}
\affiliation{Laboratory for Muon Spin Spectroscopy, Paul Scherrer Institut, 5232, Villigen PSI, Switzerland}


\author{Q. N. Meier}
\affiliation{Department of Materials, ETH Z\"urich, CH-8093 Z\"urich, Switzerland}

\author{D. J. Arseneau}
\affiliation{\textsc{Triumf},  Vancouver, BC V6T 2A3, Canada}

\author{D. L. Cortie}
\altaffiliation[Present address: ]{Institute for Superconducting and Electronic Materials, Australian Institute for Innovative Materials, University
of Wollongong, North Wollongong, NSW 2500, Australia}
\affiliation{Department of Physics and Astronomy, University of British Columbia,
Vancouver, BC V6T 1Z1, Canada }
\affiliation{Stewart Blusson Quantum Matter Institute, University of British Columbia, Vancouver, BC V6T 1Z4, Canada}
\affiliation{Department of Chemistry, University of British Columbia, Vancouver, BC, V6T 1Z1, Canada}


\author{B. Hitti}
\affiliation{\textsc{Triumf},  Vancouver, BC V6T 2A3, Canada}

\author{A. C. Y. Fang}
\altaffiliation[Present address: ]{Department of Physics, Simon Fraser University, Burnaby,  Canada V5A 1S6}
\affiliation{Department of Physics and Astronomy, University of British Columbia,
Vancouver, BC V6T 1Z1, Canada }

\author{W. A. MacFarlane}
\affiliation{Stewart Blusson Quantum Matter Institute, University of British Columbia, Vancouver, BC V6T 1Z4, Canada}
\affiliation{\textsc{Triumf},  Vancouver, BC V6T 2A3, Canada}
\affiliation{Department of Chemistry, University of British Columbia, Vancouver, BC, V6T 1Z1, Canada}

\author{R. M. L. McFadden}
\affiliation{Stewart Blusson Quantum Matter Institute, University of British Columbia, Vancouver, BC V6T 1Z4, Canada}
\affiliation{Department of Chemistry, University of British Columbia, Vancouver, BC, V6T 1Z1, Canada}

\author{G. D. Morris}
\affiliation{\textsc{Triumf},  Vancouver, BC V6T 2A3, Canada}

\author{Z. Salman}
\affiliation{Laboratory for Muon Spin Spectroscopy, Paul Scherrer Institut, 5232, Villigen PSI, Switzerland}

\author{H. Luetkens}
\affiliation{Laboratory for Muon Spin Spectroscopy, Paul Scherrer Institut, 5232, Villigen PSI, Switzerland}

\author{N. A. Spaldin}
\affiliation{Department of Materials, ETH Z\"urich, CH-8093 Z\"urich, Switzerland}

\author{M. Fechner}
\affiliation{Department of Materials, ETH Z\"urich, CH-8093 Z\"urich, Switzerland}
\affiliation{Max Planck Institute for the Structure and Dynamics of Matter, 22761 Hamburg, Germany}

\author{R. F. Kiefl}
\affiliation{Department of Physics and Astronomy, University of British Columbia,
Vancouver, BC V6T 1Z1, Canada }
\affiliation{Stewart Blusson Quantum Matter Institute, University of British Columbia, Vancouver, BC V6T 1Z4, Canada}
\affiliation{\textsc{Triumf},  Vancouver, BC V6T 2A3, Canada}


\date{\today}

\begin{abstract}

We report a comprehensive  muon spin rotation ($\mu$SR) study of the prototypical magnetoelectric  antiferromagnet Cr$_2$O$_3$.   We find the positively charged muon ($\mu^+$) occupies several distinct interstitial sites, and displays a rich dynamic behavior involving local hopping, thermally activated site transitions and the formation of a charge-neutral complex composed of a muon and an electron polaron. The discovery of such a complex has implications for the interpretation of $\mu$SR spectra in a wide range of magnetic oxides, and opens a route to study the dopant characteristics of interstitial hydrogen impurities in such materials.
We address  implications arising from implanting a $\mu^+$ into a linear magnetoelectric, and discuss the challenges of observing a local magnetoelectric effect generated by the charge of the muon.

\end{abstract}

\pacs{}

\maketitle

\section{Introduction}
Positively charged muons implanted into semiconductors and insulators often form muonium ($\mathrm{Mu}=[\mu^+e^-]$), a hydrogen-like charge-neutral  bound state. It is conventionally referred to as a \emph{paramagnetic} center as the bound electron is unpaired and its spin is decoupled from all the other electrons.
 Since the electronic structure of Mu in a solid is virtually identical to that of hydrogen, Mu has been studied extensively using muon spin spectroscopy ($\mu$SR) to  learn about interstitial hydrogen, one of the most ubiquitous defects in semiconductors. In a $\mu$SR  experiment, spin polarized muons ($\mu^+$) are implanted into the sample of interest and the subsequent time evolution of the spin polarization is observed, providing an accurate measurement of the magnetic coupling of the muon spin with its environment \cite{yaouanc2011}. 
Mu centers have been studied in a wide range of materials,  providing direct information about the electronic  structure of hydrogen defects as shallow or deep-level dopants  \cite{chow1998,cox2006,cox2006a,cox2009}.
So far, the study of such charge-neutral muon states has been limited to non-magnetic materials, since  paramagnetic Mu is widely assumed to be subject to strong depolarization in the presence of magnetic moments such as those on Cr$^{3+}$ \cite{cox2006}, and, with the 
exception of MnF$_2$ \cite{uemura1986}, no Mu has been confirmed in magnetic materials.

Here, we present strong evidence for a charge-neutral muon state in the antiferromagnet Cr$_2$O$_3$. In particular, our data, in conjunction with detailed density functional theory (DFT) calculations, make a compelling case for the existence of a muon-polaron complex, where the positive muon is bound to an oxygen and an excess electron localizes on a nearby Cr ion, changing its valence to Cr$^{2+}$ ($3d^4$). The degeneracy of the now occupied e$_g$ orbital is lifted via a lattice distortion, leading to a Jahn-Teller (JT) polaron  \cite{jahn1937,holstein1959,nickisch1983a} on the Cr ion. 
Crucially, the resulting JT-stabilized muon-polaron complex  is \emph{not paramagnetic} 
and therefore \emph{distinct from Mu}, since the bound electron is strongly coupled to the $3d$ electrons of the Cr host ion. 
Therefore, no signatures conventionally associated with a charge-neutral state are displayed, concealing  its existence. However, in spite of its inconspicuous signal, the presence of such a complex  has a significant impact on the location and stability of muon stopping sites, and the local fields  experienced there.

This discovery of a charge-neutral muon-polaron complex in Cr$_2$O$_3$ suggests that neutral charge states could form in other insulating magnetic materials as well, which has implications for the interpretation of a wide range of $\mu$SR data.  Furthermore, analogous to Mu in semiconductors, the study of muon-polaron complexes in magnetic oxides may provide detailed information on the dopant characteristics of interstitial hydrogen, a good understanding of which is crucial for a precise control of charge carriers in such materials.

Cr$_2$O$_3$ is of additional interest due to its magnetoelectric properties. Being the first material predicted  \cite{dzyaloshinskii1960} and measured  \cite{astrov1961,rado1961a} to exhibit an induced linear polarization (magnetization) in response to a magnetic (electric) field, it is widely regarded as \emph{the} prototypical linear magnetoelectric \cite{
wiegelmann1994,fiebig2005} and remains the subject of active research \cite{meier2019}, 
directed primarily at exploiting its magnetoelectric properties for device applications \cite{borisov2005,he2010,kosub2017}. 
In addition,  there are unresolved fundamental  questions  raised by   the recent prediction that an electric  charge within a linear magnetoelectric is surrounded by a monopolar magnetic field distribution, and thus is subject to a magnetic force in an external magnetic field \cite{khomskii2014}.  $\mu$SR is a unique way to investigate such predictions since the spin polarized muon acts both
as a test charge and a sensitive probe of the local magnetic field. 
However, as indicated by  studies  from the early days of  $\mu$SR \cite{ruegg1979,ruegg1980,boekema1981,boekema1983}, the  spectra in Cr$_2$O$_3$ are complex and their interpretation was inconclusive. 
Furthermore, given the weak magnetoelectric coupling in Cr$_2$O$_3$, 
only subtle changes to the local magnetic environment in response to the muon charge are expected, and a thorough understanding of the interaction between the implanted muon and host material is required as a prerequisite for the search for any muon-induced magnetoelectric effects.

The paper proceeds as follows. In Sec. \ref{sec:expdetails}, we briefly introduce the  $\mu$SR technique and summarize the experimental conditions. Next, in Sec. \ref{sec:results}, we report the results of a comprehensive $\mu$SR study of Cr$_2$O$_3$ under zero-field conditions and in applied magnetic fields. The data are presented in three parts: (1) In zero field, up to three   spin precession frequencies are observed, indicating three distinct muon environments with different internal magnetic fields $\mathbf{B}_{int}$. (2) Weak external  fields $\mathbf{B}_{ext}(\ll \mathbf{B}_{int})$ split the observed frequencies into multiplets, providing detailed information on the orientation of the internal fields. (3) Large applied fields ($\mathbf{B}_{ext}>\mathbf{B}_{int}$) corroborate the weak field results and reveal an additional frequency. 
Together, the data exhibit a rich variety of dynamic phenomena 
that we explain in terms of site metastability and muon dynamics (Sec. \ref{sec:disc}).
Most importantly, above $\sim$\,\SI{150}{K}, we observe both highly dynamic muons undergoing locally restricted hopping and muons that remain static in their site.
In order to explain this surprising behavior, 
we turn to DFT to identify candidate  muon  sites for all three environments, and conclude that the coexistence of site-stable and dynamic muons can be explained with the formation of a charge-neutral Jahn-Teller-stabilized muon-polaron complex (Sec. \ref{sec:dft}). 
Finally, in 
Sec. \ref{sec:discussion}, we discuss the implications of charge-neutral states in Cr$_2$O$_3$ and its relevance for other magnetic oxides, and possible consequences arising from implanting positively charged muons into a linear magnetoelectric.

\section{experimental details}
\label{sec:expdetails}

The \msr experiments reported on here were carried out at the Centre for Molecular and Materials Science at TRIUMF  (Vancouver, Canada), although initial spectra were taken at the GPS instrument at PSI (Villigen, Switzerland). The zero and low magnetic field measurements were taken in the LAMPF spectrometer and the high magnetic field data were acquired in  the NuTime spectrometer. 
All data were acquired with
the initial muon spin polarization $\mathbf{P}_i$ 
perpendicular to the beam direction ($\hat{z}$).

In a \msr experiment, spin polarized, positively charged muons are implanted into the sample, where they decay with a lifetime of $\tau_\mu=\SI{2.2}{\micro\s}$. The resulting decay positron is emitted preferentially along the muon spin direction, and can be detected in a time-resolved manner with plastic scintillators placed in pairs around the sample. This anisotropic positron emission introduces a spin-dependent imbalance in the count rate, causing the asymmetry signal $S(t)$, the count-normalized difference of a counter pair, to be directly proportional to the spin polarization $P(t)$ along the detector axis. A detailed description of the \msr technique can be found in Ref.  \onlinecite{yaouanc2011}.
In the presence of a magnetic field \textbf{B}, the muon spin precesses about the field direction with frequency $f=\gamma_\mu / 2\pi \cdot |\mathbf{B}|$, where $\gamma_\mu=2\pi\cdot \SI{0.01355}{ MHz/G}$ is the muon gyromagnetic ratio. This allows for a direct measure of the local magnetic field experienced by the muon.

In a crystal lattice, the charged muon usually stops in one or more distinct sites that minimize the overall energy. At a given temperature, several crystallographically distinct sites may be populated, each of which causes a different time evolution of the spin polarization. For example in magnetic materials, muons may experience different internal fields at inequivalent sites, causing spin precession at different frequencies. In this case, the observed signal $S(t)$ is a sum of several components $S_i(t)$.
In this paper, oscillatory signals  are fit to exponentially damped cosines
\begin{equation}
S_i(t) =A_i \cos(2\pi f_i t+\phi_i) \exp(-\lambda_i t), 
\label{eqn:osc_fit}
\end{equation}
where the amplitude $A_i$ is a measure of the signal weight, $f_i$ is the frequency, $\phi_i$ the initial phase, and $\lambda_i$ the relaxation rate. Non-oscillatory components are parametrized by simple exponentials of the form $S_i(t)=A_i  \exp(-\lambda_i t)$. All data is fit with the  \textsf{musrfit} analysis framework  \cite{suter2012}.

Several single crystal specimens  
sourced from SurfaceNet (Rheine, Germany) were used: a $10\times\SI{10}{mm^2}$ single crystal (C1) with the \emph{c}-axis in plane and $[11\bar{2}0]$ out of plane, and  $8\times\SI{8}{mm^2}$ (C2) and $5\times\SI{5}{mm^2}$ (C3) single crystals  with the \emph{c}-axis out of plane. The ZF data were taken on C1 with the \emph{c}-axis oriented along $\hat{x}$  to coincide with the initial spin direction. Small external fields were applied to C1 ($\mathbf{B}_{ext}\parallel[11\bar{2}0]$) and C2 ($\mathbf{B}_{ext}\parallel c$). High field experiments were carried out on C3 ($\mathbf{B}_{ext}\parallel c$).

\section{Results}
\label{sec:results}

\begin{figure}[t]
 \includegraphics[width=8.5cm]{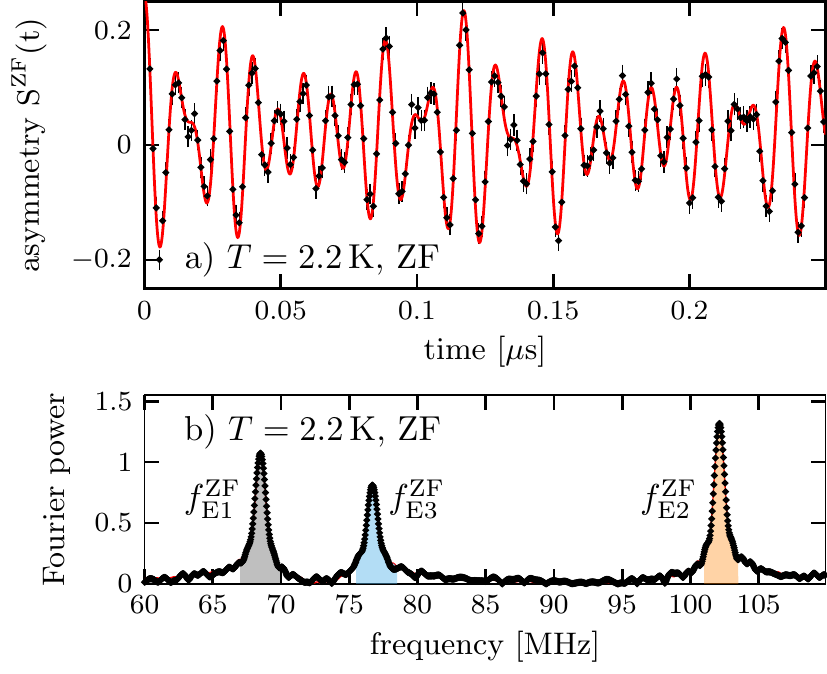}
 
 \includegraphics[width=8.5cm]{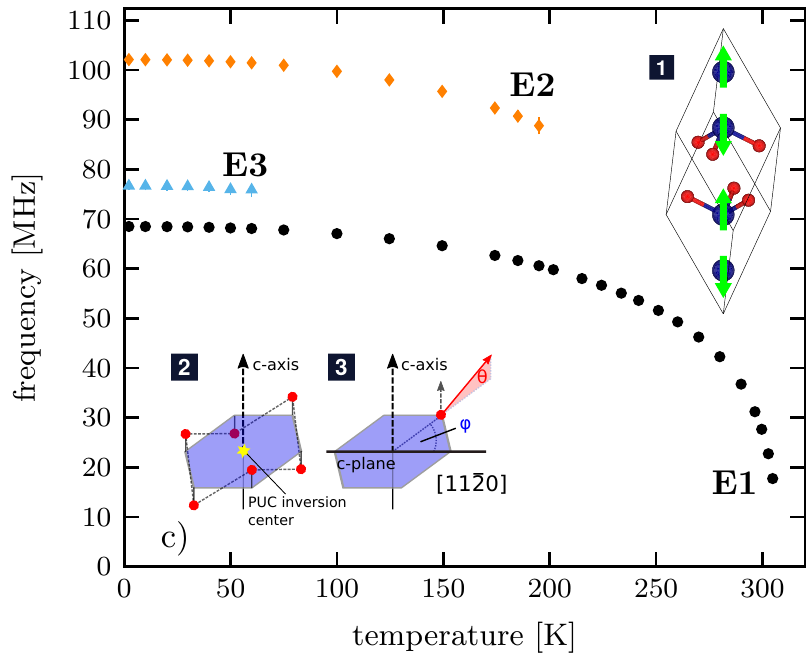}
\caption{(a) ZF \msr time-domain spectrum at $T=\SI{2.2}{K}$ and (b) its  Fourier transform. 
(c) Observed ZF precession frequencies assigned to three muon environments, E1-E3, as a function of temperature.  Insets:
(1) Primitive unit cell of Cr$_2$O$_3$ with 4 Cr atoms (blue) along the \emph{c}-axis, and 6 oxygen atoms. Below $T_N=\SI{307}{K}$, the Cr magnetic moments (green) align pairwise opposite along the \emph{c}-axis. 
 (2) Electrostatically equivalent stopping sites (red) form a hexagon when projected onto the c-plane through the inversion center. (3) Definition of angles describing the direction of the internal field at a given site.}.
\label{fig:ZFprec}
\end{figure}

The primitive unit cell  of Cr$_2$O$_3$ is rhombohedral and contains 4 Cr atoms and 6 O atoms, see inset 1 in Fig. \ref{fig:ZFprec}\,(c). The Cr  are arranged in two pairs along the rhombohedral 111-axis (\emph{c}-axis), with the oxygens forming two triangles, rotated  \SI{60}{\degree} with respect to each other, between the Cr pairs. In the absence of magnetic order, the primitive unit cell is inversion symmetric, and there is three-fold rotation symmetry around the \emph{c}-axis. 

In oxides, muons are generally  found to stop   ${\approx} \SI{1}{\angstrom}$ away from an oxygen, similar to the hydrogen in a hydroxyl OH bond \cite{denison1984,holzschuh1983a}. 
 Assuming this holds for Cr$_2$O$_3$, we can use symmetry arguments to make some general statements about potential muon stopping sites.
All six oxygens are crystallographically equivalent, thus any given muon sites close to one oxygen can be projected by either inversion or \SI{120}{\degree} rotations about \emph{c} into another equivalent site. Consequently, there are at least six (or integer multiples thereof) electrostatically equivalent stopping sites within the primitive unit cell, which, when projected onto the c-plane through the inversion center, form a hexagon, see inset 2 in Fig. \ref{fig:ZFprec}\,(c) and Fig. \ref{fig:hopsites}. 

Due to localized electrons in the Cr $3d$ shell, there is a magnetic moment associated
with each Cr. Below the N\'eel temperature $T_N=\SI{307}{K}$, those moments align pairwise opposite to each other along the \emph{c}-axis, see inset 1 in Fig. \ref{fig:ZFprec}\,(c),  causing an internal magnetic field $\mathbf{B}_{int}$ at the muon stopping sites. The magnetic structure breaks the inversion symmetry, $\mathbf{B}_{int}(\mathbf{r})=-\mathbf{B}_{int}(-\mathbf{r})$. As a consequence, the \emph{direction} of the internal fields associated with the various electrostatically equivalent sites is different.  However, as the Cr moments are  parallel to the \emph{c}, the \emph{magnitude} $|\mathbf{B}_{int}|$ at each of the sites is the same. Since only the magnitude determines the  precession frequency, muons that stop in any one of the equivalent sites in zero external field  precess with the same frequency and contribute to the same signal $S_i(t)$. From now on, we refer to an ensemble of electrostatically equivalent  sites that have the same  $|\mathbf{B}_{int}|$ as a \emph{muon environment}. Note that at a given temperature, muons may stop in \emph{different} environments with distinct $|\mathbf{B}_{int}|$.

We start with a presentation of the zero-field (ZF) results. Then, the effects of external magnetic fields $\mathbf{B}_{ext}$ are described, first for fields small compared to the internal field ($\mathbf{B}_{ext} \ll \mathbf{B}_{int}$), then for large fields ($\mathbf{B}_{ext} > \mathbf{B}_{int}$).

\subsection{Zero external field}
\label{sec:ZF}

\begin{figure}[t]
  \includegraphics[width=8.5cm]{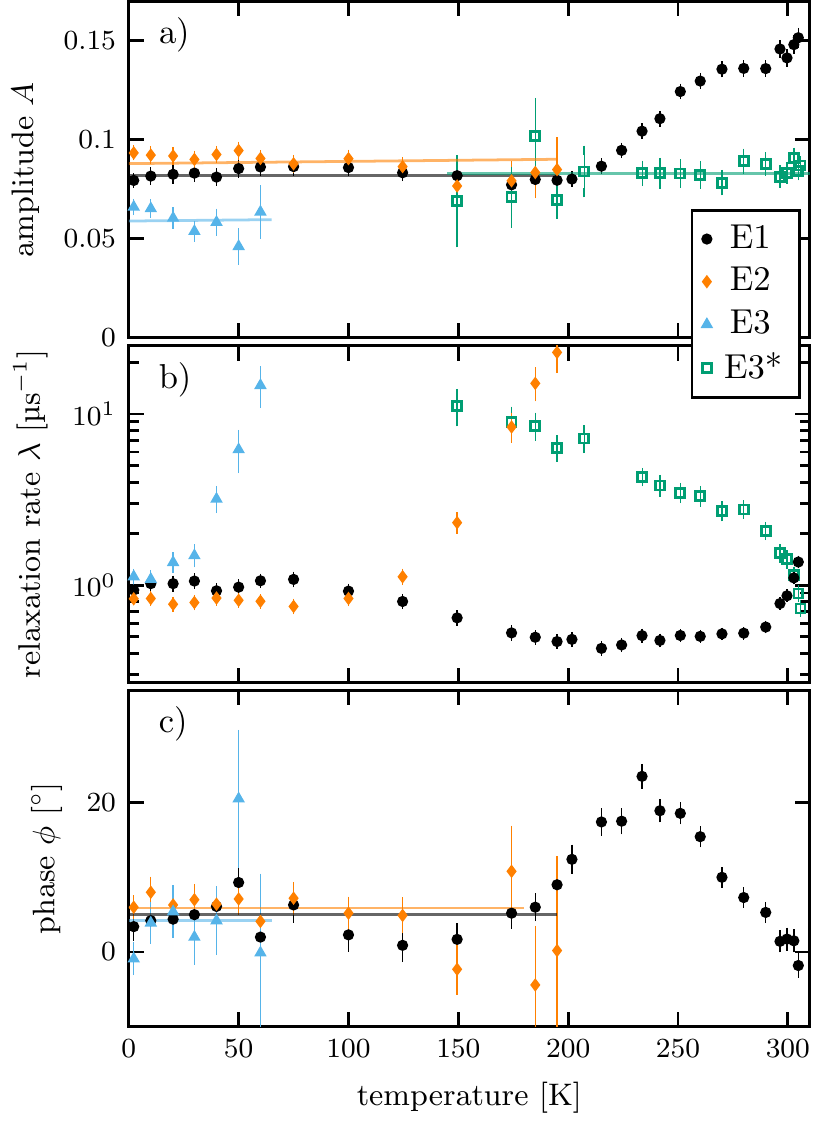}
\caption{Fit results for the three ZF oscillatory signals E1-E3  as a function of temperature: (a) amplitudes $A$ (b) relaxation rates $\lambda$ (logarithmic y-axis)  (c) phase $\phi$. Note the pronounced peak at around $\SI{240}{K}$ in the E1 phase, indicating a transition process between muon environments.  Lines are guides to the eye. Additionally, the amplitude and relaxation rate for E3* are shown, see Section \ref{sec:TF}. }
\label{fig:fitResZF}
\end{figure}

A ZF $\mu$SR spectrum showing the muon spin polarization as a function of time 
and its Fourier transform (FT) taken at $T=\SI{2.2}{K}$ are displayed in Figs. \ref{fig:ZFprec}(a) and (b). 
Three ZF precession frequencies $f^\mathrm{ZF}$ are observed, indicating three distinct muon environments, termed E1-E3. With increasing temperature, certain frequencies disappear, see Fig. \ref{fig:ZFprec}(c). The frequencies are assigned to the environments E1-E3 in order of appearance: the frequency observed up to $T_N$ is called $f^\mathrm{ZF}_\mathrm{E1}$, whereas  $f^\mathrm{ZF}_\mathrm{E2}$ can only be seen up to $\approx\SI{190}{K}$ and $f^\mathrm{ZF}_\mathrm{E3}$ up to  $\approx\SI{60}{K}$.
All  three precession frequencies increase with decreasing temperature, approximately tracking the sublattice magnetization.

The spectra contain both oscillating and non-oscillating components, and are fit with up to three  damped cosines, Eqn. (\ref{eqn:osc_fit}), a non-relaxing component  and a relaxing component (nonzero only above $\approx$\SI{160}{K}).

The fit results for the oscillatory components associated with E1-E3 are shown in Fig. \ref{fig:fitResZF}. The amplitude $A_\mathrm{E1}$ is constant up to \SI{200}{K}, above which it  increases and approximately doubles at $T_N$. Both $A_\mathrm{E2}$ and $A_\mathrm{E3}$ are approximately constant. 
The relaxation rates $\lambda_\mathrm{E2}$ and $\lambda_\mathrm{E3}$ increase sharply  when approaching the temperature where their associated ZF frequency vanishes. 
 While both  phases $\phi_\mathrm{E2}$ and $\phi_\mathrm{E3}$, shown in Fig. \ref{fig:fitResZF}(c), can be considered constant,  there is a pronounced peak in $\phi_\mathrm{E1}$ between \SI{200}{K} and $T_N$. As discussed  in Section \ref{sec:hop}, such a change in phase is indicative of a transition from another (so far unspecified) environment into E1, a hypothesis supported by the increase of $A_\mathrm{E1}$ at the same temperature.
Aside from the precession signals, there is a sizable non-oscillatory relaxing component that appears above  $\approx \SI{160}{K}$ (not shown). This is attributed to the E3* component discussed below in Section \ref{sec:TF}.

\subsection{Weak external fields}
There are two main effects caused by weak external magnetic fields ($\mathbf{B}_{ext}\ll \mathbf{B}_{int})$, (1) the degeneracy of $|\mathbf{B}_{int}|$ for electrostatically equivalent  stopping sites within one environment is lifted, and (2) a component precessing in $\mathbf{B}_{ext}$ rather than  $\mathbf{B}_{int}$ appears.

\begin{figure}[t]
 \includegraphics[width=8.5cm]{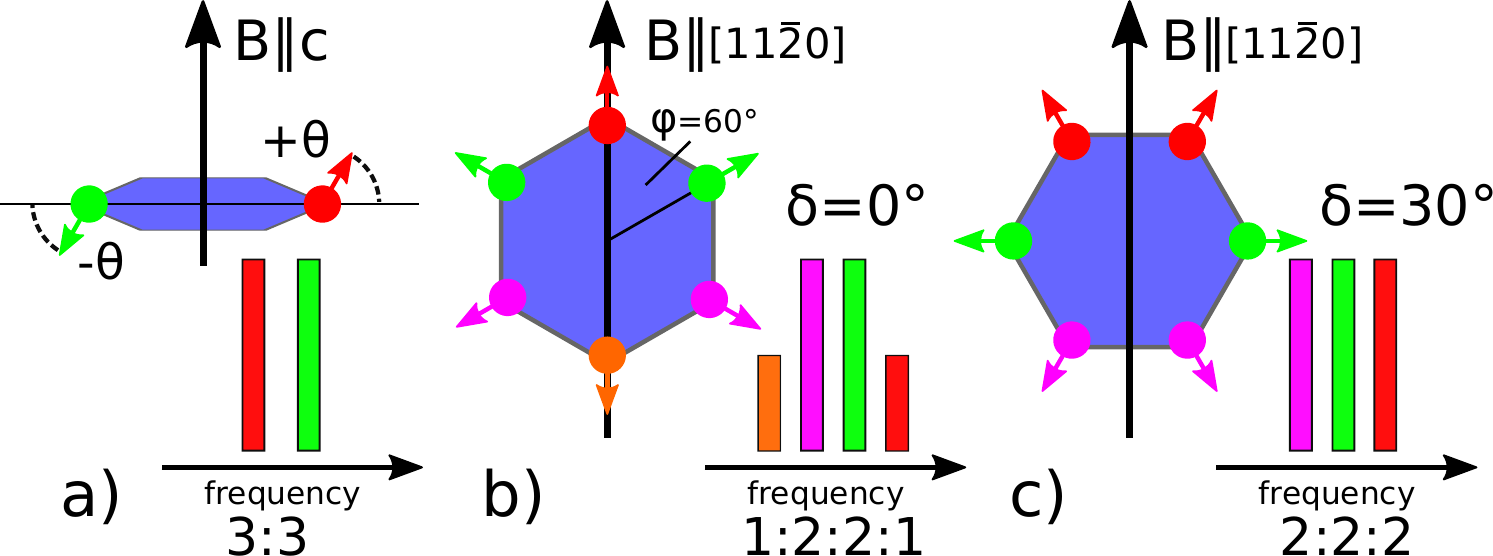}
 
 \vspace{0.3cm}

 \includegraphics[width=8.5cm]{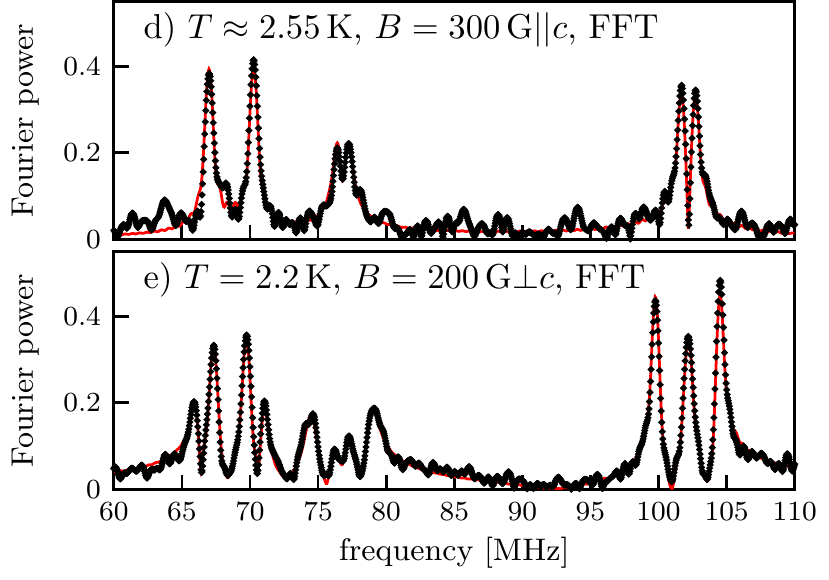}
\caption{Schematic illustration of how an external magnetic field breaks the degeneracy of $|\mathbf{B}_{int}|$ for stopping sites with the same magnitude, but different directions of $\mathbf{B}_{int}$. Only those sites shown in the same color share the same magnitude of the vector sum $|\mathbf{B}_{int}+\mathbf{B}_{ext}|$, resulting in a multiplet splitting and amplitude ratio as depicted: (a) $\mathbf{B}_{ext} || c$ causes a doublet splitting, $\mathbf{B}_{ext} \bot c$  causes (b) a quadruplet ($\delta=0$) and (c) a triplet splitting $\delta=\pm\SI{30}{\degree}$;
(d) FT of \msr spectra  in $B=\SI{300}{G} || c$ at $T={2.55}{K}$
and (e) FT of \msr spectra  in $B=\SI{200}{G}||[1120] \bot c$ at $T=\SI{2.2}{K}$ 
 }.
\label{fig:splitFFT}
\end{figure}

\subsubsection{Orientation of the internal magnetic field}
\label{sec:internalfieldsymmetry}

The internal field direction  at a stopping site can be described by two angles; $\theta$ is defined as the smallest angle enclosed by $\mathbf{B}_{int}$ and the c-plane, and $\varphi$ as the azimuthal angle (enclosed by $[11\bar{2}0]$ and the $\mathbf{B}_{int}$ projection  onto the c-plane), see insets  in Fig. \ref{fig:ZFprec}. 
From symmetry, stopping sites forming a given environment can be projected onto the c-plane to form a hexagon. The six $\varphi$ values of an environment 
are given by $\delta+\SI{0}{\degree}, \delta\pm\SI{60}{\degree}, \delta\pm\SI{120}{\degree}$ and $\delta+\SI{180}{\degree}$, with $\delta$ being the smallest angle enclosed by $[11\bar{2}0]$ and a hexagon corner.
As noted above, the internal field magnitude at the  stopping sites forming a given environment is the same, but its direction is not. While this is inconsequential in ZF, the relative orientations of $\mathbf{B}_{int}$ and $\mathbf{B}_{ext}$ matter in the presence of external fields, where the precession frequency is determined by the magnitude of the vector sum $ |\mathbf{B}_{int}+\mathbf{B}_{ext}|$. Consequently, the application of $\mathbf{B}_{ext}$ lifts the degeneracy of the precession frequencies \emph{within} an environment, causing multiplet splittings. This is schematically illustrated in Figs. \ref{fig:splitFFT} (a)-(c) both for $\mathbf{B}_{ext}||c$ and $\mathbf{B}_{ext}\bot c$.
 The FTs of the \msr spectra at low temperatures are shown in 
Fig. \ref{fig:splitFFT} for (d) $\mathbf{B}_{ext}=\SI{300}{G}\parallel c$  
\footnote{This measurement with $\mathbf{B}_{ext}=\SI{300}{G}\parallel c$ was taken in the Omni-Prime spectrometer. The Fourier transform is shown over the time range  of $0.1-\SI{2}{\micro s}$.}
and (e) $\mathbf{B}_{ext}=\SI{200}{G}\parallel [11\bar{2}0]\bot c$. Comparison with the ZF spectrum, Fig. \ref{fig:ZFprec}(b), shows that for $\mathbf{B}_{ext}\parallel c$, the E1-E3 lines split into doublets, while for $\mathbf{B}_{ext}\bot c$, more complex multiplets are observed.   The spectra were fit with up to 12 oscillatory signals, Eqn. (\ref{eqn:osc_fit}), and a small non-oscillating signal.
The obtained frequencies ($f_{exp}$) are shown in Table \ref{tbl:lowBFFTpara}  and \ref{tbl:lowBFFTbot}. Under the assumption that $\mathbf{B}_{ext}$ does not induce changes of $\mathbf{B}_{int}$, all multiplet frequencies can be consistently described by the vector sum $ |\mathbf{B}_{int}+\mathbf{B}_{ext}|$, which allows extraction of the $\theta$ and $\delta$ values describing  the orientation of $\mathbf{B}_{int}$ in E1-E3, see Table \ref{tbl:summary} for a summary  and Appendix \ref{sec:apxBint} for details. 
The obtained angles provide stringent criteria for comparison with the internal field of candidate muon sites calculated with DFT, see Section \ref{sec:dft}.

 \begin{table}[t]
 \caption{ZF precession frequencies $f^\mathrm{ZF}$, $\theta$ and $\delta$ values describing the internal field orientation for E1-E3 at $T=\SI{2.2}{K}$.
 \label{tbl:summary}}
 
 \begin{ruledtabular}
 \begin{tabular}{c c c c}
site & $f^\mathrm{ZF}$ [MHz]& $\theta$ [\si{\degree}]& $\delta$ [\si{\degree}]  \\ \hline
E1& $68.52\pm 0.01$ & $24\pm 1$ & $0\pm 3.5$\\ 
 
E2&$102.12\pm 0.01$ & $6\pm 1$ & $30\pm 3.5$  \\ 
 
 E3& $76.69\pm 0.01$ & $5\pm 1$ & $17.5\pm 2$ \\

 \end{tabular}
  \end{ruledtabular}
 \end{table}

\subsubsection{Evidence for a signal component with zero internal field }
\label{sec:TF}

\begin{figure}[t]

  \includegraphics[width=8.5cm]{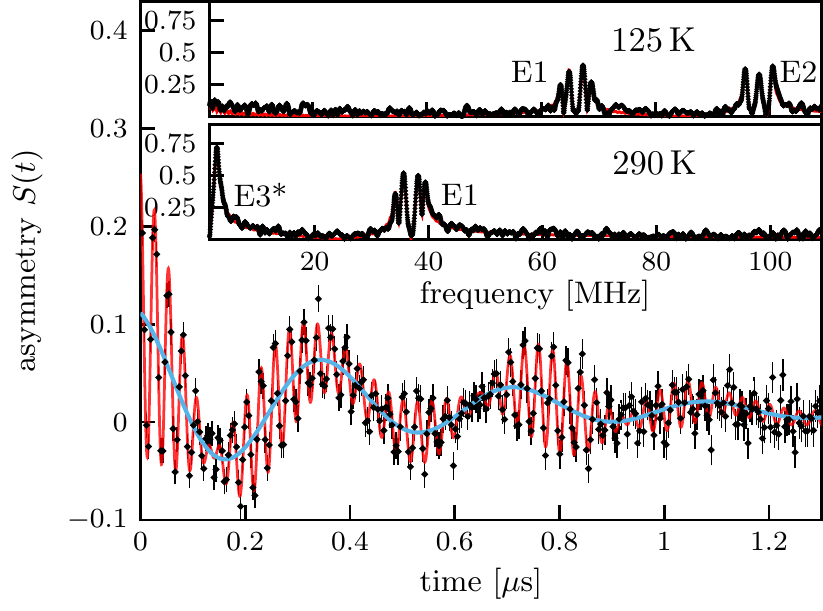}
\caption{\msr spectrum taken in $\mathbf{B}_{ext}=\SI{200}{G}\bot c$ at  $T=\SI{290}{K}$. Alongside the expected multiplet splittings, compare Fig. \ref{fig:splitFFT}\,(e), there is an additional component termed E3* that precesses at the (much lower) Larmor frequency of the applied field (blue line), indicating that some muons do not experience any internal field. Insets: FT at $T=\SI{125}{K}$ (top) and $T=\SI{290}{K}$ (bottom). The E3* line corresponding to precession in $\mathrm{B}_{ext}$ is absent at lower temperatures. }
\label{fig:Larmor270}
\end{figure}

Having discussed the effect of $\mathbf{B}_{ext}$ on the precession frequencies, we now turn our attention to the non-precessing component that appears in ZF above $\approx\SI{160}{K}$. At coinciding temperatures and in both $\mathbf{B}_{ext}\bot c$ and  $\mathbf{B}_{ext}\parallel c$ (not shown), there is a component of comparable amplitude that oscillates at the Larmor frequency of the \emph{external} field ($f_{ext}=\gamma_\mu/2\pi |\mathbf{B}_{ext}|$), which is absent below \SI{150}{K}, see  Fig. \ref{fig:Larmor270}. Spin precession about  $\mathbf{B}_{ext}$ rather than  $\mathbf{B}_{int}$, in spite of ordered Cr moments, indicates that the muons giving rise to this signal are not subject to an internal field. The temperature dependence of the amplitude and relaxation rate of this signal, termed E3* in anticipation of its interpretation in Section \ref{sec:hop},  is shown in Fig. \ref{fig:fitResZF}.

\subsection{Large external fields}
\label{sec:high}

\begin{figure}[t]
\center
            \includegraphics[]{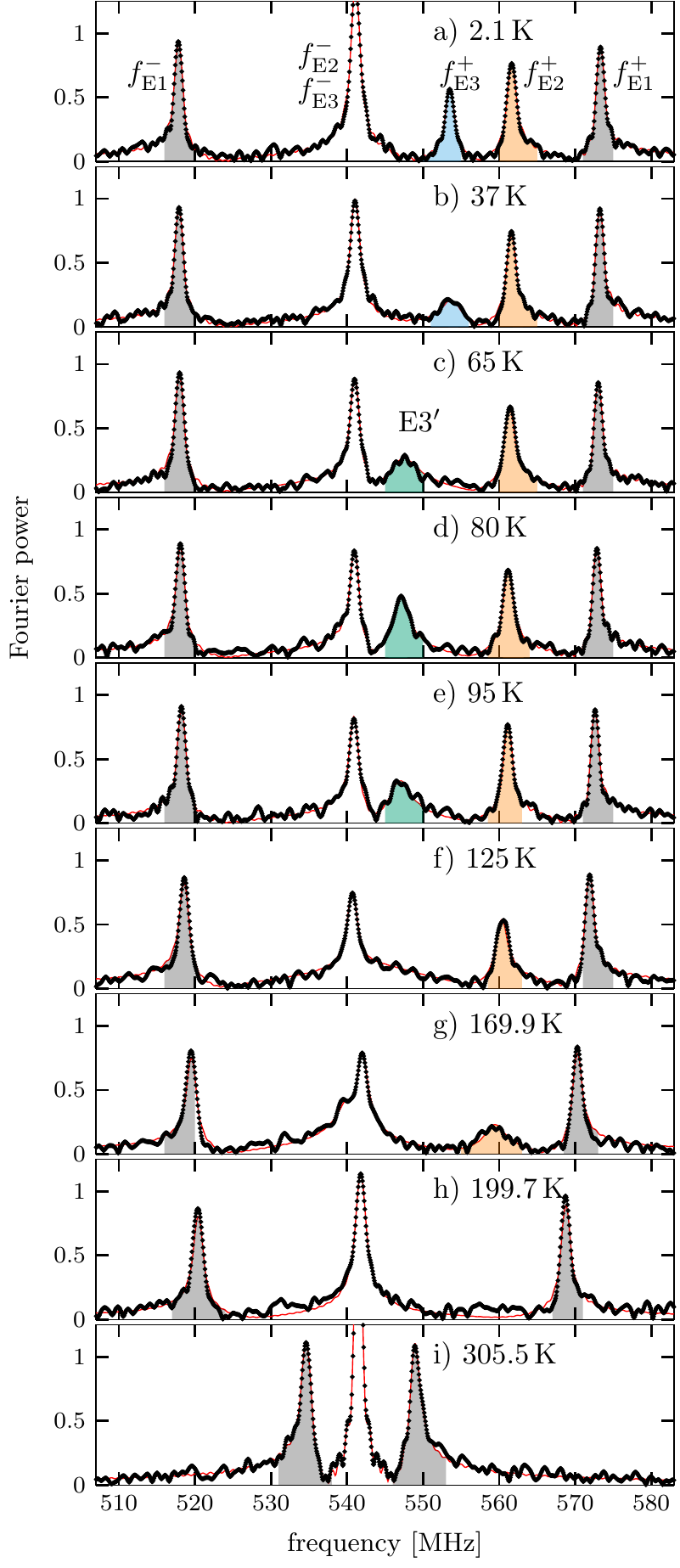}

\caption{Temperature dependence of FTs of $\mu$SR spectra taken in $\mathbf{B}_{ext}=\SI{4}{T}||c$. 
 \label{highFieldFFT}}
 \vspace{-0.2cm}

\end{figure}

\begin{figure}[t]
\center
            \includegraphics[]{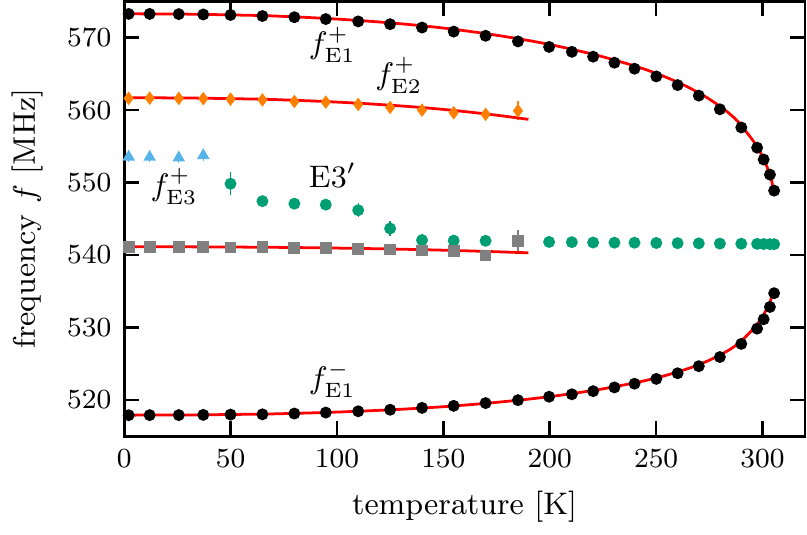}

\caption{Temperature dependence of frequencies obtained in $\mathbf{B}_{ext}=\SI{4}{T}||c$. Solid lines are calculated doublet frequencies  $f^\pm_\mathrm{E1}(T)$ and $f^\pm_\mathrm{E2}(T)$ assuming constant $\theta$. }
\label{highFieldFit}
\end{figure}

The ZF spectra indicate that the highest $\mathbf{B}_{int}$ is about \SI{0.75}{T}. Here, we apply external fields significantly higher than this.
The temperature dependence of the FTs of $\mu$SR spectra taken in $\mathbf{B}_{ext}=\SI{4}{T}||c$ is shown in Fig. \ref{highFieldFFT}. Again, a doublet splitting is expected for each ZF frequency; however , since the position of  $f^\pm$ with respect to $f^\mathrm{ZF}$ depends on the relative strength of $\mathbf{B}_{ext}$ to $\mathbf{B}_{int}$, and $\mathbf{B}_{ext}>\mathbf{B}_{int}$, $f^-$ and $f^+$ are distributed around $f_{ext}$  rather than $f^\mathrm{ZF}$, compare Figs. \ref{fig:splitFFT}(d) with \ref{highFieldFFT}(a).
At $T=\SI{2.1}{K}$, five lines are observed. They can be assigned as follows: the two outer frequencies (colored in black) compose the E1 doublet $f^\pm_\mathrm{E1}$, while the second (orange) and third (blue) highest frequencies correspond to $f^+_\mathrm{E2}$ and $f^+_\mathrm{E3}$, respectively. The remaining line (uncolored) is a superposition of both $f^-_\mathrm{E2}$ and $f^-_\mathrm{E3}$, explaining its large amplitude.
The temperature evolution follows mostly what is expected from the ZF results. With increasing temperature, the E1 doublet splitting decreases as $|\mathbf{B}_{int}|$ decreases. Above \SI{200}{K}, its amplitude becomes larger, and the line broadens approaching $T_N$.
Likewise, $f^+_\mathrm{E2}$ follows the decreasing $\mathbf{B}_{int}$, and disappears above ${\approx}\SI{185}{K}$. For all temperatures where $f^+_\mathrm{E2}$ is observed, the uncolored line has a contribution from $f^-_\mathrm{E2}$. Above $\approx$\SI{170}{K}, a large component close to $f_{ext}$ appears, and is, in accordance with Section \ref{sec:TF}, assigned to E3*. The E3 doublet is only observable below \SI{50}{K}.

Remarkably, there is an additional component between \SI{50}{K} and \SI{170}{K}, termed E3$'$ (green in Fig. \ref{highFieldFFT}). As will be discussed in Section \ref{sec:hop}, this additional line is strongly indicative of local  hopping  between adjacent E3 sites.

Details on the data analysis as well as fit results for the amplitudes can be found in Appendix \ref{apx:high}.
From the multiplet splitting, $\theta$ values matching closely those obtained in low field are extracted,  see Table \ref{tbl:baseHighField}. This indicates that even in large $\mathbf{B}_{ext}$, $\mathbf{B}_{int}$ is not significantly affected (reasonable since 
the Cr Zeeman energy in \SI{4}{T} is much smaller than the exchange coupling \cite{samuelsen1970a}), and the precession frequencies are well determined by vector addition. 
The fitted frequencies are shown in Fig. \ref{highFieldFit}. 
The red lines represent calculated doublet frequencies  $f^\pm_\mathrm{E1}(T)$ and $f^\pm_\mathrm{E2}(T)$ assuming constant $\theta$, see  Appendix \ref{apx:high} for details. There is good overall agreement with the data, indicating that $\theta$ is largely temperature independent.

\section{Experimental evidence for site metastability and dynamics}
\label{sec:disc}
Three ZF frequencies are observed, see Fig. \ref{fig:ZFprec}(c), and attributed to three distinct muon environments, E1-E3. Each environment contains a number of electrostatically equivalent sites with the same magnitude, but different directions, of  $\mathbf{B}_{int}$.
Above ${\approx}\SI{160}{K}$, a component E3* precessing in the \emph{external} rather than the internal field is observed both in low and high field, indicating an environment characterized by zero internal field.

The E2 and E3 signals disappear at different temperatures, while E1 is observed over the complete temperature range, indicating that each environment has a distinct potential energy. 
At low temperatures, E1-E3 are all populated.
Since  site populations are determined by the epithermal implantation process rather than thermodynamic equilibrium, it is possible that  a muon occupies metastable sites with higher energy than the ground state. If thermally  activated transitions to a lower energy state are inaccessible within its short lifetime, the muon may remain in the metastable site and give rise to a distinct signal \cite{browne1982,stoneham1984}. However, with increasing temperature, site changes either within one, or into another environment may become possible.

Around $\SI{180}{K}$, E1, E2 and E3* signals are observed  and account for the full signal, see Fig. \ref{fig:fitResZF}(a). Noting that the amplitudes  of both E2 and E3* are approximately temperature independent in the respective regions where they are observed, we conclude that (1) muons in E2 do not transition into E3*,  (2) thus the disappearance of E2 stems from a transition into E1 at sufficiently high temperatures and (3) consequently, by conservation of total amplitude, muons that stop in E3 below \SI{50}{K} must give rise to E3* at higher temperatures. 

In this section, we first present a model supporting the E2$\rightarrow$E1 transition. Then, the evolution of muons from E3 into E3* is discussed in terms of local muon hopping between adjacent, electrostatically equivalent E3 sites.

\subsection{E2 - E1 transition}
\label{sec:trans}
\begin{figure}[t]
\center
 \includegraphics[]{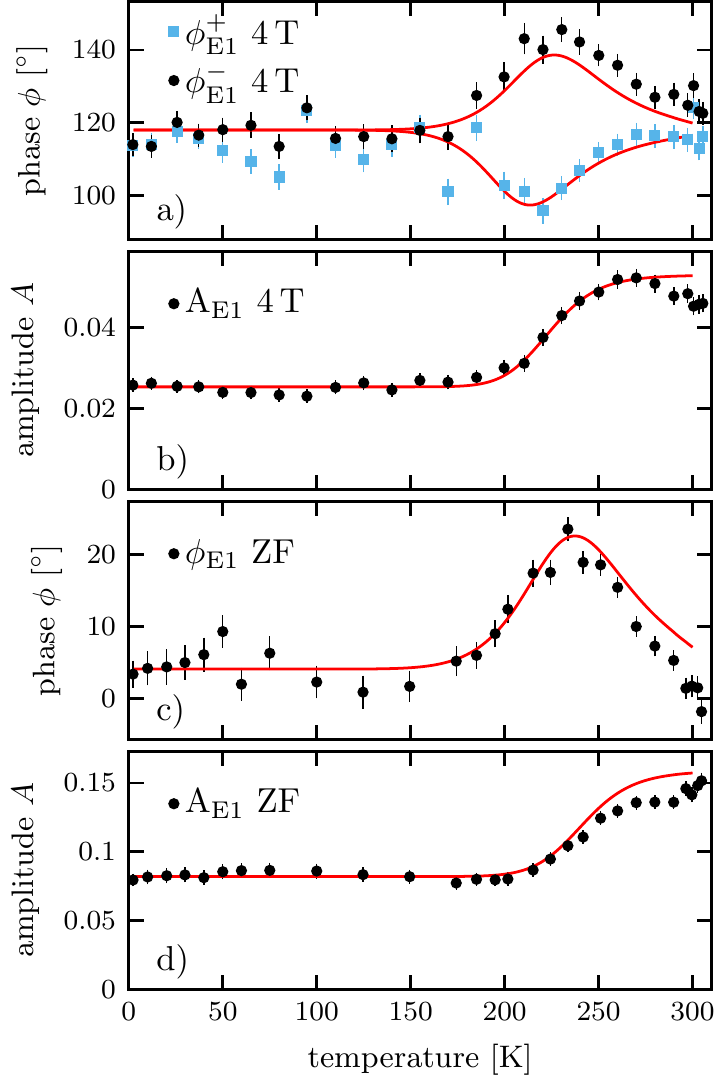}

\caption{Comparison of experimental E1 phase and amplitude with the E2$\rightarrow$E1 transition model for (a) $\phi_\mathrm{E1}$ in   \SI{4}{T}, for $f^+_\mathrm{E1}$ (blue) and $f^-_\mathrm{E1}$ (black) (b) E1 amplitude, as obtained by a shared fit of the  $f^\pm_\mathrm{E1}$ doublet. (c)  ZF $\phi_\mathrm{E1}$ and (d) ZF amplitude. Solid lines are  Eqns. (\ref{eqn:amp}) and (\ref{eqn:trans_phase}) with shared model parameters $E_a=\SI{180}{meV}$ and $\nu_0=\SI{8e11}{Hz}$.
}
\label{fig:trans}
\end{figure}

Here, we show that the disappearance of the E2 signal around \SI{200}{K}, and the subsequent increase in E1 amplitude is consistent with a metastable E2 environment that allows for transitions into E1. The following discussion is based on two assumptions: 
(1) The E2$\rightarrow$E1 transition can be described by a thermally activated, exponential rate of the form $\Lambda(T)=\nu_0 \exp(-E_a/k_BT)$, where $E_a$ and $\nu_0$ are the activation energy and attempt frequency. 
(2) At the time of implantation, the probability for the muon to initially occupy a site in any of the three environments is temperature independent (i.e. at all temperatures, the same fraction of muons start out in E1, E2 and E3). This assumption is discussed in Sec. \ref{sec:discussion}.  
While the initial fraction  starting  in E2 is independent of temperature, the actual time spent in this environment depends on  the transition rate $\Lambda(T)$.  
If $\Lambda(T)$ is much smaller than $f_\mathrm{E2}$, the  E2 muons precess for many periods with $f_\mathrm{E2}$ before transitioning, and oscillatory signals from both  E1 and E2 with amplitudes $A_\mathrm{E1}$ and $A_\mathrm{E2}$ can be detected. 
In contrast, if $\Lambda(T)$ is much larger than $f_\mathrm{E2}$, the E2 muons change to E1 before the muon spin has a chance to precess with  $f_\mathrm{E2}$, and a single oscillatory signal at $f_\mathrm{E1}$ with a combined amplitude $\mathcal{A}=A_\mathrm{E1}+A_\mathrm{E2}$ can be observed.
However, if  $\Lambda(T)$ is comparable to $f_\mathrm{E2}$, the E2 muons may precess at $f_\mathrm{E2}$ prior to the transition, and acquire a phase shift with respect to muons initially in E1. This results in a smaller apparent E1 amplitude $\mathcal{A}$ and an overall phase shift $\Phi (T)$. In Appendix \ref{apx:appE2-E1}, a model accounting for such a transition  and expressions for $\mathcal{A}(T)$ and $\Phi (T)$ are described.
In Fig. \ref{fig:trans}, the E1 phase and amplitude data of both the ZF signal and the high-field doublet are compared with the transition model, with activation energy $E_a=\SI{180}{meV}$ and attempt frequency $\nu_0=\SI{8e11}{Hz}$ being \emph{shared} parameters for the \emph{complete} data set.

There is excellent qualitative agreement between the  model and the data;  the peak in the phase, including the opposite direction for the high field doublet, and the increase in amplitude are well described both in ZF and high field  with the same parameter set. 
Thus the proposed E2$\rightarrow$E1 transition with an estimated barrier $E_a=180\pm\SI{40}{meV}$ provides a consistent explanation for the disappearance of the E2 signal, its associated increase in relaxation rate, and the subsequent increase in E1 signal amplitude.

\subsection{Local hopping}
\label{sec:hop}

\begin{figure}[b]

\hfill
\includegraphics[height=4.cm]{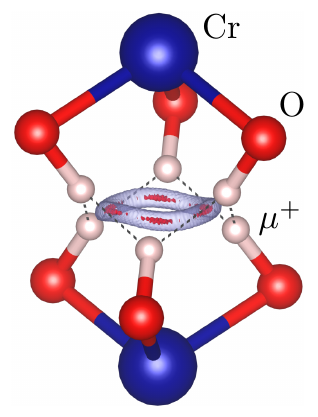} 
\hfill
\includegraphics[height=4.cm]{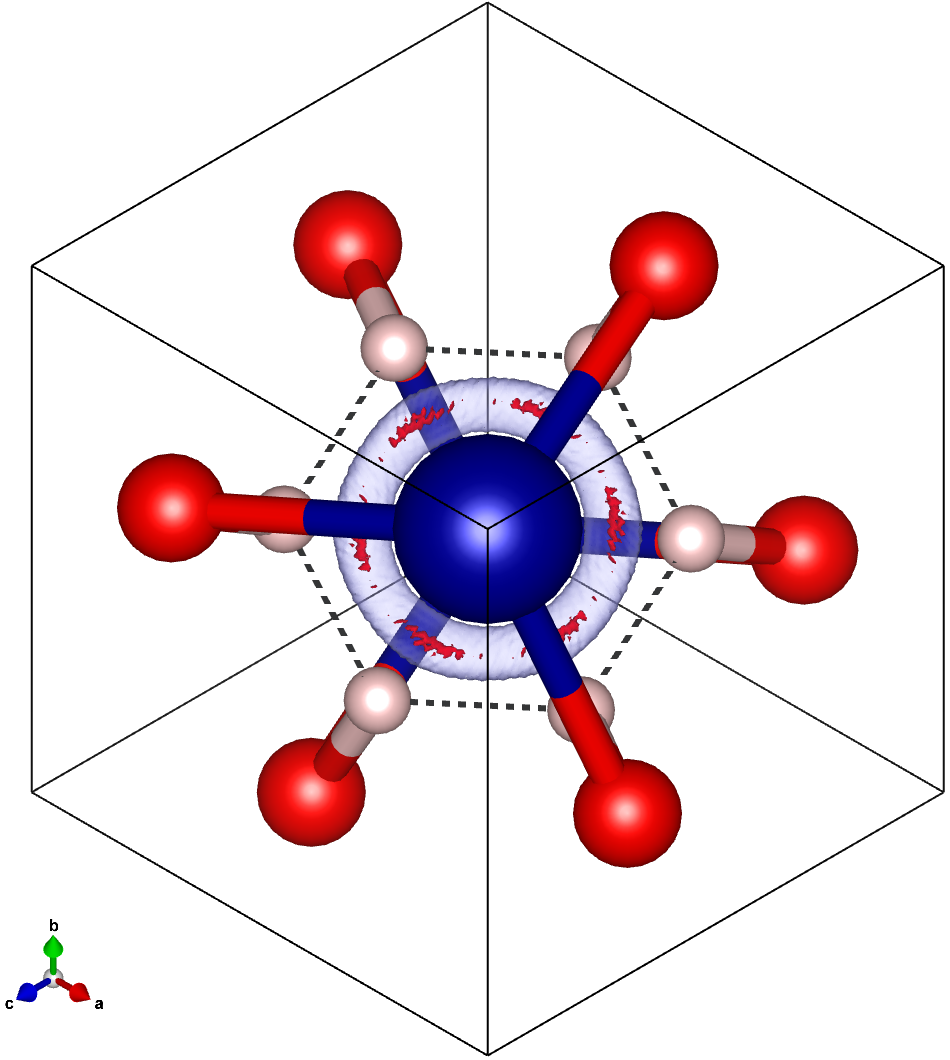}  
\hfill 
 
\caption{Example configuration of electrostatically equivalent muon sites, based solely on symmetry considerations and the constraint that muon stopping sites (white spheres) are located  \SI{1}{\angstrom} away from oxygen (red spheres). The close proximity of nearby sites 
suggests localized hopping. The electrostatic potential isosurface of the \emph{undistorted} crystal structure is shown in blue, with the red patches indicating local electrostatic minima, corresponding to the position of the so called Rodriguez (R) sites \cite{graf1978a}.}
\label{fig:hopsites}
\end{figure}

Next, we show that both the E3 frequency observed below \SI{50}{K}  and the E3* signal precessing at the Larmor frequency of the external field arise from muons in the \emph{same} environment. 
The appearance of E3* above $\approx\SI{160}{K}$, see Figs. \ref{fig:Larmor270} and \ref{highFieldFFT},  
 indicates that a fraction of muons experience no internal field. This is surprising since the simultaneous observation of the E1 signal 
clearly shows the presence of ordered Cr magnetic moments. 
Although there are high-symmetry sites along the \emph{c}-axis where the internal field precisely cancels, these sites are far ($\gg \SI{1}{\angstrom}$) from an oxygen and energetically unfavorable for an interstitial $\mu^+$ site, as confirmed by DFT in Section \ref{sec:dft}. Instead, we consider in Fig. \ref{fig:hopsites} an example configuration of one environment comprised of six electrostatically equivalent muon stopping sites (white spheres), based solely on symmetry considerations (see beginning of Section \ref{sec:results}), and a muon-oxygen distance of \SI{1}{\angstrom}.
Given the close proximity of nearby electrostatically equivalent sites, thermally activated local hopping between adjacent sites seems plausible. For sufficiently fast intra-environment hopping, the effective internal field experienced by the dynamic muon is the average over all sites. Noting  $\mathbf{B}_{int}(\mathbf{r})=-\mathbf{B}_{int}(-\mathbf{r})$ and the symmetry of equivalent sites, it becomes clear that this average is \emph{zero}. 
We hypothesize that muons stopping in E3 sites undergo such local hopping around a single hexagon at elevated temperatures, leading to the disappearance of the $f_\mathrm{E3}^\mathrm{ZF}$  signal, and the subsequent observation of the E3* signal in an external field.


 This consistently explains  the observed data: at low temperatures, muons stopping in E3 sites are quasi-static, i.e. no or only very slow hopping occurs, and each muon precesses predominately in the internal field of \emph{one} site, giving rise to   $f_\mathrm{E3}^\mathrm{ZF}$.
  Above $\approx \SI{160}{K}$, a relaxing non-oscillatory component appears in ZF, consistent with a fraction of muons that are not subject to any field. 
  In the intermediate temperature region $60-\SI{160}{K}$, no signal is observed, as the hopping is neither fast enough to efficiently average out the internal field, nor slow enough to allow for the observation of coherent E3 oscillations. External fields, both small and large, cause  multiplet splittings of  $f_\mathrm{E3}$ at low temperature, consistent with muons being quasi-static, while above $\approx \SI{160}{K}$, a signal precessing in $\mathbf{B}_{ext}$ can be observed (E3*), since $\mathbf{B}_{int}$ is averaged to zero and does not contribute to the field magnitude. 
In high field, an additional signal, E3$'$, see Fig. \ref{highFieldFFT}, is observed in the intermediate temperature region.
A Monte Carlo simulation of the muon depolarization function in $\SI{4}{T}\parallel c$, assuming local hopping, identifies E3$'$ as the average of the doublet frequencies  $f^{avg}_\mathrm{E3}=(f^+_\mathrm{E3}+f^-_\mathrm{E3})/2$, and allows for an estimate of the energy barrier between E3 sites $E_b=42\pm\SI{5}{meV}$, see Appendix \ref{apx:hop} for details on the simulation and an in-depth discussion. 
Overall, we clearly show that the E3/E3$'$/E3* signals arise  from different dynamic regimes of muons undergoing local hopping in a \emph{single} environment.


Already in the first $\mu$SR paper on antiferromagnets, local diffusion between electrostatically equivalent sites was considered a possibility in Fe$_2$O$_3$ \cite{graf1978a}.
Subsequently, local motion was speculated to occur in Cr$_2$O$_3$ \cite{boekema1983}, and is suspected \cite{duginov1994,denison1984} and observed \cite{amato2000,mulders2001,schneider1992} in a range of materials.
Here, we conclusively show that local hopping indeed occurs in Cr$_2$O$_3$ by direct observation, identification and consistent description of the distinct signals that arise as a result of restricted motion in a system with broken magnetic inversion symmetry, a hop rate changing several orders of magnitude over the observed temperature range, and various applied fields.

We further note that muons in both E1 and E2 are, apart from the E2$\rightarrow$E1 transition, site-stable, i.e. no intra-environment motion occurs. This is evident from the  pronounced multiplet splitting that is observed at all temperatures where E1 and E2 signals are detected, see Figs. \ref{fig:Larmor270} and \ref{highFieldFFT}.

\section{Identification of muon stopping sites with DFT}

\label{sec:dft}

Thus far, using simple models describing a thermally activated E2$\rightarrow$E1 transition and local hopping within the E3 environment, and without explicit knowledge of the stopping sites, we have explained the major features in the data.  
The  coexistence of site-stable and highly mobile muons is intriguing, especially since there is no evidence for interexchange between dynamic E3 muons and static E1 or E2 muons, even in the presence of the E2$\rightarrow$E1 transition. 
To gain deeper insight into this surprising behavior,  we turn to DFT to identify  muon stopping sites. 
With the recent increase in availability and capability of computing resources countering the large computational demands of first principles calculations, DFT has had great success in providing information about location and stability of muon stopping sites in  a range of materials,
and is developing into an important new tool for $\mu$SR (see Ref. \onlinecite{bonfa2016} for a review, and Refs. \onlinecite{onuorah2018,onuorah2019b} for recent developments). 

Since the inception of the $\mu$SR technique,  knowledge of the location of the muon within the sample was of key importance. The main motivation for the early $\mu$SR studies on antiferromagnets, prompted by the first observation of  ZF $\mu$SR signals in Fe$_2$O$_3$ \cite{graf1978a} and its isomorph Cr$_2$O$_3$  \cite{ruegg1979}, was to establish the muon as a sensitive and useful probe of the local magnetic properties  of the host material by determining (1) where the muon stops and what its dynamic properties are with respect to site stability and diffusion and (2)
 if, and under what conditions muonium is formed in insulating (anti)ferromagnets.
Based on simple electrostatic considerations, two sets of possible stopping sites were found for the Corundum structure, so called Rodriguez (R) sites \cite{graf1978a} located in the Cr gap close to the inversion center, see Fig. \ref{fig:hopsites}, and Bates (B) sites \cite{boekema1981} in (B0), or slightly above and below (B1) the oxygen basal plane, see Fig. 2 in Ref. \onlinecite{boekema1983}.
The internal magnetic field in these sites was estimated by summing over the dipolar contributions from surrounding Cr moments. Additionally, covalency effects were considered, and attempts to assign ZF frequencies to specific sites yielded some partial and approximate agreements, although the  overall results remained inconclusive \cite{boekema1981,boekema1983}.
No evidence for Mu or a neutral charge state was identified.

\begin{table*}[t]
  \begin{minipage}[t]{0.75\textwidth}
  \vspace{0pt}
    \begin{ruledtabular}
  \begin{tabular}{c|c|c|c|c|c|c|c|c|c}
  CS & E& Site & $d_{z^2}$ & $f_{dip}$ [\si{MHz}] & $f_{c}$ [\si{MHz}]& $f_{tot}$ [\si{MHz}] & $\theta [\si{\degree}]$ & $\varphi [\si{\degree}]$& $\Delta E [\si{meV}]$ \\  
   \hline \hline


\rowcolor{Gray}[1.5\tabcolsep]
&\textbf{E3} & D& & $94.1\pm0.0$ & $2.8\pm 4.4$ & $93.9\pm 0.4$ & $3.8\pm 2.9$ & $4.5 \pm 0.0$ &0  \\
+&& B0& & $28.2 \pm 2.2$ & $0.0 \pm 0.0$ & $28.2 \pm 2.2$ & $0.1 \pm 0.0$ & $29.5 \pm 0.0$ & 753    \\
& &C$_O$ & & $99.1 \pm 4.0$ & $49.4 \pm 14.0$ & $55.9 \pm 10.0$ & $90.0 \pm 0.0$ & $1.3 \pm 1.2$ & 1565 \\ 
&& C$_E$ &  &  $0.0 \pm 0.0$ & $0.0 \pm 0.0$ & $0.0 \pm 0.0$ & $90.0 \pm 0.0$ & $57.4 \pm 0.0$ & 1728  \\
\hline \hline 

\rowcolor{Gray}[1.5\tabcolsep]
&\textbf{E1}& D1 & $\uparrow$ & $72.4 \pm 0.3$ & $35.6 \pm 3.3$ & $78.0 \pm 1.0$ & $-22.4 \pm 2.5$ & $35.8 \pm 0.5$ & 0  \\
\rowcolor{Gray}[1.5\tabcolsep]
&\textbf{E2}& D2 & $\downarrow$ & $114.4 \pm 0.2$ & $3.9 \pm 18.4$ & $114.3 \pm 1.4$ & $0.0 \pm 8.9$ & $58.9 \pm 0.7$ & 116  \\

& & D3 &$\downarrow$& $119.0 \pm 0.8$ & $1.6 \pm 17.4$ & $118.6 \pm 6.8$ & $-17.3 \pm 7.8$ & $5.5 \pm 0.1$ & 320  \\

& & D4 &$\uparrow$ & $100.9 \pm 0.3$ & $3.8 \pm 3.2$ & $100.4 \pm 0.6$ & $5.9 \pm 2.1$ & $58.3 \pm 0.3$ & 419  \\

0& & D5 & $\uparrow$& $82.9 \pm 0.4$ & $34.7 \pm 4.1$ & $99.3 \pm 2.2$ & $-37.4 \pm 2.5$ & $39.7 \pm 0.3$ & 448  \\

& &B1 & &  $  92.1 \pm 0.2  $ & $430   \pm 30  $ & $ 470  \pm 30  $ & $ -79.4  \pm 0.7  $ & $  24.0 \pm 0.1  $ & 945 \\

& &B0 & &  $ 47.4  \pm 1.8  $ & $ 0.2  \pm 0.1  $ & $ 47.3  \pm 1.8  $ & $-17.7   \pm 0.8  $ & $ 30.3  \pm 0.0  $ & 1157 \\

&& C$_O$&  &$94.3 \pm 1.2$ & $581.3 \pm 26.7$ & $486.7 \pm 25.5$ & $-90.0 \pm 0.0$ & - & 355 \\
& &C$_E$ &  & $0.0 \pm 0.0$ & $0.0 \pm 0.0$ & $0.0 \pm 0.0$ & $90.0 \pm 0.0$ & - & 1206 \\
 \end{tabular}
  \end{ruledtabular} 
  \end{minipage}
  \hfill
  \begin{minipage}[t]{0.22\textwidth}
  \vspace{-0.3cm}
    \caption{
 DFT results for various candidate sites and two charge states (CS): dipolar $f_{dip}$ and contact contributions   $f_{c}$ giving rise to the combined $f_{tot}$, which spans the angles $\theta $ and $\varphi $ as defined in \ref{sec:internalfieldsymmetry}. $\Delta E$ is the energy relative to the ground state of each charge state respectively, and $d_{z^2}$ the spin state of the extra electron for D1$^0$-D5$^0$. Shaded rows indicate candidates for E1-E3.
 \label{tbl:dft}}
  \end{minipage}
\end{table*}

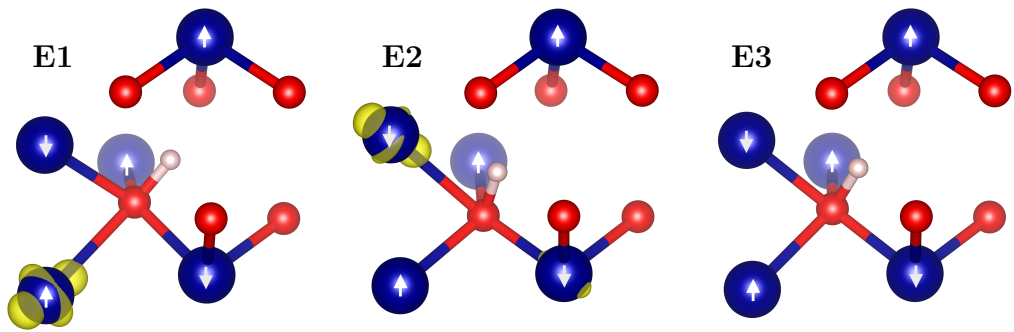
\begin{figure*}[t]
  \begin{minipage}[t]{0.77\textwidth}
    \vspace{0cm}
    \begin{tikzpicture}
  \path (0,0) node(a) { \includegraphics[height=4.3cm]{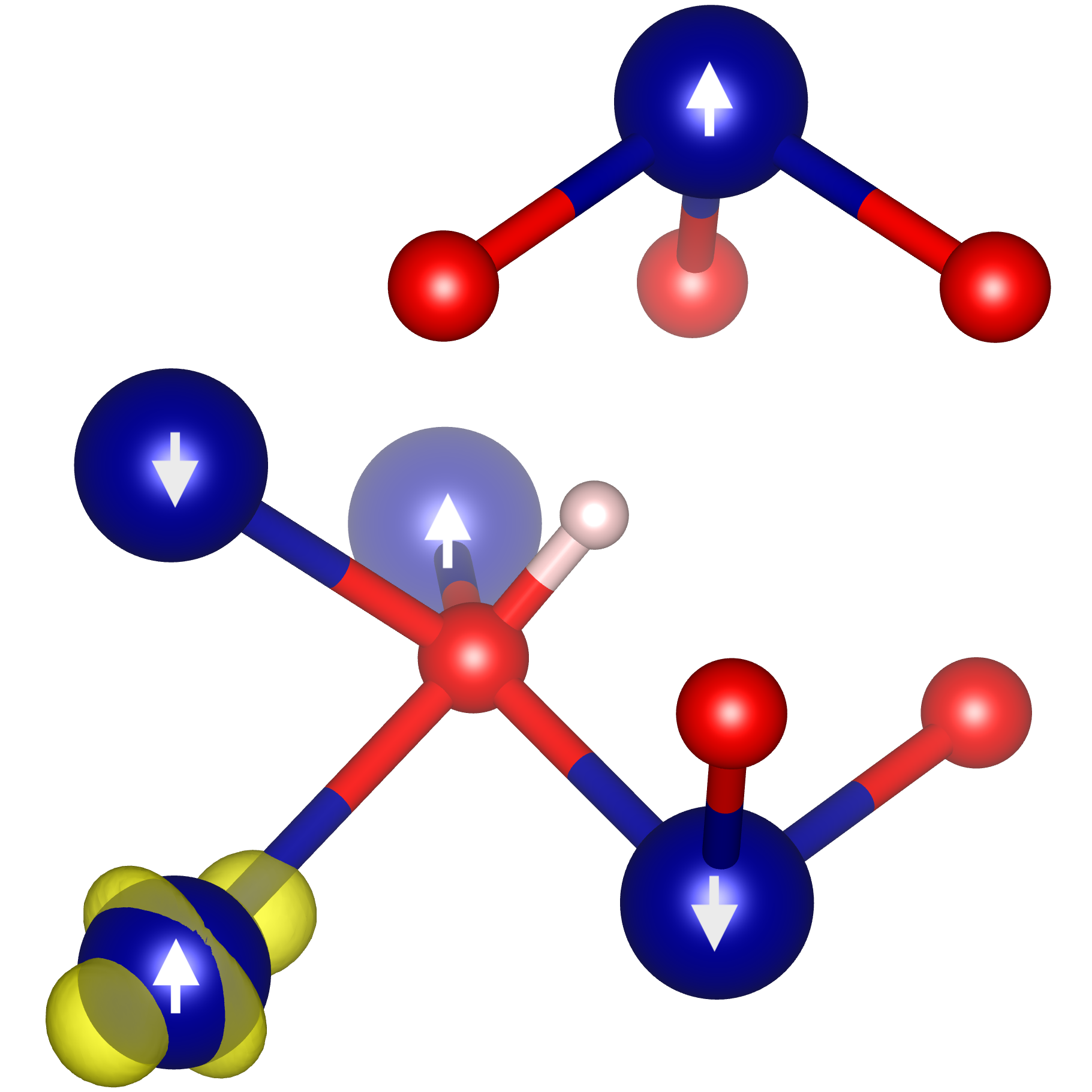} };
  \path (-1.37,1.5) node(b) {\large \textbf{E1}};
\end{tikzpicture}
\hfill
    \begin{tikzpicture}
  \path (0,0) node(a) { \includegraphics[height=4.3cm]{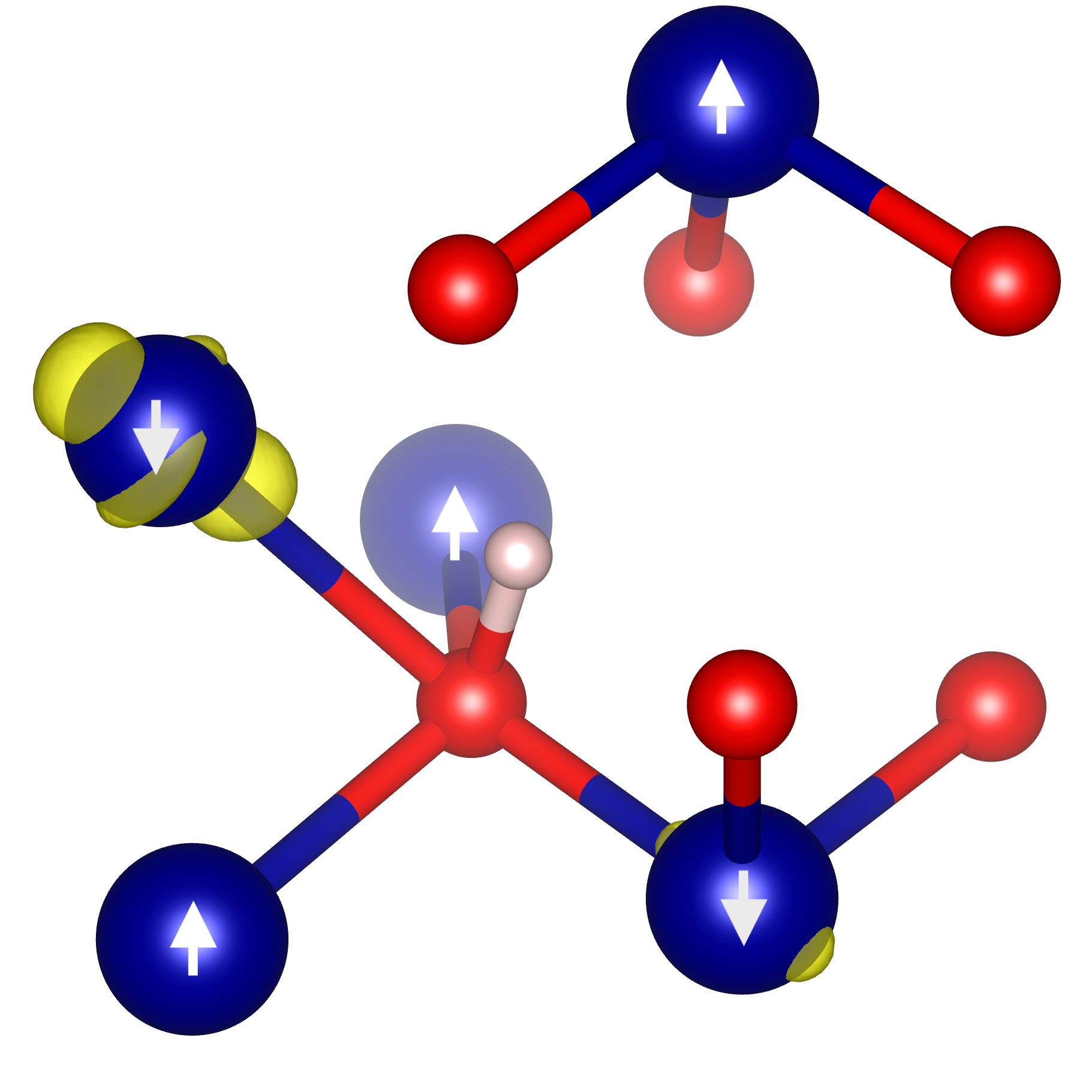} };
  \path (-1.37,1.5) node(b) {\large \textbf{E2}};
\end{tikzpicture}
\hfill
    \begin{tikzpicture}
  \path (0,0) node(a) { \includegraphics[height=4.3cm]{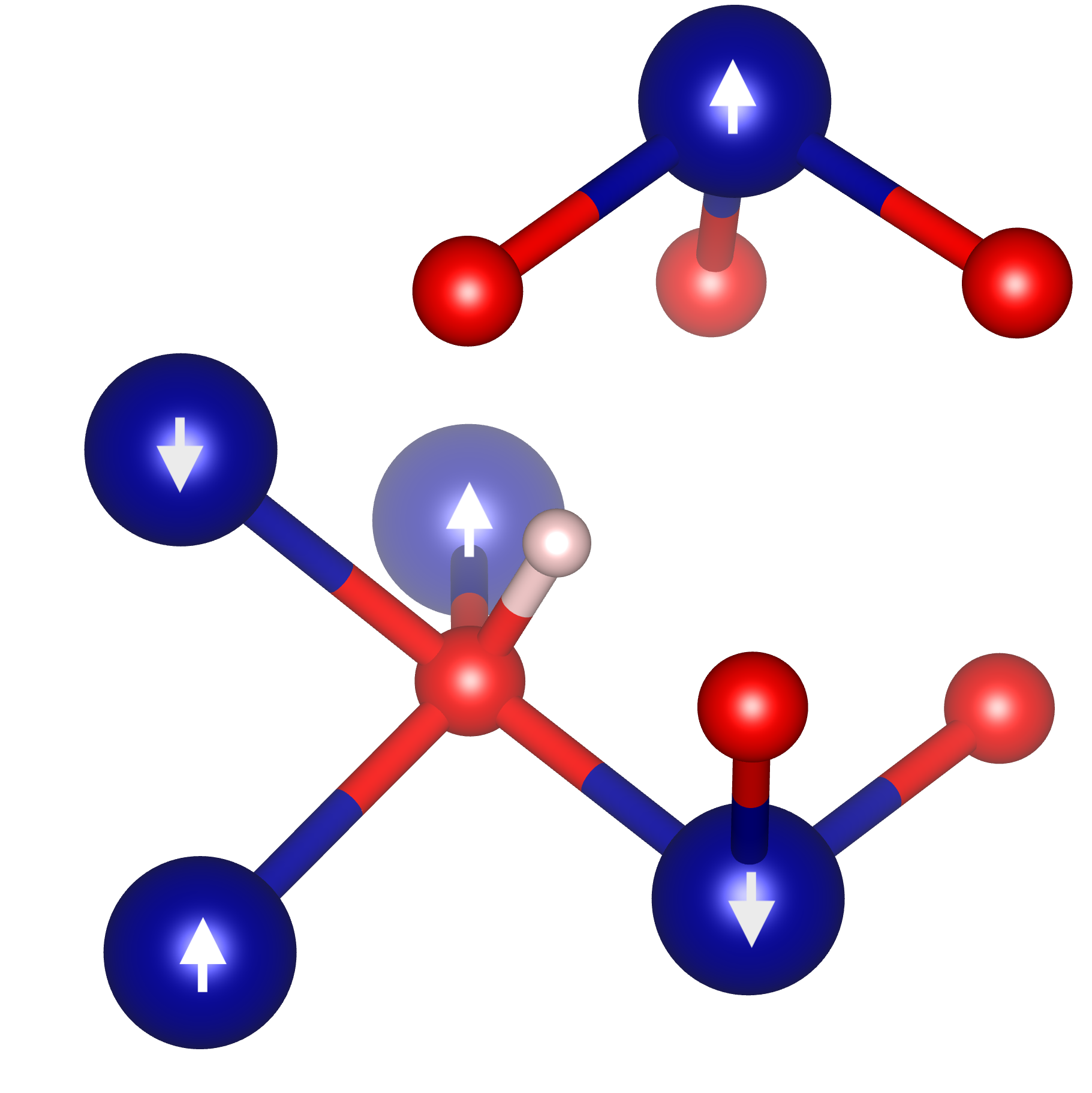} };
  \path (-1.37,1.5) node(b) {\large \textbf{E3}};
\end{tikzpicture}

  \end{minipage}\hfill
  \begin{minipage}[t]{0.22\textwidth}
    \vspace{-0.2cm}
\caption{Candidate muon stopping sites for E1-E3 (white spheres) identified using DFT. Transparency indicates a location \emph{behind} the solid atoms. Yellow isosurfaces indicate the spin density of the top-most occupied level ($d_{z^2}$) for the charge neutral E1 and E2 sites. Arrows indicate direction of magnetic moment.}
\label{fig:dftsites}
  \end{minipage}
\end{figure*}

 We have calculated the muon stopping sites in Cr$_2$O$_3$ using the Vienna ab initio Simulation Package \cite{kresse1993,kresse1994,kresse1996}, see Appendix \ref{apx:dft} for details.
 The positive muon was modeled as a hydrogen nucleus, embedded within an 80-atom  $2\times2\times2$ rhombohedral supercell (SC)  of Cr$_2$O$_3$. 
Two muon charge states were considered (1) the bare, positive muon, with a uniform charge background ensuring overall charge neutrality and (2) a neutral muon state allowing for the extra electron. 
 As a first step, promising initial muon positions were identified. Given the muon's tendency to form a muon-O bond of length $\sim \SI{1}{\angstrom}$, a set of 99 initial configurations were generated with the muon positions equidistributed on a \SI{1}{\angstrom} sphere centered on an oxygen. Since all oxygen atoms are electrostatically equivalent, it is sufficient to perform this search on a single oxygen. Initial static calculations of the Hellmann-Feynman forces allowed us to discard sites with forces larger than \SI{10}{eV/\angstrom}. For the remaining sites, the Cr$_2$O$_3$ ions were relaxed while keeping the muon fixed. This initial step of relaxing only the lattice allowed for a search for self-trapped metastable sites. 
 Finally, both the muon and lattice were fully relaxed until the Hellmann-Feynman forces were below \SI{5}{meV/\angstrom}. Additionally, muons were started out in the R and B sites, and in the unit-cell center - both the exact center (C$_E$) and slightly offset along the \emph{c}-axis (C$_O$). 
 The structure files  of all candidate  stopping sites for both charge states are available in the Supplemental Information \footnote{See Supplemental Information at \url{DOI: 10.5281/zenodo.3378994} \setcounter{footnote}{1}}. 

The total hyperfine field $\mathbf{B}_{tot}$ at each site $\mathbf{r}_\mu$ has (1) a dipolar contribution $\mathbf{B}_{dip}$ mainly from the  Cr $3d$ electrons, and (2) a Fermi contact term $
\mathbf{B}_c$ from unpaired spin density $\rho_s(\mathbf{r}_\mu)$ at the muon stopping site.
$\mathbf{B}_{dip}$ was calculated by embedding the distorted $2\times2\times2$ SC  in a superstructure of undistorted SCs and summing over the dipolar contribution from the spin density grid points. Despite a fine grid spacing of \SI{0.055}{\angstrom}, the finite grid causes artifacts for points in close proximity to  $\mathbf{r}_\mu$. This is mitigated by excluding grid points less than $R=\SI{0.5}{\angstrom}$  away from $\mathbf{r}_\mu$.
$\mathbf{B}_c$ is calculated by   \cite{onuorah2018}
\begin{equation}
\mathbf{B}_c=\tfrac{2}{3}\mu_0 \mu_B \rho_s(\mathbf{r}_\mu)\hat{c},
\end{equation} 
where $\mu_0$ is the vacuum permeability and $\mu_B$ the Bohr magneton. $\rho_s(\mathbf{r}_\mu)$ is approximated by projecting the spin density within a $R=\SI{0.5}{\angstrom}$ sphere onto an s-wave state at the $\mu^+$.

Results for both charge states (positive and neutral) are shown in Table \ref{tbl:dft}.  Calculated fields are given in units of frequency $f_i=\tfrac{\gamma_\mu}{2\pi} |\mathbf{B}_i|$. Note that $f_{tot}$ is obtained by  
\emph{vector} addition $|\mathbf{B}_{dip}+B_{c}\hat{c}|$.
The energies $\Delta E$ are given with respect to the ground state and are only comparable within a given charge state. 
The stated uncertainties are estimated by varying the sphere radius \emph{R} for both the contact and dipole term calculations in the range $0.3-\SI{0.7}{\angstrom}$.

For the positive charge state (superscripted  +), the B1$^+$ site cannot be stabilized, and muons placed in the primitive unit-cell center (C sites) can be discounted as viable muon stopping sites based on magnitude and direction of $\mathbf{B}_{int}$, and large $\Delta E$. For the same reasons, B0$^+$ is unlikely to represent a muon stopping site.  
Unless purposefully placed in a B$^+$ or C$^+$ site, muons relax into a position close to but distinct from the R-sites (the electrostatic minima of the undistorted lattice). We name this site D due to the doughnut-shaped potential energy surface formed by  electrostatically equivalent sites. The difference between D and R arises predominantly from the muon-induced lattice distortion, which was not accounted for previously.
The close proximity of adjacent D$^+$-sites makes them  excellent candidates for the E3 environment. Possible explanations for 
the discrepancy of \SI{22}{\percent}  between $f^\mathrm{ZF}_\mathrm{E3}= \SI{76.7}{MHz}$ and the calculated value are discussed below. 

While DFT calculations considering the positive muon can account for E3 muons undergoing local hopping, E1 and E2 are thus far unexplained, motivating a search for charge-neutral muon states. Results are shown in the bottom half of Table \ref{tbl:dft}. B0$^0$, B1$^0$ and C$_E^0$ can be dismissed based on large $\Delta E$, and C$_O^0$ based on the large $f_{tot}$. 
This leaves five variations of the D$^0$ site, labeled D1$^0$-D5$^0$. 
 Comparison with Table \ref{tbl:lowBFFTbot} and consideration of the calculated energies suggest that D1$^0$ and D2$^0$ are candidates for E1 and E2, respectively: the  measured and calculated frequencies of for E1 (+\SI{14}{\percent}) and  E2 (+\SI{12}{\percent}) and $\theta$ agree reasonably well and $\Delta \varphi$(E2-E1) is close to the expected \SI{30}{\degree}, compare Table \ref{tbl:summary}. 
 Additionally, consistent with the proposed E2$\rightarrow$E1 transition, D2$^0$ has a larger energy than D1$^0$. The differences between D1$^0$-D5$^0$, and further aspects of the D2$^0\rightarrow$D1$^0$ transition are discussed below.
 Fig.  \ref{fig:dftsites} shows the positions of the E1-E3 candidate stopping sites D1$^0$, D2$^0$ and D$^+$.

\begin{figure}[t]
 \includegraphics[width=8.5cm]{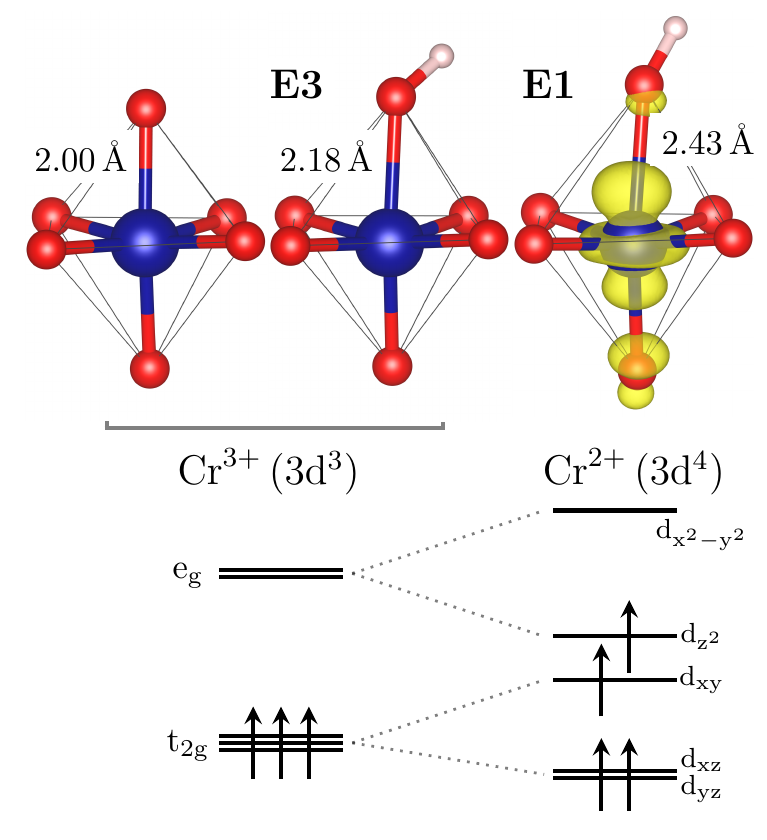} 
 
\caption{Top: Octahedrally coordinated Cr atom (1) without the muon, with (2) the positive muon and (3) a neutral charge state formed by the muon and an extra electron localized on the Cr, changing its valence state from Cr$^{3+}$ to Cr$^{2+}$. The presence of the positive muon removes some electron density from its associated Cr-O bond, causing elongation from \SI{2.00}{\angstrom} to \SI{2.18}{\angstrom}. The localization of the extra electron arising from the Coulomb attraction of the muon and the energy gain from the lattice distortion further elongates the Cr-O bond to \SI{2.43}{\angstrom} and leads to the formation of a charge-neutral muon-polaron complex. Bottom: Schematic representation of the crystal field: the occupation of the degenerate $e_g$ orbital by the extra electron leads to a Jahn-Teller distortion.  The yellow isosurface shows the  charge density of the top-most occupied band for the charge-neutral case, confirming that indeed the $d_{z^2}$ level is occupied.
 }
\label{fig:JTdist}
\end{figure}

A more detailed analysis of the charge-neutral D$^0$ states reveals that
the extra electron localizes predominately on a nearby Cr, where it changes the valence from Cr$^{3+}$ to Cr$^{2+}$, and singly occupies the initially orbitally degenerate $e_g$ orbitals. This causes a Jahn-Teller distortion \cite{jahn1937} by further elongating the Cr-O bond of the oxygen the muon is bound to from \SI{2.18}{\angstrom} for the bare muon to \SI{2.43}{\angstrom} in the charge neutral case, see Fig. \ref{fig:JTdist}. The subsequent lowering of energy, see Fig. \ref{fig:JTdist}, stabilizes this charge-neutral complex of a negatively charged JT polaron \cite{holstein1959,nickisch1983a} and positive muon.  
This proposed mechanism is supported by the extra electron occupying the lowered $e_g$ level $d_{z^2}$, 
see yellow isosurface representing the charge density of top-most occupied band in Fig. \ref{fig:JTdist}.
We note the similarity to the paramagnetic Ti-O-Mu complex recently observed in (nonmagnetic) TiO$_2$ \cite{vilao2015,shimomura2015}, where an unpaired electron sits on a nearby Ti atom, and the oxygen-bound muon forms a complex with the resulting small polaron. Crucially, however, the muon-polaron complex reported on here is not paramagnetic and therefore \emph{distinct from Mu}, since the bound electron is strongly coupled to the $3d$ electrons of the Cr host ion. This is discussed  in detail in Sec. \ref{sec:discussion}.

The D1$^0$-D5$^0$ states arise from the extra electron being localized on different Cr and small variations in the muon position. A transition between D$^0$ states is mainly characterized by a change in position of the extra electron rather than the muon. Note that the spin of the extra electron, being coupled to the $3d$ electrons of its Cr host, may be different for different D$^0$ states, as indicated in Table \ref{tbl:dft}. 
 We propose that the higher energy states D3$^0$-D5$^0$ are not occupied since they can easily transition into either D1$^0$ or D2$^0$ depending on their spin. However, going from D2$^0$ into the ground state D1$^0$ requires an electron spin flip, i.e. an additional energy barrier has to be overcome. 
 While the precise process for the E2$\rightarrow$E1 transition is still under investigation, we attribute the observation of the metastable E2 environment to the existence of such a spin barrier.

In general, the energy barrier to move the joint muon-polaron complex  is expected to be  significantly larger than for the bare muon \cite{emin1982}, providing a compelling explanation for the stability of E1 and E2, and coexistence of site-stable muons and highly mobile muons in E3.

Note that DFT local field predictions depend on the approximation of the exchange-correlation functional and value of the $\mathrm{U_{eff}}$ parameter. By comparing the LDA, PBEsol and SCAN functionals with a reasonable range of $\mathrm{U_{eff}}$ corrections, we find variations in the predicted frequency magnitudes of $\sim15$ \%, in the $\theta$ angle of $\sim 60 \%$ and in the $\phi$ angle of $\sim10$ \% with respect to the LDA+$\mathrm{U_{eff}}=4$ eV values, see Supplemental Information \cite{Note1}. 

Finally, we note that an accurate consideration of the zero-point motion of the muon is necessary to calculate the formation energies of the different charge states \cite{vandewalle2000}, and the energy barriers for intra-E3 hopping and the E2$\rightarrow$E1 transition.  While a full treatment of the quantum nature of the muon is beyond the scope of the present paper, work is currently ongoing and will be published separately.

 \section{Discussion}
\label{sec:discussion}

Paramagnetic Mu centers are expected to be subject to fast relaxation in magnetic materials \cite{cox2006}. In the previous sections, we presented strong evidence for the formation of a charge-neutral muon-polaron complex in Cr$_2$O$_3$, that, while not exhibiting signatures conventionally expected from neutral charge states, significantly influences the muon behavior and contributes a well-resolved signal. 
In particular, rather than giving rise to a well-defined spectrum  of typically two or four precession frequencies  that are determined by a spin Hamiltonian involving the muon and electron Zeeman energies and a muon-electron hyperfine  interaction \citep{yaouanc2011}, the precession signal for the muon-polaron complex  in Cr$_2$O$_3$ consists of a single frequency much like the normal positive charge state, i.e. the bare $\mu^+$ with no additional electron nearby. The reason for the different behavior compared to that of Mu is that the muon-polaron complex is  \emph{not paramagnetic}, since its bound electron is strongly coupled to the $3d$ electrons of the Cr host ion, which themselves are antiferromagnetically coupled to the ordered network of magnetic ions of the host.
We note that the bound electron is not centered on the $\mu^+$ but localizes on a nearby Cr, which is not unique to the muon-polaron complex, since similar situations are found for the  paramagnetic Mu complex in TiO$_2$ \cite{cox2006a,vilao2015,shimomura2015}, and bond-centered Mu in silicon \cite{kiefl1988}.

The relevance of the discovery of a muon-polaron complex in Cr$_2$O$_3$ may extend to other magnetic oxides, 
such as CuO \cite{duginov1994,nishiyama2001}, Fe$_2$O$_3$ and FeTiO$_3$ \cite{boekema1983}, Fe$_3$O$_4$ \cite{bimbi2008} and the orthoferrites \cite{holzschuh1983a}, which all show multiple zero-field precession frequencies that, while attributed to metastable sites, are not conclusively explained. 
In general, our study suggests that neutral charge states and their potential impact on crystal-field levels have to be carefully considered in all \emph{insulating} magnetic materials, and that detailed DFT calculations can be used to  separate  intrinsic magnetic properties from muon-induced effects.

We emphasize that the present muon-polaron complex is different from  the controversial concept of a muon-induced magnetic polaron proposed by Storchak \emph{et al} \cite{storchak2009a,kiefl2011,storchak2011,amato2014}, which invokes a localized electron bound to a muon mediating a ferromagnetic coupling between neighboring magnetic ions, resulting in a ``ferromagnetic droplet'' characterized by a gigantic local spin.

We speculate that  muon-polaron complex formation  in Cr$_2$O$_3$ occurs by a  similar mechanism to Mu formation in semiconductors \cite{cox2009}. Upon implantation, the muon slows down by creating electron-hole pairs. Towards the end of its ionization track it may capture an electron and subsequently form a charge-neutral complex. The implantation and electron capture processes are epithermal, and thus independent of sample temperature and thermodynamic equilibrium, providing justification for the previously stated hypothesis that E1-E3 are populated with the same ratio at all temperatures. Muons may stop and self-trap in metastable (charge) states with energies larger than the ground state, and only de-excite if thermally activated transitions, e.g. from E2 to E1, are accessible during the muon lifetime \cite{browne1982,stoneham1984}.

Paramagnetic Mu has been used extensively to investigate the dopant characteristics of hydrogen in a wide range of semiconductors including oxides \cite{chow1998,cox2006a,cox2006,cox2009}. This is because the electronic structure of Mu in a solid is virtually identical to that of hydrogen, aside from small differences caused by the larger zero-point motion due to the lighter muon mass.
 We propose that with the observation of a neutral charge state in  Cr$_2$O$_3$, $\mu$SR has shown its ability to investigate the behavior of interstitial hydrogen in magnetic oxides as well. A thorough understanding of unintentional hydrogen doping in such materials is crucial, since a wide range of technologically relevant fields such as dilute magnetic semiconductors for spintronics \cite{dietl2014,baqiah2016,bououdina2019,vandewalle2003} and superconductivity depend on a precise control of charge carriers in magnetic oxides, or, for example in the case of multiferroics, require low-leakage thin films \cite{spaldin2019}.

Interestingly, interstitial hydrogen is predicted to form a shallow donor state in Cr$_2$O$_3$, alongside a range of other materials including the aforementioned CuO, Fe$_2$O$_3$ and FeTiO$_3$ \citep{kilic2002}. The observation of E1 up to $T_N$ suggests that the muon-polaron complex stays intact up to room temperature, indicating that hydrogen is instead
a deep impurity. This is in contrast to an ongoing  study of muon-polaron complexes in Fe$_2$O$_3$ (to be published separately), which suggests complex ionization above \SI{200}{K}, indicating shallow donor behavior.
We note that Cr$^{2+}$ with $3d^4$ high spin is strongly JT-active, while Fe$^{2+}$ with $3d^6$ high spin is weakly JT-active, and hypothesize that donor characteristics are determined, at least in part, by the strength of the JT effect.
The role of magnetic interactions in the stabilization of the muon-polaron complex remains an open question.

The prediction of hydrogen induced \emph{n}-type conductivity in ZnO \cite{vandewalle2000} and the subsequent experimental observation of the corresponding shallow Mu donor state \cite{cox2001} prompted a search for criteria to predict dopant behavior and charge state of interstitial hydrogen and muons, leading to generalized principles  for elemental and binary semiconductors \cite{vandewalle2003} and oxides \cite{kilic2002,peacock2003,cox2006a,xiong2007}. We hope that our discovery of a JT-stabilized muon-polaron complex stimulates research to extend those principles to explicitly account for polaronic, and if required, magnetic contributions.   
Such adjusted criteria could provide valuable guidelines on whether or not neutral charge states are expected in $\mu$SR experiments on insulating magnetic materials.

Lastly, having acquired a detailed understanding of how the muon interacts with Cr$_2$O$_3$, we address possible implications arising from implanting a point charge into a linear magnetoelectric (ME) material. Khomskii predicts that a point charge inside an isotropic linear   ME is surrounded by a monopole-like magnetic texture, and subject to a force in applied magnetic fields \cite{khomskii2014}. The monopolar magnetic field distribution is only expected sufficiently far from the charge where the bulk approximation holds, while no predictions are made for the immediate core region.
The case of a muon inside Cr$_2$O$_3$ is more complex since (1) the ME coupling is not isotropic \cite{wiegelmann1994}, (2) some muons are bound in a muon-polaron complex, an effectively charge-neutral entity, and (3) the muon induces significant lattice distortions in its immediate vicinity and experiences a magnetic field dominated by the core region which cannot be treated in the bulk limit. Thus the muon may sense a change in its magnetic environment in response to the electric field of its charge, arising from changes in the position of magnetic ions and canting of magnetic moments in its immediate environement,  i.e. a  \emph{local} ME effect, however such an effect  may not necessarily follow the bulk ME coupling. Results from preliminary non-collinear DFT calculations 
 suggest only a small spin canting $<\SI{0.3}{\degree}$ in the immediate neighborhood of the muon, however the elevated ME coupling at higher temperatures is not yet taken into account. 
We note that while the ME response outside the core region is expected to have monopolar contributions only in linear ME materials, muon-induced local ME effects may occur in non-ME compounds as well. 
Finally, we comment on the prediction that a static charge inside a ME is subject to a force in a magnetic field \cite{khomskii2014}. Using a lower limit of \SI{1}{eV/\angstrom^2} for the force constant of the potential experienced by the muon bound to an oxygen, the change in position in response to a field-induced force on the muon can be estimated to be smaller than \SI{e-5}{\angstrom} in \SI{4}{T}, i.e. is negligible in the context of this experiment. 
Despite the challenges discussed above, the  investigation  of local ME effects induced by the muon in its duality as a test charge and sensitive probe for magnetism remains a fascinating area of research,
and we hope that the present paper initiates studies both experimental and theoretical in this direction.

\section{Conclusions}

In summary, we carried out a comprehensive $\mu$SR study of Cr$_2$O$_3$ under zero-field conditions and in applied magnetic fields. In zero field, we observe three spin precession frequencies, attributed to three distinct muon environments  E1-E3 with different 
internal magnetic fields. Small applied magnetic fields along various symmetry directions split the
observed frequencies into multiplets, providing detailed information on the orientation of the internal
fields. The temperature dependence reveals a rich dynamic behavior that we explain in terms of a thermally activated transition between E2 and E1, and intra-E3 local muon hopping.  Notably, we observe coexistence of highly dynamic E3 muons and site-stable muons in E1 and E2. Muon stopping sites and charge states for all three environments are determined using DFT, and the coexistence is explained by the formation of a charge-neutral, JT-stabilized muon-polaron complex. 
The identification of such a charge-neutral complex in the antiferromagnet Cr$_2$O$_3$ has implications for other magnetic oxides, since the formation of muon-polaron complexes can significantly influence the stability and location of stopping sites, but its existence may be  ``hidden" since the  behavior conventionally associated with neutral charge states is not displayed.
Furthermore, this discovery opens up a route to study interstitial hydrogen in magnetic oxides, where precise control of the carrier density may be  critical for device functionalities.
Given the technological importance of magnetic oxides, we hope this study stimulates a search for generalized principles describing the dopant behavior of hydrogen that account for polaronic, and if required, magnetic contributions.

\begin{acknowledgments}
This research was performed at the TRIUMF Centre for Materials and Molecular Science. The authors thank R. Abasalti and D. Vyas for excellent technical support. Initial spectra were taken at the Swiss Muon Source S$\mu$S at PSI.  We thank S. R. Dunsiger for critical reading of the manuscript and helpful discussions, and G. Levy for help with crystal alignment. MHD acknowledges support from a  SBQMI QuEST Fellowship, and would like to thank A. Nojeh and I. Elfimov for stimulating discussions. Financial support came from a NSERC  Discovery grant to RFK. JKS and NAS acknowledge funding from the European Research Council (ERC) under the European Union's Horizon 2020 research and innovation programme grant agreement No 810451.
Computational resources were provided by ETH Z\"urich and the Swiss National Supercomputing Centre, project ID s889.
  SH acknowledges  financial support by the Swiss National Science Foundation (SNF-Grant No. 200021-159736).
Crystal structures were drawn using \textsc{VESTA} \cite{Momma2011}.
\end{acknowledgments}

\appendix

\section{Details on internal magnetic field orientation}
\label{sec:apxBint}
Under the assumption that $\mathbf{B}_{ext}$ does not induce changes of $\mathbf{B}_{int}$,  the resulting frequency multiplets can be calculated by simple vector addition. 
For $\mathbf{B}_{ext}||c$, only  $\theta$ is relevant, and we expect the ZF frequency $f^\mathrm{ZF}$ to be split into a doublet, see Fig. \ref{fig:splitFFT}(a), with  the frequencies $f^\pm$ given by 
\begin{align}
f^\pm = \left[\left(f_{ext}\pm f^\mathrm{ZF} \sin(\theta)\right)^2+\left(f^\mathrm{ZF}\cos(\theta)\right)^2\right]^{1/2}, 
\label{eqn:lowBpara}
\end{align}
where $f_{ext}=\gamma_\mu /(2\pi)\cdot |\mathbf{B}_{ext}|$. Given the equal number of sites with $+\theta$ and $-\theta$,  equal amplitudes for both doublet lines are expected (for $\mathbf{B}_{ext}\ll \mathbf{B}_{int}$).

For $\mathbf{B}_{ext}|| [11\bar{2}0] \bot c$, the multiplet splitting,  determined by both $\varphi$ and $\theta$, is given by
\begin{align}
f(\theta,\varphi)=&\Big[(f^\mathrm{ZF}\sin(\theta))^2+(f^\mathrm{ZF}\sin(\varphi)\cos(\theta))^2
\nonumber\\+&  (f_{ext}+f^\mathrm{ZF}\cos(\varphi)\cos(\theta))^2   \Big]^{1/2}
\label{eqn:lowBbot}
\end{align}
Only for  $\delta=0$ or $ \pm \SI{30}{\degree}$, $\mathbf{B}_{ext}$ causes a multiplet splitting with less than six lines, but yields either a quadruplet for $\delta=0$ or a triplet  for $\delta=\pm \SI{30}{\degree}$, with an amplitude ratio of 1:2:2:1 and 2:2:2, respectively, see Figs. \ref{fig:splitFFT}(b) and (c).

 \begin{table}[t]
 \caption{Comparison of frequencies $f_{exp}$ measured at  \SI{2.55}{K} in C2, $\SI{300}{G}|| c$, and calculated values $f_{calc}$, obtained with Eqn.  (\ref{eqn:lowBpara})  using $f^\mathrm{ZF}$  from Fig. \ref{fig:ZFprec}(c) and by optimizing $\theta$ to minimize $|f_{calc}(\theta)-f_{exp}(\theta)|+|f_{calc}(-\theta)-f_{exp}(-\theta)|$ .  \label{tbl:lowBFFTpara}}
 
 \begin{ruledtabular}
 \begin{tabular}{c|c||c||c|c}
site & $f^\mathrm{ZF}$ [MHz]& $f_{exp}$ [MHz]& $f_{calc}$ [MHz]&$\theta$ [\si{\degree}]  \\ \hline
E1& $68.53\pm 0.01$ & $66.98\pm 0.01$ & 66.98& \textcolor{black}{-24.0} \\
 &  & $70.28\pm 0.01$& 70.28 & \textcolor{black}{+24.0} \\  \hline \hline 
 
E2&$102.12\pm 0.01$ & $101.78\pm 0.01$ & 101.78& \textcolor{black}{-6.0} \\
 &  & $102.62\pm 0.01$ &102.63& \textcolor{black}{+6.0} \\ 
  \hline \hline 
 
 E3& $76.71\pm 0.01$ & $76.46\pm 0.03$ & 76.46& -5.0 \\
 &  &  $77.18\pm0.04$& 77.17 & +5.0 \\

 \end{tabular}
  \end{ruledtabular}

 \begin{ruledtabular}
 \caption{
 Comparison of frequencies $f_{exp}$ measured at  \SI{2.2}{K} in C1, $\SI{200}{G}\bot c$, and calculated values $f_{calc}$, obtained with Eqn.  (\ref{eqn:lowBbot}) using $f^\mathrm{ZF}$ from Fig. \ref{fig:ZFprec}(c), $\theta$  from Table \ref{tbl:lowBFFTpara}, $\varphi$ given by $\SI{0}{\degree}+\delta, \pm\SI{60}{\degree}+\delta, \pm\SI{120}{\degree}+\delta$ and $\SI{180}{\degree}+\delta$ with $\delta=\SI{0}{\degree}$ for E1, $\delta=\SI{30}{\degree}$ for E2 and $\delta=\SI{17.5}{\degree}$ for E3. Bold values correspond to the average of vertically adjacent values. 
 \label{tbl:lowBFFTbot}}
 \begin{tabular}{c|c|c||c|c|c}
site & $f^\mathrm{ZF}$ [MHz]&$\theta$ [\si{\degree}] & $f_{exp}$ [MHz]& $f_{calc}$ [MHz] & $\varphi_{calc}$ [\si{\degree}]\\ \hline
E1& $68.52\pm 0.01$ & 24& $66.08\pm 0.03$ & 66.05& \textcolor{black}{180}  \\
&&&$67.38\pm 0.01$ & 67.32& \textcolor{black}{$\pm 120$} \\
&&&$69.76\pm 0.01$ & 69.80&\textcolor{black}{$\pm 60$} \\
&&&$71.00\pm 0.03$ & 71.00& \textcolor{black}{0} \\ \hline \hline

E2& $102.12\pm 0.01$ & 6& $99.80\pm 0.01$ & 99.79& \textcolor{black}{$\pm 150$}  \\
&&&$102.16\pm 0.01$ & 102.15& \textcolor{black}{$\pm 90$} \\
&&&$104.45\pm 0.01$ & 104.46&\textcolor{black}{ $\pm 30$} \\ \hline \hline

E3& $76.69\pm 0.01$ & 5 & & 74.11& \textcolor{black}{$ + 162.5$}  \\
&&& $74.51\pm 0.07$ & \textbf{74.42}& \textcolor{black}{--}  \\
&&&  & 74.72& \textcolor{black}{$- 137.5$}  \\\cline{4-6}
&&&$76.15\pm 0.09$ &76.15  & $+ 102.5$  \\\cline{4-6}
&&&$77.25\pm 0.07$ & 77.32 &$- 77.5$  \\\cline{4-6}
&&& &78.70& \textcolor{black}{$+ 42.5$}  \\
&&&$78.89\pm 0.05$ & \textbf{78.98}& \textcolor{black}{--} \\
&&& &79.27& \textcolor{black}{$- 17.5$}  

 \end{tabular}
  \end{ruledtabular}

 \begin{ruledtabular}
\caption{Comparison of frequencies $f_{exp}$ measured at  $T=\SI{2.1}{K}$, $\mathbf{B}_{ext}=\SI{4}{T}||c$, and calculated values using Eqn. (\ref{eqn:lowBpara}), $f_{ext}=\SI{541.98}{MHz}$ , and $\theta$ values that minimize $|f_{calc}(\theta)-f_{exp}(\theta)|+|f_{calc}(-\theta)-f_{exp}(-\theta)|$. Note that \SI{541.14}{MHz} is listed twice in $f_{exp}$, as $f^-_\mathrm{E2}$ and $f^-_\mathrm{E3}$ overlap. \label{tbl:baseHighField}} 
 
 \begin{tabular}{c|c||c||c|c}
site & $f^\mathrm{ZF}$ [MHz]& $f_{exp}$ [MHz]&$\theta$ [\si{\degree}] & $f_{calc}$ [MHz] \\ \hline
E1& 68.53 & $517.90\pm 0.01$ & \textcolor{black}{-24.01}& 517.90 \\
 &  & $ 573.26\pm 0.01$ & \textcolor{black}{+24.01}& 573.30 \\  \hline \hline 
 
E2&102.12 & $541.14\pm 0.01$ & \textcolor{black}{-5.88}& 541.14 \\
 &  & $561.62\pm 0.02$ & \textcolor{black}{+5.88}&561.71 \\ 
  \hline \hline 
 
 E3& 76.71 & $541.14\pm 0.01$ & -4.69& 541.14 \\
 &  &  $553.54\pm0.02$ & +4.69& 553.56 \\

 \end{tabular}
  \end{ruledtabular}

 \end{table}

For $\mathbf{B}_{ext}\parallel c$, Eqn. (\ref{eqn:lowBpara}) is used to calculate the expected doublet frequencies ($f_{calc}$) with  $f^\mathrm{ZF}$ from Fig. \ref{fig:ZFprec}(c). Minimizing $|f_{calc}(\theta)-f_{exp}(\theta)|+|f_{calc}(-\theta)-f_{exp}(-\theta)|$  yields  $\theta$ values that produce excellent agreement between $f_{calc}$ and $f_{exp}$ for E1-E3, see Table \ref{tbl:lowBFFTpara}.

Using those $\theta$ values, Eqn. (\ref{eqn:lowBbot}) is used to investigate the $\mathbf{B}_{ext}\parallel [11\bar{2}0]\bot c$ multiplet splittings. For E1 and E2, the measured frequencies are in very good agreement with the calculated values $f_{calc}$ for $\delta = 0$ and $\SI{30}{\degree}$, respectively, see Table \ref{tbl:lowBFFTbot}. 
Furthermore, the amplitudes are close to the predicted 1:2:2:1 and 2:2:2 ratio. 
For E3, $\delta =\pm\SI{17.5}{\degree}$ yields a reasonable agreement between $f_{calc}$ and $f_{exp}$, assuming that the outer two lines on either sides of the resulting sextet are not resolved, but appear at their average frequency (shown in bold in Table \ref{tbl:lowBFFTbot}) with twice the amplitude. From that, an amplitude ratio of 2:1:1:2 is expected, which is indeed observed.
Considering crystal alignment and the observed frequencies and amplitude ratios, we estimate the uncertainties to be $\pm\SI{1}{\degree}$ for all $\theta$ values, $\pm\SI{3.5}{\degree}$ for $\delta_\mathrm{E1}$ and $\delta_\mathrm{E2}$, and  $\pm\SI{2}{\degree}$ for $\delta_\mathrm{E3}$.

\section{Details on data taken in large external fields }
\label{apx:high}

The $\mu$SR spectra are analyzed over the first \SI{1}{\micro s} \footnote{A small sample misalignment of less than $\SI{1}{\degree}$ seems to cause a further splitting of the line at times longer than \SI{1}{\micro s}. This sub-splitting affects the various multiplet frequencies differently and has a small temperature dependence which is not fully understood. Limiting the analysis (and the depicted FT) to the first \SI{1}{\micro s} effectively smooths out the substructure; however a separate relaxation rate for the E1 doublet is required and attributed to broadening caused by this misalignment sub-splitting.} with the following models: Below \SI{50}{K}, where $f^-_\mathrm{E2}$ and $f^-_\mathrm{E3}$ overlap, six components are considered; all three doublets share amplitude, and the overlapping line is fit to two components with the same phase and frequency, but share amplitude and relaxation rate with $f^+_\mathrm{E2}$ and $f^+_\mathrm{E3}$, respectively. At and above \SI{50}{K}, the data is fit to up to five exponentially damped oscillatory functions, with shared amplitudes for the E1 and E2 doublets, but separate relaxation rates. 

Additionally, for the spectra at $T=50,~110-\SI{155}{K}$, the amplitude of the E3$'$ component, see Fig. \ref{highFieldFFT}, is fixed to 0.0365, a value obtained by averaging over the amplitude of surrounding temperature points. This constraint is necessary  in order to get a meaningful measure of the relaxation rate of this strongly damped component.

\begin{figure}[t]
\center
            \includegraphics[]{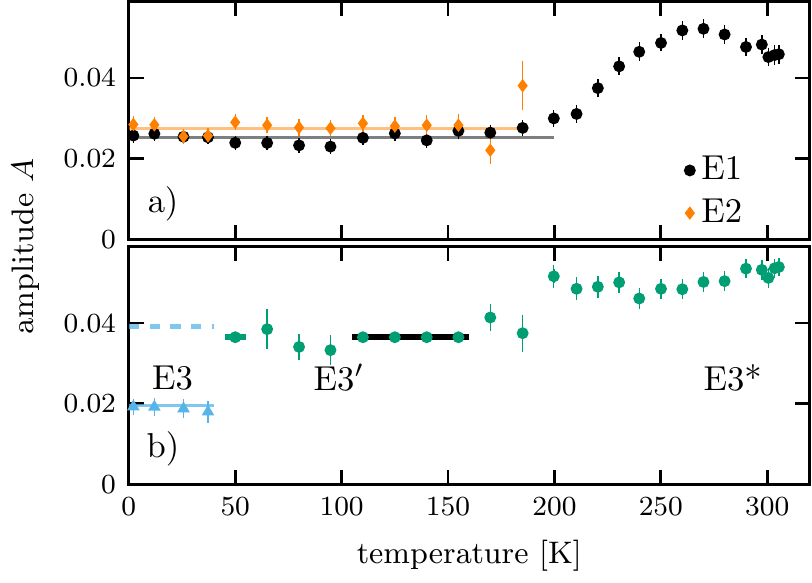}

\caption{Fit results for $\SI{4}{T}||c$ spectra:  amplitudes for (a) E1  and E2, and (b) for E3, E3$'$ and E3*. The black bar indicates that the value was fixed to 0.0365. }
\label{apxhighFieldFit}
\end{figure}

The fitted amplitudes of E1 and E2 are shown in Fig. \ref{apxhighFieldFit}(a). The temperature dependence tracks largely the ZF behavior: E2 is constant in amplitude, and E1 starts to increase above \SI{200}{K}.The displayed values are the amplitudes of one of the doublets; the total amplitude of muons in either E1 or E2 is twice that. 
 The remaining amplitudes, associated with E3, E3$'$ and E3* are displayed in Fig.  \ref{apxhighFieldFit}(b). The points with a thick horizontal bar indicate temperature points where the amplitude was constrained. The E3 doublet amplitude accounts for only one of the doublets; the dashed line indicates the total (double) value for reference.

The frequencies obtained from fitting to five exponentially damped oscillatory components are shown in Table \ref{tbl:baseHighField}. Using  Eqn.  (\ref{eqn:lowBpara}), $f_{calc}$ is calculated for the $\theta$ value that minimizes $|f_{calc}(\theta)-f_{exp}(\theta)|+|f_{calc}(-\theta)-f_{exp}(-\theta)|$, yielding very good agreement with the data, and  $\theta$ values matching closely those obtained in low field, see Table \ref{tbl:lowBFFTpara}. 

The temperature dependence of $f^\pm_\mathrm{E1}$ and $f^\pm_\mathrm{E2}$  is modeled by Eqn. (\ref{eqn:lowBpara}) with $\theta$ from Table \ref{tbl:baseHighField}, and an interpolated temperature dependence of $f^\mathrm{ZF}_\mathrm{E1}(T)$ in Fig. \ref{fig:ZFprec}(c). $f^\mathrm{ZF}_\mathrm{E2}(T)$ was further approximated by scaling $f^\mathrm{ZF}_\mathrm{E1}(T)$ by the low temperature E2/E1 frequency ratio
$f^\mathrm{ZF}_\mathrm{E2}/f^\mathrm{ZF}_\mathrm{E1}|_{\SI{2.2}{K}}=1.49$.
The red lines in Fig. \ref{highFieldFit}(c) represent the calculated doublet frequencies  $f^\pm_\mathrm{E1}(T)$ and $f^\pm_\mathrm{E2}(T)$. There is good agreement with the data, indicating that $\theta$ is largely temperature independent. A small but consistent deviation from the predicted line is observed for the E1 doublet between $200-\SI{300}{K}$, and discussed in Sec. \ref{sec:discussion}.

\section{Details on the E2$\rightarrow$E1 transition}
\label{apx:appE2-E1}

Here we describe a model for the E2$\rightarrow$E1 transition assuming  a thermally activated, exponential rate of the form $\Lambda(T)=\nu_0 \exp(-E_a/k_BT)$, where $E_a$ and $\nu_0$ are activation energy and attempt frequency. 
The following expression describes the observable signal precessing at  $f_\mathrm{E1}$ (compare \cite{meier1982,dehn2016}):
\begin{eqnarray}
S_\mathrm{E1}(t)&=&A_\mathrm{E1} \cos(2\pi f_\mathrm{E1}t)  \nonumber \\  &+& A_\mathrm{E2} \displaystyle\int^{t}_0  \Lambda \mathrm{e}^{-\Lambda t'} \cos(2\pi f_\mathrm{E1}(t-t')+2\pi f_\mathrm{E2}t')\mathrm{d}t'\nonumber \\
&=& \mathcal{A} \cos(2\pi f_\mathrm{E1} t + \Phi).
\label{eqn:trans}
\end{eqnarray}
Muons starting out in E1 are described by the first term, whereas the second  describes the E2$\rightarrow$E1 transition taking into account the phase acquired while evolving in E2. For $t\gg \Lambda^{-1}$, the resultant combined amplitude $\mathcal{A}$ and phase $\Phi$ can be expressed as
\begin{eqnarray}
\mathcal{A}&=&\sqrt{\frac{A_\mathrm{E2}^2+2A_\mathrm{E2}A_\mathrm{E1}}{\zeta(T)^2+1}+A_\mathrm{E1}^2} \label{eqn:amp}\\
\Phi&=&- \arctan \left[\frac{A_\mathrm{E2} \zeta(T)}{A_\mathrm{E2}+A_\mathrm{E1}(1+\zeta(T)^2))}\right],
\label{eqn:trans_phase}
\end{eqnarray}
where
\begin{equation}
\zeta(T)=\frac{2\pi[f_\mathrm{E1}(T)-f_\mathrm{E2}(T)]}{\nu_0 \exp(-E_a/k_BT)}.
\label{eqn:zeta}
\end{equation}
The expressions above assume that $\mathbf{B}_{int}$ in E1 and E2 are parallel and perpendicular to the initial spin polarization $\mathbf{P}_i$. In order to compare this model to data taken in both ZF and large $\mathbf{B}_{ext}$, small modifications, outlined below, are necessary.

In general, $\mathbf{B}$ is not perpendicular to the initial polarization $\mathbf{P}_i$, causing the component of $\mathbf{B}||\mathbf{P}_i$ to act as a holding field. The  polarization signal can be decomposed into oscillating and non-oscillating components  \cite{yaouanc2011}
\begin{equation}
S(t)\propto \cos(\theta)^2\cos(\gamma_\mu |\mathbf{B}|t)+\sin(\theta)^2,
\label{eqn:decomposition}
\end{equation}
where $\SI{90}{\degree}-\theta$ is the angle enclosed by $\mathbf{B}$ and $\mathbf{P}_i$.

\textbf{Zero field.} Here, $\mathbf{P}_i||c$, and for both E1 and E2,  the internal field $\mathbf{B}_{int}$ encloses an angle $\theta$ with the c-plane, see Table \ref{tbl:lowBFFTpara}. Thus $\mathbf{B}_{int}\cdot \mathbf{P}_i \neq 0$, resulting in a non-oscillatory signal component.The oscillatory $f_\mathrm{E1}$ signal amplitude represents only a fraction of $\cos(\theta_\mathrm{E1}=\SI{24}{\degree})^2=0.83$ of muons in E1, while for E2, $\cos(\theta_\mathrm{E2}=\SI{6}{\degree})^2= 0.99$, and virtually the complete E2 component is oscillating.
 Thus if the complete E2 component with observed amplitude $A_\mathrm{E2}$   coherently transfers to E1, the transferred amplitude observed at $f_\mathrm{E1}$ is only $ (\cos(\theta_\mathrm{E1})/\cos(\theta_\mathrm{E2}))^2 A_\mathrm{E2}$. Strictly, this is only valid for $\Lambda^{-1}\ll 1/f_\mathrm{E2}$; furthermore, a change $\Delta \varphi =\SI{30}{\degree}$ of the internal field direction upon transition is neglected. A calculation addressing both issues was carried out and yielded slight improvements but no major deviations from the simple model, and was not included for clarity.
The temperature dependence of $f_\mathrm{E1}^\mathrm{ZF}(T)$ and $f_\mathrm{E2}^\mathrm{ZF}(T)$ is obtained by interpolating $f_\mathrm{E1}^\mathrm{ZF}$ shown in Fig. \ref{fig:ZFprec}\,(c).

\textbf{High field.} As $\mathbf{B}_{ext}\bot \mathbf{P}_i >\mathbf{B}_{int}$, the resultant internal field is in good approximation perpendicular to $\mathbf{P}_i$. $\mathbf{B}_{ext}$ causes a frequency splitting, see  Section \ref{sec:high}. The doublet frequencies, compare red lines in Fig. \ref{highFieldFit}, are  given by Eqn. (\ref{eqn:lowBpara}). Here, only transitions without sign change of $\theta$, i.e. $f_\mathrm{E2}^+\rightarrow f_\mathrm{E1}^+$ and $f_\mathrm{E2}^-\rightarrow f_\mathrm{E1}^-$, are considered. Note that the phase  shift $\Phi$ has opposite direction for the two doublet frequencies due to a sign change of $\zeta$,  see Eqns. (\ref{eqn:trans_phase}) and (\ref{eqn:zeta}). The model predicts slightly different temperature dependences $\mathcal{A}(T)$ for the $f_\mathrm{E2}^+\rightarrow f_\mathrm{E1}^+$ and $f_\mathrm{E2}^-\rightarrow f_\mathrm{E1}^-$ transitions. Since the experimental frequency doublet was fit to a common amplitude, the model curve shown in Fig. \ref{fig:trans}(b) is the average of both contributions.

For a quantitative analysis, the following parameters (as obtained in the Sections above) were used: in ZF,  $A_\mathrm{E1}=0.082$ and  $A_\mathrm{E2}=0.090$, in high field 
 $A_\mathrm{E1}=0.0253$ and  $A_\mathrm{E2}=0.0275$. 
   The derivation of Eqns. (\ref{eqn:amp}) and (\ref{eqn:trans_phase}) does not consider initial phases ($\phi_\mathrm{E1}=\phi_\mathrm{E2}= 0$). This was accounted for by shifting the model curves by the E1 initial phase obtained at low temperatures.

 \section{Details on local hopping simulation}  
\label{apx:hop} 
 
\begin{figure}[t]
\center
            \includegraphics[]{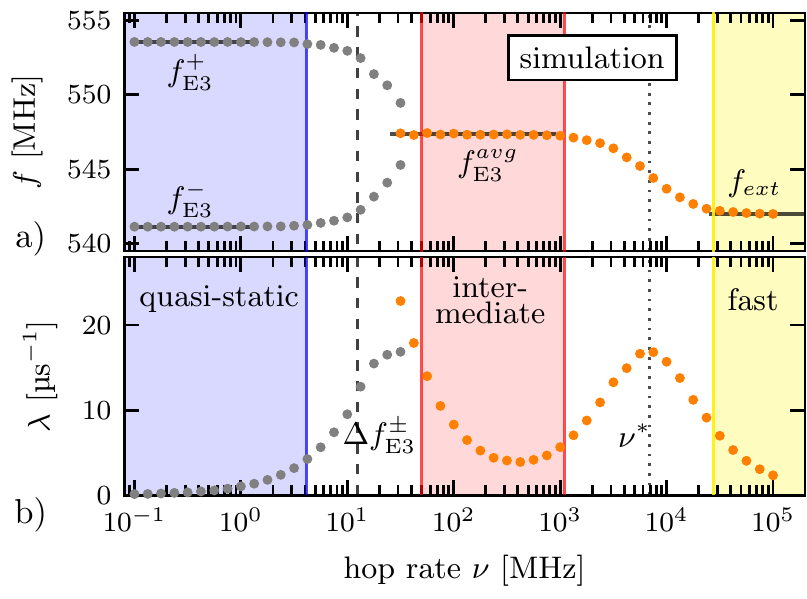}
            
                     \includegraphics[]{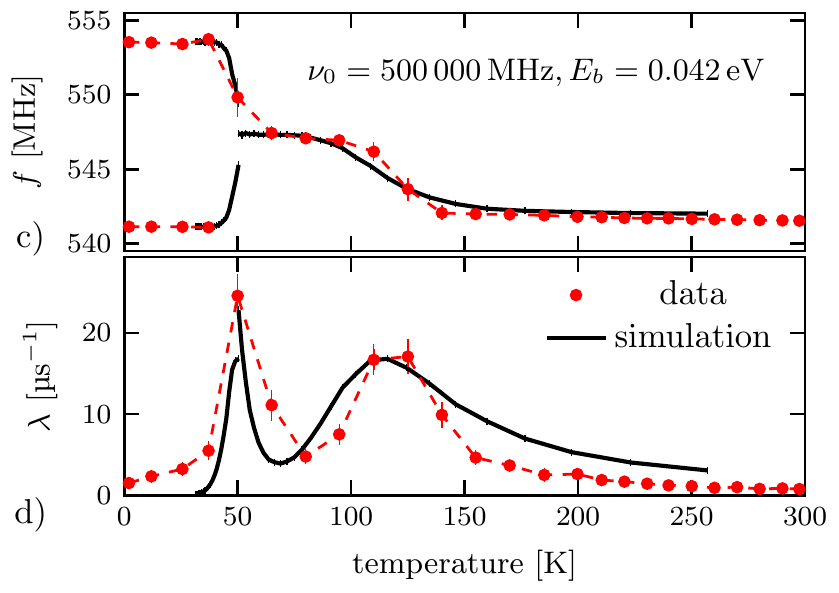}

\caption{Simulation of (a) precession frequencies and (b) relaxation rate assuming hopping between adjacent muon sites arranged on a hexagon in $\SI{4}{T}$, described by an exponential correlation time $\tau=1/\nu$, where $\nu$ is the average hop rate. There are three different regimes: quasi-static (low hop rate), where the E3 doublet is observed, intermediate hop rates, where the average of the E3 doublet is observed, and fast hopping, where internal fields are completely averaged out and a component at $f_{ext}$ is predicted. 
(c) and (d): Projection of the simulated values  with an Arrhenius activation relation onto the obtained E3/E3$'$/E3* frequencies and relaxation rate. There is excellent qualitative agreement between the model and the data (see text for details).}

\label{fig:Sim}
\end{figure}
 
 In order to investigate the E3$'$ component, and obtain a better understanding of the dynamic behavior over the full temperature range, a Monte Carlo simulation of the muon depolarization function in $\mathbf{B}_{ext}=\SI{4}{T}\parallel c$ assuming local hopping between adjacent sites was carried out. 
  Stopping sites were arranged on a hexagon, with the in-plane component of $\mathbf{B}_{int}$ pointing radially outward, and the signs of $\theta$ alternating. An exponential correlation time $\tau=1/\nu$, where $\nu$ is the average hop rate, was assumed. Polarization spectra were simulated in the range $\nu=10^{-1}-\SI{e5}{MHz}$ for \SI{1}{\micro s} with a time step $\Delta t=\SI{0.0001}{\micro s}$ and 5000 repeats, using $f_\mathrm{E3}=\SI{76.7}{MHz}$ for  $|\mathbf{B}_{int}|$, $\theta=\pm \SI{4.69}{\degree}$, and $f_{ext}=\SI{541.98}{MHz}$. Details on the general setup of the simulation can be found in Ref. \onlinecite{dehn2018a}. The simulated spectra were fit to either a single oscillatory component, Eqn. (\ref{eqn:osc_fit}), for fast hopping, or to two oscillatory components with shared relaxation rate in the quasi-static regime. The resulting precession frequencies and relaxation rates are shown in Fig. \ref{fig:Sim}(a) and (b).

For direct comparison, the simulation results are mapped onto the data assuming an Arrhenius-like activation $T=E_b/(k_B \ln[\nu_0/\nu])$, where $E_b$ is an activation energy and $\nu_0$ an attempt frequency. Fig. \ref{fig:Sim}, displaying in red the experimental \SI{4}{T} E3/E3$'$/E3* (c) frequencies and (d) relaxation rate, alongside the simulation results for $\nu_0=\SI{5e5}{MHz}$ and $E_b=\SI{42}{meV}$, shows excellent qualitative agreement. The E3$'$ component is clearly identified as $f^{avg}_\mathrm{E3}$, compare Fig. \ref{highFieldFit}(c). Note that due to the approximation of a fixed internal field, and neglect of susceptibility contributions, the model looses validity with increasing temperature. Both at the low and high temperature end of the data, the relaxation rate is not well described, indicating that a simple Arrhenius activation model is insufficient to fully describe the data.

Overall, local hopping describes the E3/E3$'$/E3* signals in the combined data set taken in ZF, small and large $\mathbf{B}_{ext}$ very well. There is convincing evidence that  E3*  arises from muons undergoing thermally activated hopping between adjacent E3 sites, causing $\mathbf{B}_{int}$ to average to zero. 
This is strongly supported by the observation of E3$'$ in large $\mathbf{B}_{ext}$ and its identification as $f^{avg}_\mathrm{E3}$. Additionally, the energy barrier $E_b=42\pm\SI{5}{meV}$ between sites is estimated.

 \section{Details on DFT calculations}
 \label{apx:dft}

DFT calculations were carried out using the Vienna ab initio Simulation Package (VASP) Version 5.4.4
\cite{kresse1993,kresse1994,kresse1996}.
 The positive muon was modeled as a hydrogen nucleus, embedded within an 80-atom  $2\times2\times2$ rhombohedral supercell (SC) of Cr$_2$O$_3$. The local spin density approximation as parameterized by Perdew and Zunger \cite{perdew1981} with an additional Hubbard-like correction (LDA+U) was used. The LDA+U correction scheme of Dudarev \emph{et al.} \cite{dudarev1998} was employed with a $\mathrm{U_{eff}}$ of 4 eV applied to the Cr $d$ states. This choice of $\mathrm{U_{eff}}$ was found to provide a good description of the crystal and electronic structure of Cr$_2$O$_3$ and is in line with values used in previous works \cite{shi2009}. 
Brillouin zone integrations were performed, using the tetrahedron method with Bl\"ochl corrections, on a $\Gamma$-centred $4\times4\times4$ Monkhorst-Pack grid \cite{monkhorst1976} for the $2\times 2\times 2$ SC. A plane-wave cutoff of 700 eV was used. With respect to a 900 eV cutoff and an $8\times8\times8\,$ \(k\)-point mesh, energy differences were found to be converged to within 1 meV/formula unit and forces to within 3 meV/\AA. The full convergence tests are available in the Supplemental Information \cite{Note1}. 
    Within the projector-augmented plane-wave (PAW) method \cite{blochl1994,kresse1999}, the 14 electrons for Cr ($3s^{2}3p^{6}3d^{4}4s^{2}$) and 6 for O ($2s^{2}2p^{4}$) were treated explicitly \footnote{The Cr, O and H PAWs are dated: 23$\mathrm{^{rd}}$ Jul. 2007, 22$\mathrm{^{nd}}$ Mar. 2012 and 6$\mathrm{^{th}}$ May 1998 respectively}. 
A collinear treatment of spins was adopted and the well established `$+-+-$' G-type antiferromagnetic order, indicated in the inset of Fig. \ref{fig:ZFprec}, confirmed. Spin-orbit coupling has previously been found to be negligible in Cr$_2$O$_3$ \cite{shi2009} and was therefore not included here.
 Supercells of up to $4\times4\times4$ were tested in order to ensure that the imposed periodic boundary conditions did not introduce artifacts. In particular, we found negligible changes in the calculated contact and dipole contributions with respect to the results shown in Table \ref{tbl:dft}. 
 

 \bibliography{full_library.bib}

\begin{thebibliography}{74}%
\makeatletter
\providecommand \@ifxundefined [1]{%
 \@ifx{#1\undefined}
}%
\providecommand \@ifnum [1]{%
 \ifnum #1\expandafter \@firstoftwo
 \else \expandafter \@secondoftwo
 \fi
}%
\providecommand \@ifx [1]{%
 \ifx #1\expandafter \@firstoftwo
 \else \expandafter \@secondoftwo
 \fi
}%
\providecommand \natexlab [1]{#1}%
\providecommand \enquote  [1]{``#1''}%
\providecommand \bibnamefont  [1]{#1}%
\providecommand \bibfnamefont [1]{#1}%
\providecommand \citenamefont [1]{#1}%
\providecommand \href@noop [0]{\@secondoftwo}%
\providecommand \href [0]{\begingroup \@sanitize@url \@href}%
\providecommand \@href[1]{\@@startlink{#1}\@@href}%
\providecommand \@@href[1]{\endgroup#1\@@endlink}%
\providecommand \@sanitize@url [0]{\catcode `\\12\catcode `\$12\catcode
  `\&12\catcode `\#12\catcode `\^12\catcode `\_12\catcode `\%12\relax}%
\providecommand \@@startlink[1]{}%
\providecommand \@@endlink[0]{}%
\providecommand \url  [0]{\begingroup\@sanitize@url \@url }%
\providecommand \@url [1]{\endgroup\@href {#1}{\urlprefix }}%
\providecommand \urlprefix  [0]{URL }%
\providecommand \Eprint [0]{\href }%
\providecommand \doibase [0]{http://dx.doi.org/}%
\providecommand \selectlanguage [0]{\@gobble}%
\providecommand \bibinfo  [0]{\@secondoftwo}%
\providecommand \bibfield  [0]{\@secondoftwo}%
\providecommand \translation [1]{[#1]}%
\providecommand \BibitemOpen [0]{}%
\providecommand \bibitemStop [0]{}%
\providecommand \bibitemNoStop [0]{.\EOS\space}%
\providecommand \EOS [0]{\spacefactor3000\relax}%
\providecommand \BibitemShut  [1]{\csname bibitem#1\endcsname}%
\let\auto@bib@innerbib\@empty
\bibitem [{\citenamefont {Yaouanc}\ and\ \citenamefont {{Dalmas de
  R{\'e}otier}}(2011)}]{yaouanc2011}%
  \BibitemOpen
  \bibfield  {author} {\bibinfo {author} {\bibfnamefont {A.}~\bibnamefont
  {Yaouanc}}\ and\ \bibinfo {author} {\bibfnamefont {P.}~\bibnamefont {{Dalmas
  de R{\'e}otier}}},\ }\href@noop {} {\emph {\bibinfo {title} {Muon {{Spin
  Rotation}}, {{Relaxation}}, and {{Resonance}}}}}\ (\bibinfo  {publisher}
  {{Oxford University Press}},\ \bibinfo {year} {2011})\BibitemShut {NoStop}%
\bibitem [{\citenamefont {Chow}\ \emph {et~al.}(1998)\citenamefont {Chow},
  \citenamefont {Hitti},\ and\ \citenamefont {Kiefl}}]{chow1998}%
  \BibitemOpen
  \bibfield  {author} {\bibinfo {author} {\bibfnamefont {K.}~\bibnamefont
  {Chow}}, \bibinfo {author} {\bibfnamefont {B.}~\bibnamefont {Hitti}}, \ and\
  \bibinfo {author} {\bibfnamefont {R.}~\bibnamefont {Kiefl}},\ }in\ \href
  {\doibase 10.1016/S0080-8784(08)63056-2} {\emph {\bibinfo {booktitle}
  {Semiconductors and {{Semimetals}}}}},\ Vol.~\bibinfo {volume} {51}\
  (\bibinfo  {publisher} {{Elsevier}},\ \bibinfo {year} {1998})\ pp.\ \bibinfo
  {pages} {137--207}\BibitemShut {NoStop}%
\bibitem [{\citenamefont {Cox}\ \emph {et~al.}(2006{\natexlab{a}})\citenamefont
  {Cox}, \citenamefont {Lord}, \citenamefont {Cottrell}, \citenamefont {Gil},
  \citenamefont {Alberto}, \citenamefont {Keren}, \citenamefont {Prabhakaran},
  \citenamefont {Scheuermann},\ and\ \citenamefont {Stoykov}}]{cox2006}%
  \BibitemOpen
  \bibfield  {author} {\bibinfo {author} {\bibfnamefont {S.~F.~J.}\
  \bibnamefont {Cox}}, \bibinfo {author} {\bibfnamefont {J.~S.}\ \bibnamefont
  {Lord}}, \bibinfo {author} {\bibfnamefont {S.~P.}\ \bibnamefont {Cottrell}},
  \bibinfo {author} {\bibfnamefont {J.~M.}\ \bibnamefont {Gil}}, \bibinfo
  {author} {\bibfnamefont {H.~V.}\ \bibnamefont {Alberto}}, \bibinfo {author}
  {\bibfnamefont {A.}~\bibnamefont {Keren}}, \bibinfo {author} {\bibfnamefont
  {D.}~\bibnamefont {Prabhakaran}}, \bibinfo {author} {\bibfnamefont
  {R.}~\bibnamefont {Scheuermann}}, \ and\ \bibinfo {author} {\bibfnamefont
  {A.}~\bibnamefont {Stoykov}},\ }\href
  {http://stacks.iop.org/0953-8984/18/i=3/a=021?key=crossref.59e013daa07017ade61b623a7580f343}
  {\bibfield  {journal} {\bibinfo  {journal} {J. Phys. Condens. Matter}\
  }\textbf {\bibinfo {volume} {18}},\ \bibinfo {pages} {1061} (\bibinfo {year}
  {2006}{\natexlab{a}})}\BibitemShut {NoStop}%
\bibitem [{\citenamefont {Cox}\ \emph {et~al.}(2006{\natexlab{b}})\citenamefont
  {Cox}, \citenamefont {Gavartin}, \citenamefont {Lord}, \citenamefont
  {Cottrell}, \citenamefont {Gil}, \citenamefont {Alberto}, \citenamefont
  {Duarte}, \citenamefont {Vil{\~a}o}, \citenamefont {de~Campos}, \citenamefont
  {Keeble}, \citenamefont {Davis}, \citenamefont {Charlton},\ and\
  \citenamefont {van~der Werf}}]{cox2006a}%
  \BibitemOpen
  \bibfield  {author} {\bibinfo {author} {\bibfnamefont {S.~F.~J.}\
  \bibnamefont {Cox}}, \bibinfo {author} {\bibfnamefont {J.~L.}\ \bibnamefont
  {Gavartin}}, \bibinfo {author} {\bibfnamefont {J.~S.}\ \bibnamefont {Lord}},
  \bibinfo {author} {\bibfnamefont {S.~P.}\ \bibnamefont {Cottrell}}, \bibinfo
  {author} {\bibfnamefont {J.~M.}\ \bibnamefont {Gil}}, \bibinfo {author}
  {\bibfnamefont {H.~V.}\ \bibnamefont {Alberto}}, \bibinfo {author}
  {\bibfnamefont {J.~P.}\ \bibnamefont {Duarte}}, \bibinfo {author}
  {\bibfnamefont {R.~C.}\ \bibnamefont {Vil{\~a}o}}, \bibinfo {author}
  {\bibfnamefont {N.~A.}\ \bibnamefont {de~Campos}}, \bibinfo {author}
  {\bibfnamefont {D.~J.}\ \bibnamefont {Keeble}}, \bibinfo {author}
  {\bibfnamefont {E.~A.}\ \bibnamefont {Davis}}, \bibinfo {author}
  {\bibfnamefont {M.}~\bibnamefont {Charlton}}, \ and\ \bibinfo {author}
  {\bibfnamefont {D.~P.}\ \bibnamefont {van~der Werf}},\ }\href
  {http://stacks.iop.org/0953-8984/18/i=3/a=022?key=crossref.c25d9eb258bc1efcdd69b6f624b69b0a}
  {\bibfield  {journal} {\bibinfo  {journal} {J. Phys. Condens. Matter}\
  }\textbf {\bibinfo {volume} {18}},\ \bibinfo {pages} {1079} (\bibinfo {year}
  {2006}{\natexlab{b}})}\BibitemShut {NoStop}%
\bibitem [{\citenamefont {Cox}(2009)}]{cox2009}%
  \BibitemOpen
  \bibfield  {author} {\bibinfo {author} {\bibfnamefont {S.~F.~J.}\
  \bibnamefont {Cox}},\ }\href
  {http://stacks.iop.org/0034-4885/72/i=11/a=116501?key=crossref.ce6d2acc8d75ace555b554ef00e9e661}
  {\bibfield  {journal} {\bibinfo  {journal} {Rep. Prog. Phys.}\ }\textbf
  {\bibinfo {volume} {72}},\ \bibinfo {pages} {116501} (\bibinfo {year}
  {2009})}\BibitemShut {NoStop}%
\bibitem [{\citenamefont {Uemura}\ \emph {et~al.}(1986)\citenamefont {Uemura},
  \citenamefont {Keitel}, \citenamefont {Senba}, \citenamefont {Kiefl},
  \citenamefont {Kreitzman}, \citenamefont {Noakes}, \citenamefont {Brewer},
  \citenamefont {Harshman}, \citenamefont {Ansaldo}, \citenamefont {Crowe},
  \citenamefont {Portis},\ and\ \citenamefont {Jaccarino}}]{uemura1986}%
  \BibitemOpen
  \bibfield  {author} {\bibinfo {author} {\bibfnamefont {Y.~J.}\ \bibnamefont
  {Uemura}}, \bibinfo {author} {\bibfnamefont {R.}~\bibnamefont {Keitel}},
  \bibinfo {author} {\bibfnamefont {M.}~\bibnamefont {Senba}}, \bibinfo
  {author} {\bibfnamefont {R.~F.}\ \bibnamefont {Kiefl}}, \bibinfo {author}
  {\bibfnamefont {S.~R.}\ \bibnamefont {Kreitzman}}, \bibinfo {author}
  {\bibfnamefont {D.~R.}\ \bibnamefont {Noakes}}, \bibinfo {author}
  {\bibfnamefont {J.~H.}\ \bibnamefont {Brewer}}, \bibinfo {author}
  {\bibfnamefont {D.}~\bibnamefont {Harshman}}, \bibinfo {author}
  {\bibfnamefont {E.~J.}\ \bibnamefont {Ansaldo}}, \bibinfo {author}
  {\bibfnamefont {K.~M.}\ \bibnamefont {Crowe}}, \bibinfo {author}
  {\bibfnamefont {A.~M.}\ \bibnamefont {Portis}}, \ and\ \bibinfo {author}
  {\bibfnamefont {V.}~\bibnamefont {Jaccarino}},\ }\href@noop {} {\bibfield
  {journal} {\bibinfo  {journal} {Hyperfine Interact.}\ }\textbf {\bibinfo
  {volume} {31}},\ \bibinfo {pages} {313} (\bibinfo {year} {1986})}\BibitemShut
  {NoStop}%
\bibitem [{\citenamefont {Jahn}\ and\ \citenamefont {Teller}(1937)}]{jahn1937}%
  \BibitemOpen
  \bibfield  {author} {\bibinfo {author} {\bibfnamefont {H.~A.}\ \bibnamefont
  {Jahn}}\ and\ \bibinfo {author} {\bibfnamefont {E.}~\bibnamefont {Teller}},\
  }\href@noop {} {\bibfield  {journal} {\bibinfo  {journal} {Proc Roy Soc A}\
  }\textbf {\bibinfo {volume} {161}},\ \bibinfo {pages} {16} (\bibinfo {year}
  {1937})}\BibitemShut {NoStop}%
\bibitem [{\citenamefont {Holstein}(1959)}]{holstein1959}%
  \BibitemOpen
  \bibfield  {author} {\bibinfo {author} {\bibfnamefont {T.}~\bibnamefont
  {Holstein}},\ }\href
  {https://linkinghub.elsevier.com/retrieve/pii/0003491659900028} {\bibfield
  {journal} {\bibinfo  {journal} {Ann. Phys.}\ }\textbf {\bibinfo {volume}
  {8}},\ \bibinfo {pages} {325} (\bibinfo {year} {1959})}\BibitemShut {NoStop}%
\bibitem [{\citenamefont {Nickisch}\ \emph {et~al.}(1983)\citenamefont
  {Nickisch}, \citenamefont {Thomas},\ and\ \citenamefont
  {H{\"o}ck}}]{nickisch1983a}%
  \BibitemOpen
  \bibfield  {author} {\bibinfo {author} {\bibfnamefont {H.}~\bibnamefont
  {Nickisch}}, \bibinfo {author} {\bibfnamefont {H.}~\bibnamefont {Thomas}}, \
  and\ \bibinfo {author} {\bibfnamefont {K.-H.}\ \bibnamefont {H{\"o}ck}},\
  }\href {https://www.e-periodica.ch/digbib/view?pid=hpa-001:1983:56::1253}
  {\bibfield  {journal} {\bibinfo  {journal} {Birkh{\"a}user}\ } (\bibinfo
  {year} {1983})}\BibitemShut {NoStop}%
\bibitem [{\citenamefont {Dzyaloshinskii}(1960)}]{dzyaloshinskii1960}%
  \BibitemOpen
  \bibfield  {author} {\bibinfo {author} {\bibfnamefont {I.~E.}\ \bibnamefont
  {Dzyaloshinskii}},\ }\href@noop {} {\bibfield  {journal} {\bibinfo  {journal}
  {Sov Phys JETP}\ }\textbf {\bibinfo {volume} {10}},\ \bibinfo {pages} {628}
  (\bibinfo {year} {1960})}\BibitemShut {NoStop}%
\bibitem [{\citenamefont {Astrov}(1961)}]{astrov1961}%
  \BibitemOpen
  \bibfield  {author} {\bibinfo {author} {\bibfnamefont {D.~N.}\ \bibnamefont
  {Astrov}},\ }\href {http://www.jetp.ac.ru/cgi-bin/dn/e_013_04_0729.pdf}
  {\bibfield  {journal} {\bibinfo  {journal} {Sov Phys JETP}\ }\textbf
  {\bibinfo {volume} {13}},\ \bibinfo {pages} {729} (\bibinfo {year}
  {1961})}\BibitemShut {NoStop}%
\bibitem [{\citenamefont {Rado}\ and\ \citenamefont {Folen}(1961)}]{rado1961a}%
  \BibitemOpen
  \bibfield  {author} {\bibinfo {author} {\bibfnamefont {G.~T.}\ \bibnamefont
  {Rado}}\ and\ \bibinfo {author} {\bibfnamefont {V.~J.}\ \bibnamefont
  {Folen}},\ }\href {https://link.aps.org/doi/10.1103/PhysRevLett.7.310}
  {\bibfield  {journal} {\bibinfo  {journal} {Phys. Rev. Lett.}\ }\textbf
  {\bibinfo {volume} {7}},\ \bibinfo {pages} {310} (\bibinfo {year}
  {1961})}\BibitemShut {NoStop}%
\bibitem [{\citenamefont {Wiegelmann}\ \emph {et~al.}(1994)\citenamefont
  {Wiegelmann}, \citenamefont {Jansen}, \citenamefont {Wyder}, \citenamefont
  {Rivera},\ and\ \citenamefont {Schmid}}]{wiegelmann1994}%
  \BibitemOpen
  \bibfield  {author} {\bibinfo {author} {\bibfnamefont {H.}~\bibnamefont
  {Wiegelmann}}, \bibinfo {author} {\bibfnamefont {A.~G.~M.}\ \bibnamefont
  {Jansen}}, \bibinfo {author} {\bibfnamefont {P.}~\bibnamefont {Wyder}},
  \bibinfo {author} {\bibfnamefont {J.-P.}\ \bibnamefont {Rivera}}, \ and\
  \bibinfo {author} {\bibfnamefont {H.}~\bibnamefont {Schmid}},\ }\href
  {http://www.tandfonline.com/doi/abs/10.1080/00150199408245099} {\bibfield
  {journal} {\bibinfo  {journal} {Ferroelectrics}\ }\textbf {\bibinfo {volume}
  {162}},\ \bibinfo {pages} {141} (\bibinfo {year} {1994})}\BibitemShut
  {NoStop}%
\bibitem [{\citenamefont {Fiebig}(2005)}]{fiebig2005}%
  \BibitemOpen
  \bibfield  {author} {\bibinfo {author} {\bibfnamefont {M.}~\bibnamefont
  {Fiebig}},\ }\href
  {http://stacks.iop.org/0022-3727/38/i=8/a=R01?key=crossref.debaec7d983bfa8c9af8c02008b6e621}
  {\bibfield  {journal} {\bibinfo  {journal} {J. Phys. Appl. Phys.}\ }\textbf
  {\bibinfo {volume} {38}},\ \bibinfo {pages} {R123} (\bibinfo {year}
  {2005})}\BibitemShut {NoStop}%
\bibitem [{\citenamefont {Meier}\ \emph {et~al.}(2019)\citenamefont {Meier},
  \citenamefont {Fechner}, \citenamefont {Nozaki}, \citenamefont {Sahashi},
  \citenamefont {Salman}, \citenamefont {Prokscha}, \citenamefont {Suter},
  \citenamefont {Schoenherr}, \citenamefont {Lilienblum}, \citenamefont
  {Borisov}, \citenamefont {Dzyaloshinskii}, \citenamefont {Fiebig},
  \citenamefont {Luetkens},\ and\ \citenamefont {Spaldin}}]{meier2019}%
  \BibitemOpen
  \bibfield  {author} {\bibinfo {author} {\bibfnamefont {Q.~N.}\ \bibnamefont
  {Meier}}, \bibinfo {author} {\bibfnamefont {M.}~\bibnamefont {Fechner}},
  \bibinfo {author} {\bibfnamefont {T.}~\bibnamefont {Nozaki}}, \bibinfo
  {author} {\bibfnamefont {M.}~\bibnamefont {Sahashi}}, \bibinfo {author}
  {\bibfnamefont {Z.}~\bibnamefont {Salman}}, \bibinfo {author} {\bibfnamefont
  {T.}~\bibnamefont {Prokscha}}, \bibinfo {author} {\bibfnamefont
  {A.}~\bibnamefont {Suter}}, \bibinfo {author} {\bibfnamefont
  {P.}~\bibnamefont {Schoenherr}}, \bibinfo {author} {\bibfnamefont
  {M.}~\bibnamefont {Lilienblum}}, \bibinfo {author} {\bibfnamefont
  {P.}~\bibnamefont {Borisov}}, \bibinfo {author} {\bibfnamefont {I.~E.}\
  \bibnamefont {Dzyaloshinskii}}, \bibinfo {author} {\bibfnamefont
  {M.}~\bibnamefont {Fiebig}}, \bibinfo {author} {\bibfnamefont
  {H.}~\bibnamefont {Luetkens}}, \ and\ \bibinfo {author} {\bibfnamefont
  {N.~A.}\ \bibnamefont {Spaldin}},\ }\href
  {https://link.aps.org/doi/10.1103/PhysRevX.9.011011} {\bibfield  {journal}
  {\bibinfo  {journal} {Phys. Rev. X}\ }\textbf {\bibinfo {volume} {9}}
  (\bibinfo {year} {2019})}\BibitemShut {NoStop}%
\bibitem [{\citenamefont {Borisov}\ \emph {et~al.}(2005)\citenamefont
  {Borisov}, \citenamefont {Hochstrat}, \citenamefont {Chen}, \citenamefont
  {Kleemann},\ and\ \citenamefont {Binek}}]{borisov2005}%
  \BibitemOpen
  \bibfield  {author} {\bibinfo {author} {\bibfnamefont {P.}~\bibnamefont
  {Borisov}}, \bibinfo {author} {\bibfnamefont {A.}~\bibnamefont {Hochstrat}},
  \bibinfo {author} {\bibfnamefont {X.}~\bibnamefont {Chen}}, \bibinfo {author}
  {\bibfnamefont {W.}~\bibnamefont {Kleemann}}, \ and\ \bibinfo {author}
  {\bibfnamefont {C.}~\bibnamefont {Binek}},\ }\href
  {https://link.aps.org/doi/10.1103/PhysRevLett.94.117203} {\bibfield
  {journal} {\bibinfo  {journal} {Phys. Rev. Lett.}\ }\textbf {\bibinfo
  {volume} {94}} (\bibinfo {year} {2005})}\BibitemShut {NoStop}%
\bibitem [{\citenamefont {He}\ \emph {et~al.}(2010)\citenamefont {He},
  \citenamefont {Wang}, \citenamefont {Wu}, \citenamefont {Caruso},
  \citenamefont {Vescovo}, \citenamefont {Belashchenko}, \citenamefont
  {Dowben},\ and\ \citenamefont {Binek}}]{he2010}%
  \BibitemOpen
  \bibfield  {author} {\bibinfo {author} {\bibfnamefont {X.}~\bibnamefont
  {He}}, \bibinfo {author} {\bibfnamefont {Y.}~\bibnamefont {Wang}}, \bibinfo
  {author} {\bibfnamefont {N.}~\bibnamefont {Wu}}, \bibinfo {author}
  {\bibfnamefont {A.~N.}\ \bibnamefont {Caruso}}, \bibinfo {author}
  {\bibfnamefont {E.}~\bibnamefont {Vescovo}}, \bibinfo {author} {\bibfnamefont
  {K.~D.}\ \bibnamefont {Belashchenko}}, \bibinfo {author} {\bibfnamefont
  {P.~A.}\ \bibnamefont {Dowben}}, \ and\ \bibinfo {author} {\bibfnamefont
  {C.}~\bibnamefont {Binek}},\ }\href {http://www.nature.com/articles/nmat2785}
  {\bibfield  {journal} {\bibinfo  {journal} {Nat. Mater.}\ }\textbf {\bibinfo
  {volume} {9}},\ \bibinfo {pages} {579} (\bibinfo {year} {2010})}\BibitemShut
  {NoStop}%
\bibitem [{\citenamefont {Kosub}\ \emph {et~al.}(2017)\citenamefont {Kosub},
  \citenamefont {Kopte}, \citenamefont {H{\"u}hne}, \citenamefont {Appel},
  \citenamefont {Shields}, \citenamefont {Maletinsky}, \citenamefont
  {H{\"u}bner}, \citenamefont {Liedke}, \citenamefont {Fassbender},
  \citenamefont {Schmidt},\ and\ \citenamefont {Makarov}}]{kosub2017}%
  \BibitemOpen
  \bibfield  {author} {\bibinfo {author} {\bibfnamefont {T.}~\bibnamefont
  {Kosub}}, \bibinfo {author} {\bibfnamefont {M.}~\bibnamefont {Kopte}},
  \bibinfo {author} {\bibfnamefont {R.}~\bibnamefont {H{\"u}hne}}, \bibinfo
  {author} {\bibfnamefont {P.}~\bibnamefont {Appel}}, \bibinfo {author}
  {\bibfnamefont {B.}~\bibnamefont {Shields}}, \bibinfo {author} {\bibfnamefont
  {P.}~\bibnamefont {Maletinsky}}, \bibinfo {author} {\bibfnamefont
  {R.}~\bibnamefont {H{\"u}bner}}, \bibinfo {author} {\bibfnamefont {M.~O.}\
  \bibnamefont {Liedke}}, \bibinfo {author} {\bibfnamefont {J.}~\bibnamefont
  {Fassbender}}, \bibinfo {author} {\bibfnamefont {O.~G.}\ \bibnamefont
  {Schmidt}}, \ and\ \bibinfo {author} {\bibfnamefont {D.}~\bibnamefont
  {Makarov}},\ }\href {http://www.nature.com/doifinder/10.1038/ncomms13985}
  {\bibfield  {journal} {\bibinfo  {journal} {Nat. Commun.}\ }\textbf {\bibinfo
  {volume} {8}},\ \bibinfo {pages} {13985} (\bibinfo {year}
  {2017})}\BibitemShut {NoStop}%
\bibitem [{\citenamefont {Khomskii}(2014)}]{khomskii2014}%
  \BibitemOpen
  \bibfield  {author} {\bibinfo {author} {\bibfnamefont {D.~I.}\ \bibnamefont
  {Khomskii}},\ }\href {http://www.nature.com/doifinder/10.1038/ncomms5793}
  {\bibfield  {journal} {\bibinfo  {journal} {Nat. Commun.}\ }\textbf {\bibinfo
  {volume} {5}},\ \bibinfo {pages} {4793} (\bibinfo {year} {2014})}\BibitemShut
  {NoStop}%
\bibitem [{\citenamefont {R{\"u}egg}\ \emph {et~al.}(1979)\citenamefont
  {R{\"u}egg}, \citenamefont {Boekema}, \citenamefont {Hofmann}, \citenamefont
  {K{\"u}ndig},\ and\ \citenamefont {Meier}}]{ruegg1979}%
  \BibitemOpen
  \bibfield  {author} {\bibinfo {author} {\bibfnamefont {K.}~\bibnamefont
  {R{\"u}egg}}, \bibinfo {author} {\bibfnamefont {C.}~\bibnamefont {Boekema}},
  \bibinfo {author} {\bibfnamefont {W.}~\bibnamefont {Hofmann}}, \bibinfo
  {author} {\bibfnamefont {W.}~\bibnamefont {K{\"u}ndig}}, \ and\ \bibinfo
  {author} {\bibfnamefont {P.~F.}\ \bibnamefont {Meier}},\ }\href@noop {}
  {\bibfield  {journal} {\bibinfo  {journal} {Hyperfine Interact.}\ }\textbf
  {\bibinfo {volume} {6}},\ \bibinfo {pages} {99} (\bibinfo {year}
  {1979})}\BibitemShut {NoStop}%
\bibitem [{\citenamefont {R{\"u}egg}\ \emph {et~al.}(1980)\citenamefont
  {R{\"u}egg}, \citenamefont {Boekema}, \citenamefont {Denison}, \citenamefont
  {Hoffmann},\ and\ \citenamefont {K{\"u}ndig}}]{ruegg1980}%
  \BibitemOpen
  \bibfield  {author} {\bibinfo {author} {\bibfnamefont {K.}~\bibnamefont
  {R{\"u}egg}}, \bibinfo {author} {\bibfnamefont {C.}~\bibnamefont {Boekema}},
  \bibinfo {author} {\bibfnamefont {A.}~\bibnamefont {Denison}}, \bibinfo
  {author} {\bibfnamefont {W.}~\bibnamefont {Hoffmann}}, \ and\ \bibinfo
  {author} {\bibfnamefont {W.}~\bibnamefont {K{\"u}ndig}},\ }\href@noop {}
  {\bibfield  {journal} {\bibinfo  {journal} {J. Magn. Magn. Mater.}\ }\textbf
  {\bibinfo {volume} {15-18}},\ \bibinfo {pages} {669} (\bibinfo {year}
  {1980})}\BibitemShut {NoStop}%
\bibitem [{\citenamefont {Boekema}\ \emph {et~al.}(1981)\citenamefont
  {Boekema}, \citenamefont {R{\"u}egg},\ and\ \citenamefont
  {Hofmann}}]{boekema1981}%
  \BibitemOpen
  \bibfield  {author} {\bibinfo {author} {\bibfnamefont {C.}~\bibnamefont
  {Boekema}}, \bibinfo {author} {\bibfnamefont {K.}~\bibnamefont {R{\"u}egg}},
  \ and\ \bibinfo {author} {\bibfnamefont {W.~P.}\ \bibnamefont {Hofmann}},\
  }\href {http://www.springerlink.com/index/Q73Q98U713X87648.pdf} {\bibfield
  {journal} {\bibinfo  {journal} {Hyperfine Interact.}\ }\textbf {\bibinfo
  {volume} {8}},\ \bibinfo {pages} {609} (\bibinfo {year} {1981})}\BibitemShut
  {NoStop}%
\bibitem [{\citenamefont {Boekema}\ \emph {et~al.}(1983)\citenamefont
  {Boekema}, \citenamefont {Denison},\ and\ \citenamefont
  {R{\"u}egg}}]{boekema1983}%
  \BibitemOpen
  \bibfield  {author} {\bibinfo {author} {\bibfnamefont {C.}~\bibnamefont
  {Boekema}}, \bibinfo {author} {\bibfnamefont {A.~B.}\ \bibnamefont
  {Denison}}, \ and\ \bibinfo {author} {\bibfnamefont {K.~J.}\ \bibnamefont
  {R{\"u}egg}},\ }\href@noop {} {\bibfield  {journal} {\bibinfo  {journal} {J.
  Magn. Magn. Mater.}\ }\textbf {\bibinfo {volume} {36}},\ \bibinfo {pages}
  {111} (\bibinfo {year} {1983})}\BibitemShut {NoStop}%
\bibitem [{\citenamefont {Suter}\ and\ \citenamefont
  {Wojek}(2012)}]{suter2012}%
  \BibitemOpen
  \bibfield  {author} {\bibinfo {author} {\bibfnamefont {A.}~\bibnamefont
  {Suter}}\ and\ \bibinfo {author} {\bibfnamefont {B.}~\bibnamefont {Wojek}},\
  }\href {http://linkinghub.elsevier.com/retrieve/pii/S187538921201228X}
  {\bibfield  {journal} {\bibinfo  {journal} {Phys. Procedia}\ }\textbf
  {\bibinfo {volume} {30}},\ \bibinfo {pages} {69} (\bibinfo {year}
  {2012})}\BibitemShut {NoStop}%
\bibitem [{\citenamefont {Denison}(1984)}]{denison1984}%
  \BibitemOpen
  \bibfield  {author} {\bibinfo {author} {\bibfnamefont {A.~B.}\ \bibnamefont
  {Denison}},\ }\href {http://aip.scitation.org/doi/10.1063/1.333635}
  {\bibfield  {journal} {\bibinfo  {journal} {J. Appl. Phys.}\ }\textbf
  {\bibinfo {volume} {55}},\ \bibinfo {pages} {2278} (\bibinfo {year}
  {1984})}\BibitemShut {NoStop}%
\bibitem [{\citenamefont {Holzschuh}\ \emph {et~al.}(1983)\citenamefont
  {Holzschuh}, \citenamefont {Denison}, \citenamefont {K{\"u}ndig},
  \citenamefont {Meier},\ and\ \citenamefont {Patterson}}]{holzschuh1983a}%
  \BibitemOpen
  \bibfield  {author} {\bibinfo {author} {\bibfnamefont {E.}~\bibnamefont
  {Holzschuh}}, \bibinfo {author} {\bibfnamefont {A.~B.}\ \bibnamefont
  {Denison}}, \bibinfo {author} {\bibfnamefont {W.}~\bibnamefont {K{\"u}ndig}},
  \bibinfo {author} {\bibfnamefont {P.~F.}\ \bibnamefont {Meier}}, \ and\
  \bibinfo {author} {\bibfnamefont {B.~D.}\ \bibnamefont {Patterson}},\ }\href
  {https://link.aps.org/doi/10.1103/PhysRevB.27.5294} {\bibfield  {journal}
  {\bibinfo  {journal} {Phys. Rev. B}\ }\textbf {\bibinfo {volume} {27}},\
  \bibinfo {pages} {5294} (\bibinfo {year} {1983})}\BibitemShut {NoStop}%
\bibitem [{Note1()}]{Note1}%
  \BibitemOpen
  \bibinfo {note} {This measurement with $\protect \mathbf {B}_{ext}=\SI
  {300}{G}\parallel c$ was taken in the Omni-Prime spectrometer. The Fourier
  transform is shown over the time range of $0.1-\SI {2}{\micro
  s}$.}\BibitemShut {Stop}%
\bibitem [{\citenamefont {Samuelsen}\ \emph {et~al.}(1970)\citenamefont
  {Samuelsen}, \citenamefont {Hutchings},\ and\ \citenamefont
  {Shirane}}]{samuelsen1970a}%
  \BibitemOpen
  \bibfield  {author} {\bibinfo {author} {\bibfnamefont {E.}~\bibnamefont
  {Samuelsen}}, \bibinfo {author} {\bibfnamefont {M.}~\bibnamefont
  {Hutchings}}, \ and\ \bibinfo {author} {\bibfnamefont {G.}~\bibnamefont
  {Shirane}},\ }\href
  {https://linkinghub.elsevier.com/retrieve/pii/0031891470901588} {\bibfield
  {journal} {\bibinfo  {journal} {Physica}\ }\textbf {\bibinfo {volume} {48}},\
  \bibinfo {pages} {13} (\bibinfo {year} {1970})}\BibitemShut {NoStop}%
\bibitem [{\citenamefont {Browne}\ and\ \citenamefont
  {Stoneham}(1982)}]{browne1982}%
  \BibitemOpen
  \bibfield  {author} {\bibinfo {author} {\bibfnamefont {A.~M.}\ \bibnamefont
  {Browne}}\ and\ \bibinfo {author} {\bibfnamefont {A.~M.}\ \bibnamefont
  {Stoneham}},\ }\href
  {http://stacks.iop.org/0022-3719/15/i=12/a=018?key=crossref.f173bbe4103d3bc818b2d4833bc5a221}
  {\bibfield  {journal} {\bibinfo  {journal} {J. Phys. C Solid State Phys.}\
  }\textbf {\bibinfo {volume} {15}},\ \bibinfo {pages} {2709} (\bibinfo {year}
  {1982})}\BibitemShut {NoStop}%
\bibitem [{\citenamefont {Stoneham}(1984)}]{stoneham1984}%
  \BibitemOpen
  \bibfield  {author} {\bibinfo {author} {\bibfnamefont {A.~M.}\ \bibnamefont
  {Stoneham}},\ }\href {http://link.springer.com/10.1007/BF02065886} {\bibfield
   {journal} {\bibinfo  {journal} {Hyperfine Interact.}\ }\textbf {\bibinfo
  {volume} {17}},\ \bibinfo {pages} {53} (\bibinfo {year} {1984})}\BibitemShut
  {NoStop}%
\bibitem [{\citenamefont {Graf}\ \emph {et~al.}(1978)\citenamefont {Graf},
  \citenamefont {Hofmann}, \citenamefont {K{\"u}ndig}, \citenamefont {Meier},
  \citenamefont {Patterson}, \citenamefont {Reichhart},\ and\ \citenamefont
  {Rodriguez}}]{graf1978a}%
  \BibitemOpen
  \bibfield  {author} {\bibinfo {author} {\bibfnamefont {H.}~\bibnamefont
  {Graf}}, \bibinfo {author} {\bibfnamefont {W.}~\bibnamefont {Hofmann}},
  \bibinfo {author} {\bibfnamefont {W.}~\bibnamefont {K{\"u}ndig}}, \bibinfo
  {author} {\bibfnamefont {P.~F.}\ \bibnamefont {Meier}}, \bibinfo {author}
  {\bibfnamefont {B.~D.}\ \bibnamefont {Patterson}}, \bibinfo {author}
  {\bibfnamefont {W.}~\bibnamefont {Reichhart}}, \ and\ \bibinfo {author}
  {\bibfnamefont {A.}~\bibnamefont {Rodriguez}},\ }\href@noop {} {\bibfield
  {journal} {\bibinfo  {journal} {Hyperfine Interact.}\ }\textbf {\bibinfo
  {volume} {4}},\ \bibinfo {pages} {452} (\bibinfo {year} {1978})}\BibitemShut
  {NoStop}%
\bibitem [{\citenamefont {Duginov}\ \emph {et~al.}(1994)\citenamefont
  {Duginov}, \citenamefont {Grebinnik}, \citenamefont {Gritsaj}, \citenamefont
  {Mamedov}, \citenamefont {Olshevsky}, \citenamefont {Pomjakushin},
  \citenamefont {Zhukov}, \citenamefont {Kirillov}, \citenamefont {Krivosheev},
  \citenamefont {Pirogov},\ and\ \citenamefont {Ponomarev}}]{duginov1994}%
  \BibitemOpen
  \bibfield  {author} {\bibinfo {author} {\bibfnamefont {V.~N.}\ \bibnamefont
  {Duginov}}, \bibinfo {author} {\bibfnamefont {V.~G.}\ \bibnamefont
  {Grebinnik}}, \bibinfo {author} {\bibfnamefont {K.~I.}\ \bibnamefont
  {Gritsaj}}, \bibinfo {author} {\bibfnamefont {T.~N.}\ \bibnamefont
  {Mamedov}}, \bibinfo {author} {\bibfnamefont {V.~G.}\ \bibnamefont
  {Olshevsky}}, \bibinfo {author} {\bibfnamefont {V.~Y.}\ \bibnamefont
  {Pomjakushin}}, \bibinfo {author} {\bibfnamefont {V.~A.}\ \bibnamefont
  {Zhukov}}, \bibinfo {author} {\bibfnamefont {B.~F.}\ \bibnamefont
  {Kirillov}}, \bibinfo {author} {\bibfnamefont {I.~A.}\ \bibnamefont
  {Krivosheev}}, \bibinfo {author} {\bibfnamefont {A.~V.}\ \bibnamefont
  {Pirogov}}, \ and\ \bibinfo {author} {\bibfnamefont {A.~N.}\ \bibnamefont
  {Ponomarev}},\ }\href@noop {} {\bibfield  {journal} {\bibinfo  {journal}
  {Hyperfine Interact.}\ }\textbf {\bibinfo {volume} {85}},\ \bibinfo {pages}
  {317} (\bibinfo {year} {1994})}\BibitemShut {NoStop}%
\bibitem [{\citenamefont {Amato}\ \emph {et~al.}(2000)\citenamefont {Amato},
  \citenamefont {Andreica}, \citenamefont {Gygax}, \citenamefont {Pinkpank},
  \citenamefont {Sato}, \citenamefont {Schenck},\ and\ \citenamefont
  {Solt}}]{amato2000}%
  \BibitemOpen
  \bibfield  {author} {\bibinfo {author} {\bibfnamefont {A.}~\bibnamefont
  {Amato}}, \bibinfo {author} {\bibfnamefont {D.}~\bibnamefont {Andreica}},
  \bibinfo {author} {\bibfnamefont {F.}~\bibnamefont {Gygax}}, \bibinfo
  {author} {\bibfnamefont {M.}~\bibnamefont {Pinkpank}}, \bibinfo {author}
  {\bibfnamefont {N.}~\bibnamefont {Sato}}, \bibinfo {author} {\bibfnamefont
  {A.}~\bibnamefont {Schenck}}, \ and\ \bibinfo {author} {\bibfnamefont
  {G.}~\bibnamefont {Solt}},\ }\href
  {https://linkinghub.elsevier.com/retrieve/pii/S0921452600004324} {\bibfield
  {journal} {\bibinfo  {journal} {Phys. B Condens. Matter}\ }\textbf {\bibinfo
  {volume} {289-290}},\ \bibinfo {pages} {447} (\bibinfo {year}
  {2000})}\BibitemShut {NoStop}%
\bibitem [{\citenamefont {Mulders}\ \emph {et~al.}(2001)\citenamefont
  {Mulders}, \citenamefont {Gubbens}, \citenamefont {Kaiser}, \citenamefont
  {Amato}, \citenamefont {Gygax}, \citenamefont {Schenck}, \citenamefont
  {{Dalmas de R{\'e}otier}}, \citenamefont {Yaouanc}, \citenamefont {Buschow},
  \citenamefont {Kayzel},\ and\ \citenamefont {Menovsky}}]{mulders2001}%
  \BibitemOpen
  \bibfield  {author} {\bibinfo {author} {\bibfnamefont {A.~M.}\ \bibnamefont
  {Mulders}}, \bibinfo {author} {\bibfnamefont {P.~C.~M.}\ \bibnamefont
  {Gubbens}}, \bibinfo {author} {\bibfnamefont {C.}~\bibnamefont {Kaiser}},
  \bibinfo {author} {\bibfnamefont {A.}~\bibnamefont {Amato}}, \bibinfo
  {author} {\bibfnamefont {F.}~\bibnamefont {Gygax}}, \bibinfo {author}
  {\bibfnamefont {A.}~\bibnamefont {Schenck}}, \bibinfo {author} {\bibfnamefont
  {P.}~\bibnamefont {{Dalmas de R{\'e}otier}}}, \bibinfo {author}
  {\bibfnamefont {A.}~\bibnamefont {Yaouanc}}, \bibinfo {author} {\bibfnamefont
  {K.}~\bibnamefont {Buschow}}, \bibinfo {author} {\bibfnamefont
  {F.}~\bibnamefont {Kayzel}}, \ and\ \bibinfo {author} {\bibfnamefont
  {A.}~\bibnamefont {Menovsky}},\ }\href
  {http://link.springer.com/10.1023/A:1012220909947} {\bibfield  {journal}
  {\bibinfo  {journal} {Hyperfine Interact.}\ }\textbf {\bibinfo {volume}
  {133}},\ \bibinfo {pages} {197} (\bibinfo {year} {2001})}\BibitemShut
  {NoStop}%
\bibitem [{\citenamefont {Schneider}\ \emph {et~al.}(1992)\citenamefont
  {Schneider}, \citenamefont {Kiefl}, \citenamefont {Chow}, \citenamefont
  {Cox}, \citenamefont {Dodds}, \citenamefont {DuVarney}, \citenamefont
  {Estle}, \citenamefont {Kadono}, \citenamefont {Kreitzman}, \citenamefont
  {Lichti},\ and\ \citenamefont {Schwab}}]{schneider1992}%
  \BibitemOpen
  \bibfield  {author} {\bibinfo {author} {\bibfnamefont {J.~W.}\ \bibnamefont
  {Schneider}}, \bibinfo {author} {\bibfnamefont {R.~F.}\ \bibnamefont
  {Kiefl}}, \bibinfo {author} {\bibfnamefont {K.}~\bibnamefont {Chow}},
  \bibinfo {author} {\bibfnamefont {S.~F.~J.}\ \bibnamefont {Cox}}, \bibinfo
  {author} {\bibfnamefont {S.~A.}\ \bibnamefont {Dodds}}, \bibinfo {author}
  {\bibfnamefont {R.~C.}\ \bibnamefont {DuVarney}}, \bibinfo {author}
  {\bibfnamefont {T.~L.}\ \bibnamefont {Estle}}, \bibinfo {author}
  {\bibfnamefont {R.}~\bibnamefont {Kadono}}, \bibinfo {author} {\bibfnamefont
  {S.~R.}\ \bibnamefont {Kreitzman}}, \bibinfo {author} {\bibfnamefont {R.~L.}\
  \bibnamefont {Lichti}}, \ and\ \bibinfo {author} {\bibfnamefont
  {C.}~\bibnamefont {Schwab}},\ }\href
  {https://link.aps.org/doi/10.1103/PhysRevLett.68.3196} {\bibfield  {journal}
  {\bibinfo  {journal} {Phys. Rev. Lett.}\ }\textbf {\bibinfo {volume} {68}},\
  \bibinfo {pages} {3196} (\bibinfo {year} {1992})}\BibitemShut {NoStop}%
\bibitem [{\citenamefont {Bonf{\`a}}\ and\ \citenamefont
  {De~Renzi}(2016)}]{bonfa2016}%
  \BibitemOpen
  \bibfield  {author} {\bibinfo {author} {\bibfnamefont {P.}~\bibnamefont
  {Bonf{\`a}}}\ and\ \bibinfo {author} {\bibfnamefont {R.}~\bibnamefont
  {De~Renzi}},\ }\href@noop {} {\bibfield  {journal} {\bibinfo  {journal} {J.
  Phys. Soc. Jpn.}\ }\textbf {\bibinfo {volume} {85}},\ \bibinfo {pages}
  {091014} (\bibinfo {year} {2016})}\BibitemShut {NoStop}%
\bibitem [{\citenamefont {Onuorah}\ \emph {et~al.}(2018)\citenamefont
  {Onuorah}, \citenamefont {Bonf{\`a}},\ and\ \citenamefont
  {De~Renzi}}]{onuorah2018}%
  \BibitemOpen
  \bibfield  {author} {\bibinfo {author} {\bibfnamefont {I.~J.}\ \bibnamefont
  {Onuorah}}, \bibinfo {author} {\bibfnamefont {P.}~\bibnamefont {Bonf{\`a}}},
  \ and\ \bibinfo {author} {\bibfnamefont {R.}~\bibnamefont {De~Renzi}},\
  }\href {https://link.aps.org/doi/10.1103/PhysRevB.97.174414} {\bibfield
  {journal} {\bibinfo  {journal} {Phys. Rev. B}\ }\textbf {\bibinfo {volume}
  {97}} (\bibinfo {year} {2018})}\BibitemShut {NoStop}%
\bibitem [{\citenamefont {Onuorah}\ \emph {et~al.}(2019)\citenamefont
  {Onuorah}, \citenamefont {Bonf{\`a}}, \citenamefont {De~Renzi}, \citenamefont
  {Monacelli}, \citenamefont {Mauri}, \citenamefont {Calandra},\ and\
  \citenamefont {Errea}}]{onuorah2019b}%
  \BibitemOpen
  \bibfield  {author} {\bibinfo {author} {\bibfnamefont {I.~J.}\ \bibnamefont
  {Onuorah}}, \bibinfo {author} {\bibfnamefont {P.}~\bibnamefont {Bonf{\`a}}},
  \bibinfo {author} {\bibfnamefont {R.}~\bibnamefont {De~Renzi}}, \bibinfo
  {author} {\bibfnamefont {L.}~\bibnamefont {Monacelli}}, \bibinfo {author}
  {\bibfnamefont {F.}~\bibnamefont {Mauri}}, \bibinfo {author} {\bibfnamefont
  {M.}~\bibnamefont {Calandra}}, \ and\ \bibinfo {author} {\bibfnamefont
  {I.}~\bibnamefont {Errea}},\ }\href {http://arxiv.org/abs/1904.11913}
  {\bibfield  {journal} {\bibinfo  {journal} {Phys. Rev. Materials}\ }\textbf
  {\bibinfo {volume} {3}},\ \bibinfo {pages} {073804} (\bibinfo {year}
  {2019})},\ \Eprint {http://arxiv.org/abs/1904.11913} {arXiv:1904.11913}
  \BibitemShut {NoStop}%
\bibitem [{\citenamefont {Kresse}\ and\ \citenamefont
  {Hafner}(1993)}]{kresse1993}%
  \BibitemOpen
  \bibfield  {author} {\bibinfo {author} {\bibfnamefont {G.}~\bibnamefont
  {Kresse}}\ and\ \bibinfo {author} {\bibfnamefont {J.}~\bibnamefont
  {Hafner}},\ }\href {https://link.aps.org/doi/10.1103/PhysRevB.47.558}
  {\bibfield  {journal} {\bibinfo  {journal} {Phys. Rev. B}\ }\textbf {\bibinfo
  {volume} {47}},\ \bibinfo {pages} {558} (\bibinfo {year} {1993})}\BibitemShut
  {NoStop}%
\bibitem [{\citenamefont {Kresse}\ and\ \citenamefont
  {Hafner}(1994)}]{kresse1994}%
  \BibitemOpen
  \bibfield  {author} {\bibinfo {author} {\bibfnamefont {G.}~\bibnamefont
  {Kresse}}\ and\ \bibinfo {author} {\bibfnamefont {J.}~\bibnamefont
  {Hafner}},\ }\href {https://link.aps.org/doi/10.1103/PhysRevB.49.14251}
  {\bibfield  {journal} {\bibinfo  {journal} {Phys. Rev. B}\ }\textbf {\bibinfo
  {volume} {49}},\ \bibinfo {pages} {14251} (\bibinfo {year}
  {1994})}\BibitemShut {NoStop}%
\bibitem [{\citenamefont {Kresse}\ and\ \citenamefont
  {Furthm{\"u}ller}(1996)}]{kresse1996}%
  \BibitemOpen
  \bibfield  {author} {\bibinfo {author} {\bibfnamefont {G.}~\bibnamefont
  {Kresse}}\ and\ \bibinfo {author} {\bibfnamefont {J.}~\bibnamefont
  {Furthm{\"u}ller}},\ }\href
  {https://link.aps.org/doi/10.1103/PhysRevB.54.11169} {\bibfield  {journal}
  {\bibinfo  {journal} {Phys. Rev. B}\ }\textbf {\bibinfo {volume} {54}},\
  \bibinfo {pages} {11169} (\bibinfo {year} {1996})}\BibitemShut {NoStop}%
\bibitem [{Note2()}]{Note2}%
  \BibitemOpen
  \bibinfo {note} {See Supplemental Information at \protect \url {DOI:
  10.5281/zenodo.3378994} \global \c@footnote 1\relax}\BibitemShut {NoStop}%
\bibitem [{\citenamefont {Vil{\~a}o}\ \emph {et~al.}(2015)\citenamefont
  {Vil{\~a}o}, \citenamefont {Vieira}, \citenamefont {Alberto}, \citenamefont
  {Gil}, \citenamefont {Weidinger}, \citenamefont {Lichti}, \citenamefont
  {Baker}, \citenamefont {Mengyan},\ and\ \citenamefont {Lord}}]{vilao2015}%
  \BibitemOpen
  \bibfield  {author} {\bibinfo {author} {\bibfnamefont {R.~C.}\ \bibnamefont
  {Vil{\~a}o}}, \bibinfo {author} {\bibfnamefont {R.~B.~L.}\ \bibnamefont
  {Vieira}}, \bibinfo {author} {\bibfnamefont {H.~V.}\ \bibnamefont {Alberto}},
  \bibinfo {author} {\bibfnamefont {J.~M.}\ \bibnamefont {Gil}}, \bibinfo
  {author} {\bibfnamefont {A.}~\bibnamefont {Weidinger}}, \bibinfo {author}
  {\bibfnamefont {R.~L.}\ \bibnamefont {Lichti}}, \bibinfo {author}
  {\bibfnamefont {B.~B.}\ \bibnamefont {Baker}}, \bibinfo {author}
  {\bibfnamefont {P.~W.}\ \bibnamefont {Mengyan}}, \ and\ \bibinfo {author}
  {\bibfnamefont {J.~S.}\ \bibnamefont {Lord}},\ }\href
  {https://link.aps.org/doi/10.1103/PhysRevB.92.081202} {\bibfield  {journal}
  {\bibinfo  {journal} {Phys. Rev. B}\ }\textbf {\bibinfo {volume} {92}}
  (\bibinfo {year} {2015})}\BibitemShut {NoStop}%
\bibitem [{\citenamefont {Shimomura}\ \emph {et~al.}(2015)\citenamefont
  {Shimomura}, \citenamefont {Kadono}, \citenamefont {Koda}, \citenamefont
  {Nishiyama},\ and\ \citenamefont {Mihara}}]{shimomura2015}%
  \BibitemOpen
  \bibfield  {author} {\bibinfo {author} {\bibfnamefont {K.}~\bibnamefont
  {Shimomura}}, \bibinfo {author} {\bibfnamefont {R.}~\bibnamefont {Kadono}},
  \bibinfo {author} {\bibfnamefont {A.}~\bibnamefont {Koda}}, \bibinfo {author}
  {\bibfnamefont {K.}~\bibnamefont {Nishiyama}}, \ and\ \bibinfo {author}
  {\bibfnamefont {M.}~\bibnamefont {Mihara}},\ }\href
  {https://link.aps.org/doi/10.1103/PhysRevB.92.075203} {\bibfield  {journal}
  {\bibinfo  {journal} {Phys. Rev. B}\ }\textbf {\bibinfo {volume} {92}}
  (\bibinfo {year} {2015})}\BibitemShut {NoStop}%
\bibitem [{\citenamefont {Emin}(1982)}]{emin1982}%
  \BibitemOpen
  \bibfield  {author} {\bibinfo {author} {\bibfnamefont {D.}~\bibnamefont
  {Emin}},\ }\href {http://physicstoday.scitation.org/doi/10.1063/1.2938044}
  {\bibfield  {journal} {\bibinfo  {journal} {Phys. Today}\ }\textbf {\bibinfo
  {volume} {35}},\ \bibinfo {pages} {34} (\bibinfo {year} {1982})}\BibitemShut
  {NoStop}%
\bibitem [{\citenamefont {{Van de Walle}}(2000)}]{vandewalle2000}%
  \BibitemOpen
  \bibfield  {author} {\bibinfo {author} {\bibfnamefont {C.~G.}\ \bibnamefont
  {{Van de Walle}}},\ }\href
  {https://link.aps.org/doi/10.1103/PhysRevLett.85.1012} {\bibfield  {journal}
  {\bibinfo  {journal} {Phys. Rev. Lett.}\ }\textbf {\bibinfo {volume} {85}},\
  \bibinfo {pages} {1012} (\bibinfo {year} {2000})}\BibitemShut {NoStop}%
\bibitem [{\citenamefont {Kiefl}\ \emph {et~al.}(1988)\citenamefont {Kiefl},
  \citenamefont {Celio}, \citenamefont {Estle}, \citenamefont {Kreitzman},
  \citenamefont {Luke}, \citenamefont {Riseman},\ and\ \citenamefont
  {Ansaldo}}]{kiefl1988}%
  \BibitemOpen
  \bibfield  {author} {\bibinfo {author} {\bibfnamefont {R.~F.}\ \bibnamefont
  {Kiefl}}, \bibinfo {author} {\bibfnamefont {M.}~\bibnamefont {Celio}},
  \bibinfo {author} {\bibfnamefont {T.~L.}\ \bibnamefont {Estle}}, \bibinfo
  {author} {\bibfnamefont {S.~R.}\ \bibnamefont {Kreitzman}}, \bibinfo {author}
  {\bibfnamefont {G.~M.}\ \bibnamefont {Luke}}, \bibinfo {author}
  {\bibfnamefont {T.~M.}\ \bibnamefont {Riseman}}, \ and\ \bibinfo {author}
  {\bibfnamefont {E.~J.}\ \bibnamefont {Ansaldo}},\ }\href
  {https://link.aps.org/doi/10.1103/PhysRevLett.60.224} {\bibfield  {journal}
  {\bibinfo  {journal} {Phys. Rev. Lett.}\ }\textbf {\bibinfo {volume} {60}},\
  \bibinfo {pages} {224} (\bibinfo {year} {1988})}\BibitemShut {NoStop}%
\bibitem [{\citenamefont {Nishiyama}(2001)}]{nishiyama2001}%
  \BibitemOpen
  \bibfield  {author} {\bibinfo {author} {\bibfnamefont {K.}~\bibnamefont
  {Nishiyama}},\ }\href@noop {} {\bibfield  {journal} {\bibinfo  {journal}
  {Hyperfine Interact.}\ }\textbf {\bibinfo {volume} {135/137}},\ \bibinfo
  {pages} {289} (\bibinfo {year} {2001})}\BibitemShut {NoStop}%
\bibitem [{\citenamefont {Bimbi}\ \emph {et~al.}(2008)\citenamefont {Bimbi},
  \citenamefont {Allodi}, \citenamefont {De~Renzi}, \citenamefont {Mazzoli},\
  and\ \citenamefont {Berger}}]{bimbi2008}%
  \BibitemOpen
  \bibfield  {author} {\bibinfo {author} {\bibfnamefont {M.}~\bibnamefont
  {Bimbi}}, \bibinfo {author} {\bibfnamefont {G.}~\bibnamefont {Allodi}},
  \bibinfo {author} {\bibfnamefont {R.}~\bibnamefont {De~Renzi}}, \bibinfo
  {author} {\bibfnamefont {C.}~\bibnamefont {Mazzoli}}, \ and\ \bibinfo
  {author} {\bibfnamefont {H.}~\bibnamefont {Berger}},\ }\href
  {https://link.aps.org/doi/10.1103/PhysRevB.77.045115} {\bibfield  {journal}
  {\bibinfo  {journal} {Phys. Rev. B}\ }\textbf {\bibinfo {volume} {77}}
  (\bibinfo {year} {2008})}\BibitemShut {NoStop}%
\bibitem [{\citenamefont {Storchak}\ \emph {et~al.}(2009)\citenamefont
  {Storchak}, \citenamefont {Parfenov}, \citenamefont {Brewer}, \citenamefont
  {Russo}, \citenamefont {Stubbs}, \citenamefont {Lichti}, \citenamefont
  {Eshchenko}, \citenamefont {Morenzoni}, \citenamefont {Aminov}, \citenamefont
  {Zlomanov}, \citenamefont {Vinokurov}, \citenamefont {Kallaher},\ and\
  \citenamefont {{von Moln{\'a}r}}}]{storchak2009a}%
  \BibitemOpen
  \bibfield  {author} {\bibinfo {author} {\bibfnamefont {V.~G.}\ \bibnamefont
  {Storchak}}, \bibinfo {author} {\bibfnamefont {O.~E.}\ \bibnamefont
  {Parfenov}}, \bibinfo {author} {\bibfnamefont {J.~H.}\ \bibnamefont
  {Brewer}}, \bibinfo {author} {\bibfnamefont {P.~L.}\ \bibnamefont {Russo}},
  \bibinfo {author} {\bibfnamefont {S.~L.}\ \bibnamefont {Stubbs}}, \bibinfo
  {author} {\bibfnamefont {R.~L.}\ \bibnamefont {Lichti}}, \bibinfo {author}
  {\bibfnamefont {D.~G.}\ \bibnamefont {Eshchenko}}, \bibinfo {author}
  {\bibfnamefont {E.}~\bibnamefont {Morenzoni}}, \bibinfo {author}
  {\bibfnamefont {T.~G.}\ \bibnamefont {Aminov}}, \bibinfo {author}
  {\bibfnamefont {V.~P.}\ \bibnamefont {Zlomanov}}, \bibinfo {author}
  {\bibfnamefont {A.~A.}\ \bibnamefont {Vinokurov}}, \bibinfo {author}
  {\bibfnamefont {R.~L.}\ \bibnamefont {Kallaher}}, \ and\ \bibinfo {author}
  {\bibfnamefont {S.}~\bibnamefont {{von Moln{\'a}r}}},\ }\href
  {https://link.aps.org/doi/10.1103/PhysRevB.80.235203} {\bibfield  {journal}
  {\bibinfo  {journal} {Phys. Rev. B}\ }\textbf {\bibinfo {volume} {80}}
  (\bibinfo {year} {2009})}\BibitemShut {NoStop}%
\bibitem [{\citenamefont {Kiefl}(2011)}]{kiefl2011}%
  \BibitemOpen
  \bibfield  {author} {\bibinfo {author} {\bibfnamefont {R.~F.}\ \bibnamefont
  {Kiefl}},\ }\href {https://link.aps.org/doi/10.1103/PhysRevB.83.077201}
  {\bibfield  {journal} {\bibinfo  {journal} {Phys. Rev. B}\ }\textbf {\bibinfo
  {volume} {83}} (\bibinfo {year} {2011})}\BibitemShut {NoStop}%
\bibitem [{\citenamefont {Storchak}\ \emph {et~al.}(2011)\citenamefont
  {Storchak}, \citenamefont {Brewer}, \citenamefont {Lichti}, \citenamefont
  {Lograsso},\ and\ \citenamefont {Schlagel}}]{storchak2011}%
  \BibitemOpen
  \bibfield  {author} {\bibinfo {author} {\bibfnamefont {V.~G.}\ \bibnamefont
  {Storchak}}, \bibinfo {author} {\bibfnamefont {J.~H.}\ \bibnamefont
  {Brewer}}, \bibinfo {author} {\bibfnamefont {R.~L.}\ \bibnamefont {Lichti}},
  \bibinfo {author} {\bibfnamefont {T.~A.}\ \bibnamefont {Lograsso}}, \ and\
  \bibinfo {author} {\bibfnamefont {D.~L.}\ \bibnamefont {Schlagel}},\ }\href
  {https://link.aps.org/doi/10.1103/PhysRevB.83.140404} {\bibfield  {journal}
  {\bibinfo  {journal} {Phys. Rev. B}\ }\textbf {\bibinfo {volume} {83}}
  (\bibinfo {year} {2011})}\BibitemShut {NoStop}%
\bibitem [{\citenamefont {Amato}\ \emph {et~al.}(2014)\citenamefont {Amato},
  \citenamefont {{Dalmas de Reotier}}, \citenamefont {Andreica}, \citenamefont
  {Yaouanc}, \citenamefont {Suter}, \citenamefont {Lapertot}, \citenamefont
  {Pop}, \citenamefont {Morenzoni}, \citenamefont {Bonf{\`a}}, \citenamefont
  {Bernardini},\ and\ \citenamefont {De~Renzi}}]{amato2014}%
  \BibitemOpen
  \bibfield  {author} {\bibinfo {author} {\bibfnamefont {A.}~\bibnamefont
  {Amato}}, \bibinfo {author} {\bibfnamefont {P.}~\bibnamefont {{Dalmas de
  Reotier}}}, \bibinfo {author} {\bibfnamefont {D.}~\bibnamefont {Andreica}},
  \bibinfo {author} {\bibfnamefont {A.}~\bibnamefont {Yaouanc}}, \bibinfo
  {author} {\bibfnamefont {A.}~\bibnamefont {Suter}}, \bibinfo {author}
  {\bibfnamefont {G.}~\bibnamefont {Lapertot}}, \bibinfo {author}
  {\bibfnamefont {I.~M.}\ \bibnamefont {Pop}}, \bibinfo {author} {\bibfnamefont
  {E.}~\bibnamefont {Morenzoni}}, \bibinfo {author} {\bibfnamefont
  {P.}~\bibnamefont {Bonf{\`a}}}, \bibinfo {author} {\bibfnamefont
  {F.}~\bibnamefont {Bernardini}}, \ and\ \bibinfo {author} {\bibfnamefont
  {R.}~\bibnamefont {De~Renzi}},\ }\href
  {https://link.aps.org/doi/10.1103/PhysRevB.89.184425} {\bibfield  {journal}
  {\bibinfo  {journal} {Phys. Rev. B}\ }\textbf {\bibinfo {volume} {89}}
  (\bibinfo {year} {2014})}\BibitemShut {NoStop}%
\bibitem [{\citenamefont {Dietl}\ and\ \citenamefont {Ohno}(2014)}]{dietl2014}%
  \BibitemOpen
  \bibfield  {author} {\bibinfo {author} {\bibfnamefont {T.}~\bibnamefont
  {Dietl}}\ and\ \bibinfo {author} {\bibfnamefont {H.}~\bibnamefont {Ohno}},\
  }\href {https://link.aps.org/doi/10.1103/RevModPhys.86.187} {\bibfield
  {journal} {\bibinfo  {journal} {Rev. Mod. Phys.}\ }\textbf {\bibinfo {volume}
  {86}},\ \bibinfo {pages} {187} (\bibinfo {year} {2014})}\BibitemShut
  {NoStop}%
\bibitem [{\citenamefont {Baqiah}\ \emph {et~al.}(2016)\citenamefont {Baqiah},
  \citenamefont {Ibrahim}, \citenamefont {Halim}, \citenamefont {Chen},
  \citenamefont {Lim},\ and\ \citenamefont {Kechik}}]{baqiah2016}%
  \BibitemOpen
  \bibfield  {author} {\bibinfo {author} {\bibfnamefont {H.}~\bibnamefont
  {Baqiah}}, \bibinfo {author} {\bibfnamefont {N.}~\bibnamefont {Ibrahim}},
  \bibinfo {author} {\bibfnamefont {S.}~\bibnamefont {Halim}}, \bibinfo
  {author} {\bibfnamefont {S.}~\bibnamefont {Chen}}, \bibinfo {author}
  {\bibfnamefont {K.}~\bibnamefont {Lim}}, \ and\ \bibinfo {author}
  {\bibfnamefont {M.~A.}\ \bibnamefont {Kechik}},\ }\href
  {https://linkinghub.elsevier.com/retrieve/pii/S0304885315306491} {\bibfield
  {journal} {\bibinfo  {journal} {J. Magn. Magn. Mater.}\ }\textbf {\bibinfo
  {volume} {401}},\ \bibinfo {pages} {102} (\bibinfo {year}
  {2016})}\BibitemShut {NoStop}%
\bibitem [{\citenamefont {Bououdina}\ \emph {et~al.}(2019)\citenamefont
  {Bououdina}, \citenamefont {Dakhel}, \citenamefont {Jaafar}, \citenamefont
  {Dai},\ and\ \citenamefont {Song}}]{bououdina2019}%
  \BibitemOpen
  \bibfield  {author} {\bibinfo {author} {\bibfnamefont {M.}~\bibnamefont
  {Bououdina}}, \bibinfo {author} {\bibfnamefont {A.}~\bibnamefont {Dakhel}},
  \bibinfo {author} {\bibfnamefont {A.}~\bibnamefont {Jaafar}}, \bibinfo
  {author} {\bibfnamefont {J.}~\bibnamefont {Dai}}, \ and\ \bibinfo {author}
  {\bibfnamefont {Y.}~\bibnamefont {Song}},\ }\href
  {https://linkinghub.elsevier.com/retrieve/pii/S092583881833980X} {\bibfield
  {journal} {\bibinfo  {journal} {J. Alloys Compd.}\ }\textbf {\bibinfo
  {volume} {776}},\ \bibinfo {pages} {575} (\bibinfo {year}
  {2019})}\BibitemShut {NoStop}%
\bibitem [{\citenamefont {{Van de Walle}}\ and\ \citenamefont
  {Neugebauer}(2003)}]{vandewalle2003}%
  \BibitemOpen
  \bibfield  {author} {\bibinfo {author} {\bibfnamefont {C.~G.}\ \bibnamefont
  {{Van de Walle}}}\ and\ \bibinfo {author} {\bibfnamefont {J.}~\bibnamefont
  {Neugebauer}},\ }\href {http://www.nature.com/articles/nature01665}
  {\bibfield  {journal} {\bibinfo  {journal} {Nature}\ }\textbf {\bibinfo
  {volume} {423}},\ \bibinfo {pages} {626} (\bibinfo {year}
  {2003})}\BibitemShut {NoStop}%
\bibitem [{\citenamefont {Spaldin}\ and\ \citenamefont
  {Ramesh}(2019)}]{spaldin2019}%
  \BibitemOpen
  \bibfield  {author} {\bibinfo {author} {\bibfnamefont {N.~A.}\ \bibnamefont
  {Spaldin}}\ and\ \bibinfo {author} {\bibfnamefont {R.}~\bibnamefont
  {Ramesh}},\ }\href {http://www.nature.com/articles/s41563-018-0275-2}
  {\bibfield  {journal} {\bibinfo  {journal} {Nat. Mater.}\ }\textbf {\bibinfo
  {volume} {18}},\ \bibinfo {pages} {203} (\bibinfo {year} {2019})}\BibitemShut
  {NoStop}%
\bibitem [{\citenamefont {K\i{}l\i{\c c}}\ and\ \citenamefont
  {Zunger}(2002)}]{kilic2002}%
  \BibitemOpen
  \bibfield  {author} {\bibinfo {author} {\bibfnamefont {{\c C}.}~\bibnamefont
  {K\i{}l\i{\c c}}}\ and\ \bibinfo {author} {\bibfnamefont {A.}~\bibnamefont
  {Zunger}},\ }\href {http://aip.scitation.org/doi/10.1063/1.1482783}
  {\bibfield  {journal} {\bibinfo  {journal} {Appl. Phys. Lett.}\ }\textbf
  {\bibinfo {volume} {81}},\ \bibinfo {pages} {73} (\bibinfo {year}
  {2002})}\BibitemShut {NoStop}%
\bibitem [{\citenamefont {Cox}\ \emph {et~al.}(2001)\citenamefont {Cox},
  \citenamefont {Davis}, \citenamefont {Cottrell}, \citenamefont {King},
  \citenamefont {Lord}, \citenamefont {Gil}, \citenamefont {Alberto},
  \citenamefont {Vil{\~a}o}, \citenamefont {Piroto~Duarte}, \citenamefont
  {{Ayres de Campos}}, \citenamefont {Weidinger}, \citenamefont {Lichti},\ and\
  \citenamefont {Irvine}}]{cox2001}%
  \BibitemOpen
  \bibfield  {author} {\bibinfo {author} {\bibfnamefont {S.~F.~J.}\
  \bibnamefont {Cox}}, \bibinfo {author} {\bibfnamefont {E.~A.}\ \bibnamefont
  {Davis}}, \bibinfo {author} {\bibfnamefont {S.~P.}\ \bibnamefont {Cottrell}},
  \bibinfo {author} {\bibfnamefont {P.~J.~C.}\ \bibnamefont {King}}, \bibinfo
  {author} {\bibfnamefont {J.~S.}\ \bibnamefont {Lord}}, \bibinfo {author}
  {\bibfnamefont {J.~M.}\ \bibnamefont {Gil}}, \bibinfo {author} {\bibfnamefont
  {H.~V.}\ \bibnamefont {Alberto}}, \bibinfo {author} {\bibfnamefont {R.~C.}\
  \bibnamefont {Vil{\~a}o}}, \bibinfo {author} {\bibfnamefont {J.}~\bibnamefont
  {Piroto~Duarte}}, \bibinfo {author} {\bibfnamefont {N.}~\bibnamefont {{Ayres
  de Campos}}}, \bibinfo {author} {\bibfnamefont {A.}~\bibnamefont
  {Weidinger}}, \bibinfo {author} {\bibfnamefont {R.~L.}\ \bibnamefont
  {Lichti}}, \ and\ \bibinfo {author} {\bibfnamefont {S.~J.~C.}\ \bibnamefont
  {Irvine}},\ }\href {https://link.aps.org/doi/10.1103/PhysRevLett.86.2601}
  {\bibfield  {journal} {\bibinfo  {journal} {Phys. Rev. Lett.}\ }\textbf
  {\bibinfo {volume} {86}},\ \bibinfo {pages} {2601} (\bibinfo {year}
  {2001})}\BibitemShut {NoStop}%
\bibitem [{\citenamefont {Peacock}\ and\ \citenamefont
  {Robertson}(2003)}]{peacock2003}%
  \BibitemOpen
  \bibfield  {author} {\bibinfo {author} {\bibfnamefont {P.~W.}\ \bibnamefont
  {Peacock}}\ and\ \bibinfo {author} {\bibfnamefont {J.}~\bibnamefont
  {Robertson}},\ }\href {http://aip.scitation.org/doi/10.1063/1.1609245}
  {\bibfield  {journal} {\bibinfo  {journal} {Appl. Phys. Lett.}\ }\textbf
  {\bibinfo {volume} {83}},\ \bibinfo {pages} {2025} (\bibinfo {year}
  {2003})}\BibitemShut {NoStop}%
\bibitem [{\citenamefont {Xiong}\ \emph {et~al.}(2007)\citenamefont {Xiong},
  \citenamefont {Robertson},\ and\ \citenamefont {Clark}}]{xiong2007}%
  \BibitemOpen
  \bibfield  {author} {\bibinfo {author} {\bibfnamefont {K.}~\bibnamefont
  {Xiong}}, \bibinfo {author} {\bibfnamefont {J.}~\bibnamefont {Robertson}}, \
  and\ \bibinfo {author} {\bibfnamefont {S.~J.}\ \bibnamefont {Clark}},\ }\href
  {http://aip.scitation.org/doi/10.1063/1.2798910} {\bibfield  {journal}
  {\bibinfo  {journal} {J. Appl. Phys.}\ }\textbf {\bibinfo {volume} {102}},\
  \bibinfo {pages} {083710} (\bibinfo {year} {2007})}\BibitemShut {NoStop}%
\bibitem [{\citenamefont {Momma}\ and\ \citenamefont
  {Izumi}(2011)}]{Momma2011}%
  \BibitemOpen
  \bibfield  {author} {\bibinfo {author} {\bibfnamefont {K.}~\bibnamefont
  {Momma}}\ and\ \bibinfo {author} {\bibfnamefont {F.}~\bibnamefont {Izumi}},\
  }\href {http://scripts.iucr.org/cgi-bin/paper?S0021889811038970} {\bibfield
  {journal} {\bibinfo  {journal} {J. Appl. Crystallogr.}\ }\textbf {\bibinfo
  {volume} {44}},\ \bibinfo {pages} {1272} (\bibinfo {year}
  {2011})}\BibitemShut {NoStop}%
\bibitem [{Note3()}]{Note3}%
  \BibitemOpen
  \bibinfo {note} {A small sample misalignment of less than $\SI {1}{\degree }$
  seems to cause a further splitting of the line at times longer than \SI
  {1}{\micro s}. This sub-splitting affects the various multiplet frequencies
  differently and has a small temperature dependence which is not fully
  understood. Limiting the analysis (and the depicted FT) to the first \SI
  {1}{\micro s} effectively smooths out the substructure; however a separate
  relaxation rate for the E1 doublet is required and attributed to broadening
  caused by this misalignment sub-splitting.}\BibitemShut {Stop}%
\bibitem [{\citenamefont {Meier}(1982)}]{meier1982}%
  \BibitemOpen
  \bibfield  {author} {\bibinfo {author} {\bibfnamefont {P.~F.}\ \bibnamefont
  {Meier}},\ }\href@noop {} {\bibfield  {journal} {\bibinfo  {journal} {Phys.
  Rev. A}\ }\textbf {\bibinfo {volume} {25}},\ \bibinfo {pages} {1287}
  (\bibinfo {year} {1982})}\BibitemShut {NoStop}%
\bibitem [{\citenamefont {Dehn}\ \emph {et~al.}(2016)\citenamefont {Dehn},
  \citenamefont {Arseneau}, \citenamefont {B{\"o}ni}, \citenamefont {Bridges},
  \citenamefont {Buck}, \citenamefont {Cortie}, \citenamefont {Fleming},
  \citenamefont {Kelly}, \citenamefont {MacFarlane}, \citenamefont
  {MacLachlan}, \citenamefont {McFadden}, \citenamefont {Morris}, \citenamefont
  {Wang}, \citenamefont {Xiao}, \citenamefont {Zamarion},\ and\ \citenamefont
  {Kiefl}}]{dehn2016}%
  \BibitemOpen
  \bibfield  {author} {\bibinfo {author} {\bibfnamefont {M.~H.}\ \bibnamefont
  {Dehn}}, \bibinfo {author} {\bibfnamefont {D.~J.}\ \bibnamefont {Arseneau}},
  \bibinfo {author} {\bibfnamefont {P.}~\bibnamefont {B{\"o}ni}}, \bibinfo
  {author} {\bibfnamefont {M.~D.}\ \bibnamefont {Bridges}}, \bibinfo {author}
  {\bibfnamefont {T.}~\bibnamefont {Buck}}, \bibinfo {author} {\bibfnamefont
  {D.~L.}\ \bibnamefont {Cortie}}, \bibinfo {author} {\bibfnamefont {D.~G.}\
  \bibnamefont {Fleming}}, \bibinfo {author} {\bibfnamefont {J.~A.}\
  \bibnamefont {Kelly}}, \bibinfo {author} {\bibfnamefont {W.~A.}\ \bibnamefont
  {MacFarlane}}, \bibinfo {author} {\bibfnamefont {M.~J.}\ \bibnamefont
  {MacLachlan}}, \bibinfo {author} {\bibfnamefont {R.~M.~L.}\ \bibnamefont
  {McFadden}}, \bibinfo {author} {\bibfnamefont {G.~D.}\ \bibnamefont
  {Morris}}, \bibinfo {author} {\bibfnamefont {P.-X.}\ \bibnamefont {Wang}},
  \bibinfo {author} {\bibfnamefont {J.}~\bibnamefont {Xiao}}, \bibinfo {author}
  {\bibfnamefont {V.~M.}\ \bibnamefont {Zamarion}}, \ and\ \bibinfo {author}
  {\bibfnamefont {R.~F.}\ \bibnamefont {Kiefl}},\ }\href
  {http://aip.scitation.org/doi/10.1063/1.4967460} {\bibfield  {journal}
  {\bibinfo  {journal} {J. Chem. Phys.}\ }\textbf {\bibinfo {volume} {145}},\
  \bibinfo {pages} {181102} (\bibinfo {year} {2016})}\BibitemShut {NoStop}%
\bibitem [{\citenamefont {Dehn}\ \emph {et~al.}(2018)\citenamefont {Dehn},
  \citenamefont {Fleming}, \citenamefont {MacFarlane}, \citenamefont
  {MacLachlan}, \citenamefont {Zamarion},\ and\ \citenamefont
  {Kiefl}}]{dehn2018a}%
  \BibitemOpen
  \bibfield  {author} {\bibinfo {author} {\bibfnamefont {M.~H.}\ \bibnamefont
  {Dehn}}, \bibinfo {author} {\bibfnamefont {D.~G.}\ \bibnamefont {Fleming}},
  \bibinfo {author} {\bibfnamefont {W.~A.}\ \bibnamefont {MacFarlane}},
  \bibinfo {author} {\bibfnamefont {M.~J.}\ \bibnamefont {MacLachlan}},
  \bibinfo {author} {\bibfnamefont {V.~M.}\ \bibnamefont {Zamarion}}, \ and\
  \bibinfo {author} {\bibfnamefont {R.~F.}\ \bibnamefont {Kiefl}},\ }\href
  {http://journals.jps.jp/doi/10.7566/JPSCP.21.011032} {\bibfield  {journal}
  {\bibinfo  {journal} {Proc 14th Int Conf Muon Spin Rotat. Relax. Reson. J.
  Phys. Soc. Jpn.}\ }\textbf {\bibinfo {volume} {21}},\ \bibinfo {pages}
  {011032} (\bibinfo {year} {2018})}\BibitemShut {NoStop}%
\bibitem [{\citenamefont {Perdew}\ and\ \citenamefont
  {Zunger}(1981)}]{perdew1981}%
  \BibitemOpen
  \bibfield  {author} {\bibinfo {author} {\bibfnamefont {J.~P.}\ \bibnamefont
  {Perdew}}\ and\ \bibinfo {author} {\bibfnamefont {A.}~\bibnamefont
  {Zunger}},\ }\href {https://link.aps.org/doi/10.1103/PhysRevB.23.5048}
  {\bibfield  {journal} {\bibinfo  {journal} {Phys. Rev. B}\ }\textbf {\bibinfo
  {volume} {23}},\ \bibinfo {pages} {5048} (\bibinfo {year}
  {1981})}\BibitemShut {NoStop}%
\bibitem [{\citenamefont {Dudarev}\ \emph {et~al.}(1998)\citenamefont
  {Dudarev}, \citenamefont {Botton}, \citenamefont {Savrasov}, \citenamefont
  {Humphreys},\ and\ \citenamefont {Sutton}}]{dudarev1998}%
  \BibitemOpen
  \bibfield  {author} {\bibinfo {author} {\bibfnamefont {S.~L.}\ \bibnamefont
  {Dudarev}}, \bibinfo {author} {\bibfnamefont {G.~A.}\ \bibnamefont {Botton}},
  \bibinfo {author} {\bibfnamefont {S.~Y.}\ \bibnamefont {Savrasov}}, \bibinfo
  {author} {\bibfnamefont {C.~J.}\ \bibnamefont {Humphreys}}, \ and\ \bibinfo
  {author} {\bibfnamefont {A.~P.}\ \bibnamefont {Sutton}},\ }\href
  {https://link.aps.org/doi/10.1103/PhysRevB.57.1505} {\bibfield  {journal}
  {\bibinfo  {journal} {Phys. Rev. B}\ }\textbf {\bibinfo {volume} {57}},\
  \bibinfo {pages} {1505} (\bibinfo {year} {1998})}\BibitemShut {NoStop}%
\bibitem [{\citenamefont {Shi}\ \emph {et~al.}(2009)\citenamefont {Shi},
  \citenamefont {Wysocki},\ and\ \citenamefont {Belashchenko}}]{shi2009}%
  \BibitemOpen
  \bibfield  {author} {\bibinfo {author} {\bibfnamefont {S.}~\bibnamefont
  {Shi}}, \bibinfo {author} {\bibfnamefont {A.~L.}\ \bibnamefont {Wysocki}}, \
  and\ \bibinfo {author} {\bibfnamefont {K.~D.}\ \bibnamefont {Belashchenko}},\
  }\href {https://link.aps.org/doi/10.1103/PhysRevB.79.104404} {\bibfield
  {journal} {\bibinfo  {journal} {Phys. Rev. B}\ }\textbf {\bibinfo {volume}
  {79}} (\bibinfo {year} {2009})}\BibitemShut {NoStop}%
\bibitem [{\citenamefont {Monkhorst}\ and\ \citenamefont
  {Pack}(1976)}]{monkhorst1976}%
  \BibitemOpen
  \bibfield  {author} {\bibinfo {author} {\bibfnamefont {H.~J.}\ \bibnamefont
  {Monkhorst}}\ and\ \bibinfo {author} {\bibfnamefont {J.~D.}\ \bibnamefont
  {Pack}},\ }\href {https://link.aps.org/doi/10.1103/PhysRevB.13.5188}
  {\bibfield  {journal} {\bibinfo  {journal} {Phys. Rev. B}\ }\textbf {\bibinfo
  {volume} {13}},\ \bibinfo {pages} {5188} (\bibinfo {year}
  {1976})}\BibitemShut {NoStop}%
\bibitem [{\citenamefont {Bl{\"o}chl}(1994)}]{blochl1994}%
  \BibitemOpen
  \bibfield  {author} {\bibinfo {author} {\bibfnamefont {P.~E.}\ \bibnamefont
  {Bl{\"o}chl}},\ }\href {https://link.aps.org/doi/10.1103/PhysRevB.50.17953}
  {\bibfield  {journal} {\bibinfo  {journal} {Phys. Rev. B}\ }\textbf {\bibinfo
  {volume} {50}},\ \bibinfo {pages} {17953} (\bibinfo {year}
  {1994})}\BibitemShut {NoStop}%
\bibitem [{\citenamefont {Kresse}\ and\ \citenamefont
  {Joubert}(1999)}]{kresse1999}%
  \BibitemOpen
  \bibfield  {author} {\bibinfo {author} {\bibfnamefont {G.}~\bibnamefont
  {Kresse}}\ and\ \bibinfo {author} {\bibfnamefont {D.}~\bibnamefont
  {Joubert}},\ }\href {https://link.aps.org/doi/10.1103/PhysRevB.59.1758}
  {\bibfield  {journal} {\bibinfo  {journal} {Phys. Rev. B}\ }\textbf {\bibinfo
  {volume} {59}},\ \bibinfo {pages} {1758} (\bibinfo {year}
  {1999})}\BibitemShut {NoStop}%
\bibitem [{Note4()}]{Note4}%
  \BibitemOpen
  \bibinfo {note} {The Cr, O and H PAWs are dated: 23$\protect \mathrm {^{rd}}$
  Jul. 2007, 22$\protect \mathrm {^{nd}}$ Mar. 2012 and 6$\protect \mathrm
  {^{th}}$ May 1998 respectively}\BibitemShut {NoStop}%
\end{thebibliography}%


\end{document}